\documentclass{article}

\usepackage{fullpage}
\usepackage{graphicx}
\usepackage{hyperref}
\usepackage{bbm}
\usepackage{amsmath}
\usepackage{amssymb}
\usepackage{xparse}
\usepackage{natbib}
\usepackage{enumitem}
\usepackage[T1]{fontenc}
\bibpunct{[}{]}{,}{n}{,}{;}

\newcounter{arXivVersion}
\def\clap#1{\hbox to 0pt{\hss#1\hss}}

\def\mathrlap{\mathpalette\mathrlapinternal}

\def\mathrlapinternal#1#2{%
\rlap{$\mathsurround=0pt#1{#2}$}}

\title{Anomaly Detection in Networks with Application to Financial Transaction Networks}
\author{
{Andrew Elliott\footnote{Corresponding author: 
ande.elliott@gmail.com}
}
\\
{The Alan Turing Institute; Dept. of Statistics, University of Oxford}
  \and
{Mihai Cucuringu}
\\
{The Alan Turing Institute; Dept. of Statistics, University of Oxford}
  \and
{Milton Martinez Luaces}
\\
{Accenture Plc}
  \and
{Paul Reidy}
\\
{Accenture Plc}
\and
{Gesine Reinert}
\\
{The Alan Turing Institute; Dept. of Statistics, University of Oxford}
}

\begin{document}

\maketitle

\begin{abstract} 
{
Detecting financial fraud is a global challenge, with financial fraud losses estimated to at least GBP 768 million in the UK in 2015. Financial transactions can be represented as networks, with accounts as nodes and a directed weighted edge between a pair of nodes when one transfers money to the other. The weight of the edge denotes the transaction amount between the two parties. The task considered in this paper can be cast as a special instance of detecting anomalies in networks.

\smallskip 
Examples of anomalies in networks depend on the type of network, but may
include structures such as long paths of large transaction amounts, rings of large payments, and cliques of accounts which send money to each other. There are a number of methods available which detect such specific structures in networks. In this paper, we introduce a method which is able to detect previously unspecified anomalies in networks. The method is mainly based on a combination of features from network comparison and spectral analysis. Combining network comparison and spectral analysis tools with various local statistics, altogether give rise to a total of 140 main features. We then use a simple sum ({\sc Feature sum}), as well as a random forest  method ({\sc Random forest}), in order to classify nodes as normal or anomalous.

\smallskip 
We test the method first on synthetic networks which we generated, and second, on a set of synthetic networks generated without the methods team knowing the generating mechanism during the method development stage. The first set of synthetic networks was split into a training, respectively  test, set of size 70\%, respectively 30\%, of the networks. 
The resulting classifier was then applied to the second set of synthetic
networks as well.

\smallskip 
We compare our method with Oddball, a widely used method for anomaly detection in networks, as well as to random classification. While Oddball outperforms random classification, both the  {\sc Feature sum} and {\sc Random forest} methods outperform Oddball. On the test set, the {\sc Random forest} outperforms {\sc Feature sum}, whereas on the second  synthetic data set, initially {\sc Feature sum} tends to pick
up more anomalies compared to {\sc Random forest}, with this behaviour reversing for lower-scoring anomalies. In all cases, the top 2 percent of flagged anomalies contained, on average, over 90 percent of the planted anomalies.  
}
\newline
{{\bf Keywords:} {\it Anomaly Detection, Financial Transaction Networks, Network Comparison, Spectral Methods}}
\end{abstract}

\section{Introduction}

Financial fraud is a global challenge.  
According to \cite{fraudthefacts}, financial fraud losses  in the UK payment industry across payment cards, remote banking and cheques amounted to  \pounds 768.8 million in 2016, an increase of 2\% compared to 2015.

Some of this fraudulent behaviour, such as money laundering, may manifest itself through
unusual patterns in financial transaction networks. In such networks, customers are nodes, and two nodes $u$ and $v$ are linked by a directed edge if there is a money transfer from $u$ to $v$; the edge is
annotated with the transferred amount as its weight. Unusual patterns could
include long paths of transactions with large weights (denoted hereafter by \textit{long heavy
paths}) and large cliques. The patterns of financial fraud are highly variable,
and many patterns are likely to remain undetected to date. Moreover, the number of cases is small compared to the number of legitimate transactions, and fraudsters aim to change their behaviour to avoid detection. Similar challenges arise in networks for cyber security~\cite{neilsubgraphScan}. In this work, we develop a method to uncover such unknown patterns.

Fraud detection and the related problem of anomaly detection are well
studied problems; some reviews can be found in
\cite{akoglu2015graph,bolton2002statistical,delamaire2009credit,ngai2011application,savage2014anomaly,west2014intelligent}. 
Popular techniques which have been applied to the task of uncovering money laundering and crime span a wide range of topics, including  autoencoders~\cite{paulaetal,schreyer2017detection},
SVMs~\cite{tang2005developing}, social network analysis~\cite{drezewski20128} and rule-based approaches~\cite{senator1995financial}. Furthermore, anomaly detection is performed on many different types of data including (but not limited to), unstructured data~\cite{zhang2003applying}, financial records~\cite{paulaetal}, social network data~\cite{drezewski20128}, time series data~\cite{larik2011606} and investment bank
data~\cite{le2010data}. Some anomaly detection methods look for attempts to conceal
behaviour as normal, such as \cite{chen2012community} and \cite{EberleHolder1}.

Several approaches for anomaly detection exploit the network information. For
example, \cite{fronzetticolladon201749} uses network statistics and a logit model to
detect fraud in a factor house.
In a different line of work, \cite{Savage2016DetectionOM} starts with a weighted network based on multiple evidence sources, constructs a $3$-hop snowball sampled network, and uses features of these networks in a machine learning framework to uncover suspicious transactions.

Many general network anomaly detection methods rely on  community detection in networks.
In \cite{chen2012community}, the authors declare an anomaly if the temporal evolution of network communities change sufficiently, whereas \cite{gao2013multi} compares the spectral embedding (a form of community
detection) of individuals in different data sources in a cross data source
embedding, declaring an anomaly if the embeddings deviate substantially.
In a related fashion, \cite{larik2011606} clusters accounts and then constructs a score based on the deviations of an account from the mean performance in the cluster. 
Furthermore, network comparison methods have been employed for anomaly detection. 
For example, the approach in~\cite{pincombe2005anomaly} uses network comparison measures, and compares each network in a time series to their immediate neighbours, fitting an ARMA process to the resultant time series. The authors identify anomalies by looking for deviations from the model.

There is also a large literature on spectral localisation and general spectral approaches for anomaly detection in networks. 
Spectral localisation is defined as the phenomenon in which a large amount of the mass of an eigenvector is placed on a small number of its entries \cite{belgium,eigenvector_localization_networks}. When considering eigenvectors of various matrix operators derived from the graph adjacency matrix, the corresponding nodes on which the eigenvector localised can be regarded as 
{different} when compared to the rest of the nodes in the network, and constitute good candidates for the anomaly detection task.
For example,
Miller~\emph{et al.}~\cite{miller2010subgraph, miller2013efficient, miller2011eigenspace, miller2011matched, miller2015spectral,miller2010toward}
develop a series of methods to uncover anomalies using spectral features of the
modularity matrix. The approach that is most relevant to this work arises 
in~\cite{miller2010subgraph}, in which eigenvectors are assessed by their deviation in the z-score of the $\ell_1$ norm.  In~\cite{miller2015spectral}, the authors extend these methods to use the $\ell_1$ norm for eigenvectors of sparse PCA, which performs well at the cost of being more computationally intensive. A different spectral approach, proposed by Tong and Lin~\cite{tonglin}, is based on matrix factorisation in bipartite networks and leverages the intuition that the nodes and edges which are badly represented by the factorisation should be considered as anomalies.

In this paper, the standard network anomaly detection method which we shall employ for a comparison with our results is Oddball, by Akoglu~\emph{et al.}~\cite{akoglu2010oddball}, which exploits power-law like relations between different network node level statistics. This is a widely used method which provides publicly available code; it detects several types of known local anomalies, see Subsection~\ref{oddball} for full details.

\begin{figure}[hbt!]
\centering
\includegraphics[width=0.63\textwidth]{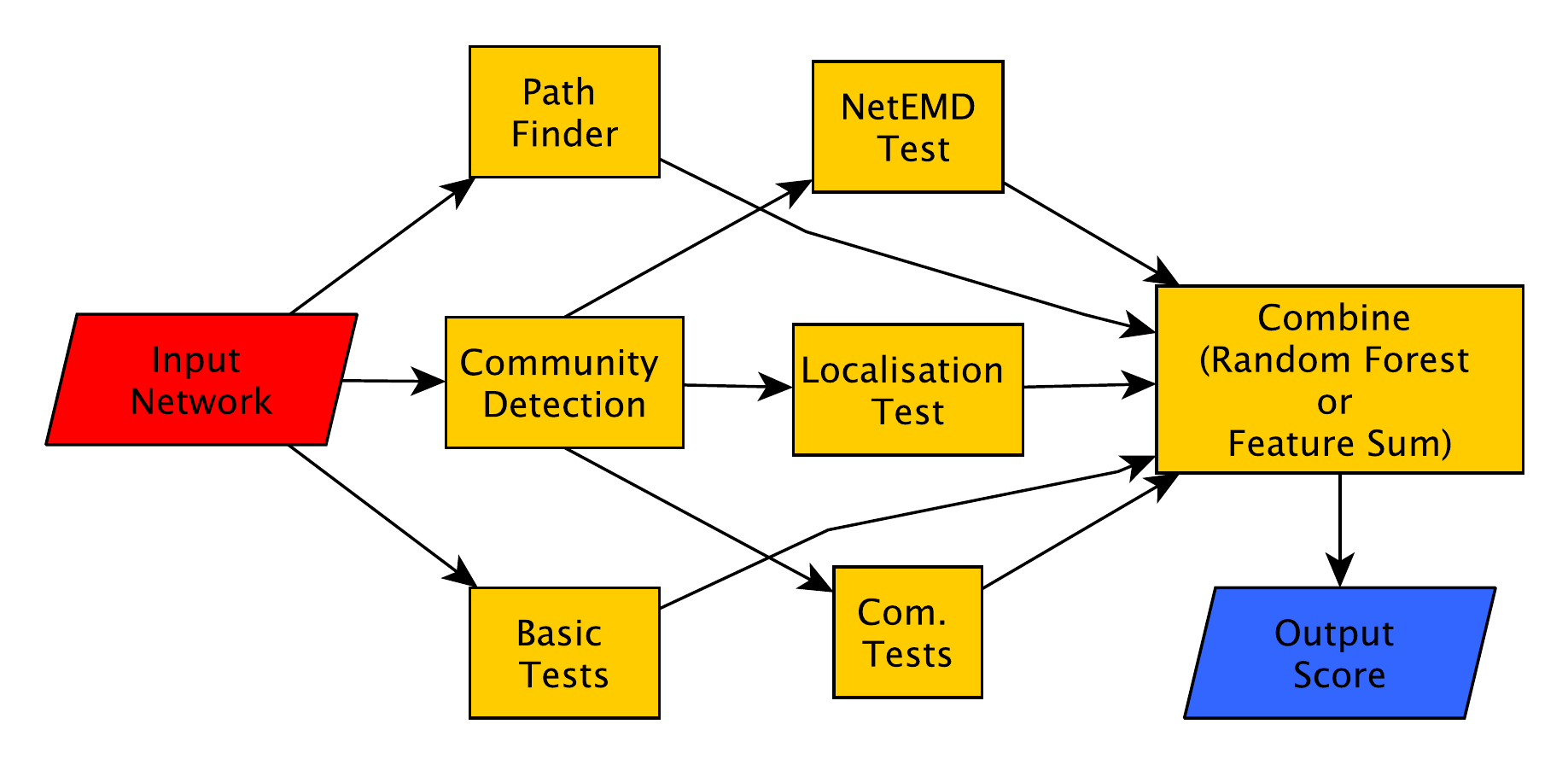}
\caption{
\label{fig:process} 
The complete pipeline; red parallelograms represent the input,
    blue parallelograms represent the output scores, and the yellow rectangles
    represent processes which are defined in the following sections: 
{\bf Basic Tests }
Sec.~\ref{sec:basicDetection}, 
{\bf Community Detection 
}
Sec.~\ref{sec:communityDetection},
{\bf Path Finder 
}
Sec.~\ref{sec:PathFinder},
{\bf Com. Tests 
}
Sec.~\ref{sec:communityDetection},
{\bf Localisation Test 
}
Sec.~\ref{sec:SpecLocal},
{\bf NetEMD Test }
Sec.~\ref{sec:NetEMDMod},
{\bf Combine ({\sc Random forest}/{\sc Feature sum}) }
Sec.~\ref{sec:combineScore}.}
\end{figure}

Our method is based on a combination of several approaches, namely an extension of the NetEMD network comparison method from \cite{wegner2017identifying}, an application of new and existing spectral localisation statistics, as well as several other modules as detailed in our pipeline in Figure~\ref{fig:process}. We combine these two main approaches with community detection, which is used to decompose the network into smaller but denser subnetworks, thus allowing for parallel computation of the code and rendering the approach computationally scalable. Community detection has additional advantages, as it increases the detectability of the embedded structures in the resultant subnetworks (see Sec.~\ref{sec:communityDetection} for details), and it allows for the assignment of an anomaly score directly to each of the communities. 
However, the split of the network into communities has as drawback the fact that long paths may be cut (i.e., split over two or more clusters), and thus no longer detectable. To alleviate this risk, we modify our community detection approach in an attempt to avoid cutting through long heavy paths (see Sec.~\ref{sec:communityDetection} for details), and we enrich our method by a component which is specifically designed to find paths in networks.

The two sets of spectral features developed in this paper are,
firstly, an adaptation of the network comparison framework NetEMD to use
eigenvectors. The second set of features is based on powers of the
entries of eigenvectors, as well as their signs and magnitudes, see  Sec.~\ref{sec:SpecLocal}. These two sets of features may be of
interest beyond the application considered in this paper.

Altogether, our combined method calculates 140 main features for each network, and uses these features to pinpoint anomalies. These features are processed in two different ways. Our {\sc{Feature sum}} method aggregates the features as an unweighted sum. Our {\sc{Random forest}} method is constructed by performing training and feature selection on the $140$ main features using a generating model for networks with randomly added anomalies, which is developed in Sec.~\ref{weightERgraph}. As these planted anomalies are inspired by anti-money laundering, they focus on structures with heavy edges. For the synthetic models, we consider parameter regimes which ensure that in our simulated networks without anomalies, the expected count for each particular anomaly would be less than 1.

To understand what is driving our performance, we also explore the features which are selected by the random forest. We find that different parameter regions favour different features, with features based on geometric average of edge weights and short paths dominating the space of selected features.

We apply our method to a test data set derived from our model, and to a data set which is provided by the Accenture team. The first data set is generated in the same way as the data set which is used to train the {\sc Random forest} method, while the second data set is generated using a different  model. On both data sets, our methods outperform random classification, and they perform favourably against Oddball from \cite{akoglu2010oddball}. While the {\sc Random forest} method outperforms the {\sc Feature sum} method, in the second data set the {\sc Random forest} is outperformed by the {\sc Feature sum} method. Furthermore, on the second data set, both methods uncover on average $92.3$\% of the anomalous nodes within the top $1024$ nodes, in a network of size $n=55000$. 

Finally, we investigate the nodes that are identified as anomalies but were not inserted in the graph in the Accenture model, using the {\sc Random Forest} method. We empirically observe that the structure around many of the nodes have a heavy structure, including a reasonable number of heavy connections to the node of interest, and in some cases, a flow structure. 
Therefore, anomalies are detected without having prior explicit knowledge of their structure.

This paper is structured as follows. In Section~\ref{models}, the two  synthetic network data generation methods are described.  Section~\ref{sec:overviewOfPipeline} develops our new anomaly detection framework. We first perform a set of basic tests which look for deviations in the degree or the local edge weights of the nodes in the network, as  described in Section~\ref{sec:nodeLevelStats}. In  Section~\ref{sec:communityDetection}, the network is divided into communities, within which we look for anomalies using modules relying on (i)  basic features of the communities (Section~\ref{sec:communityDetection}), (ii) localisation (Section~\ref{sec:SpecLocal}), and (iii) NetEMD (Section~\ref{sec:NetEMDMod}). We complement these methods with an approach that searches for paths and path-like structures directly in the whole network (Section~\ref{sec:PathFinder}), before finally combining the individual contributions either via a random forest, as described in Section~\ref{sec:combineScore}, or by summing up all the features, without any prior feature selection. The results on the two synthetic data sets can be found in Section~\ref{results}. The paper ends with a discussion of the results in Section~\ref{sec:discussion}. The appendix contains implementation details, calculations regarding regions of detectability, and a proof regarding the validity of an eigenvector score. It also contains a list of all the features considered, additional details on the feature selection step, in-sample performance for the {\sc Random Forest}, and more detailed Oddball results.

The code relies on NetworkX~\cite{networkx}, Scikit-learn~\cite{sklearn},
Cython~\cite{cython}, Matplotlib~\cite{matplotlib}, Scipy~\cite{scipy}, NumPy~\cite{numpy}, Rpy2~\cite{rpy2} and the
NetEMD network comparison package~\cite{netemdgithub} from~\cite{netemdSoftwarePaper}.

\ifnum\value{arXivVersion}>0 {
Availability: The software for this paper can be found at \url{https://sites.google.com/site/elliottande/anomalydetection}.
} \else {
Availability: The software for this paper can be found at {https://sites.google.com/site/ elliottande/anomalydetection}.
}\fi

\section{Network Models and Network Data}\label{models}

In the following, {\it at random} means {\it uniformly at random}, unless otherwise stated, 
and $p_{val}$ represent a $p$-value. 

\subsection{Synthetic Data Set: Weighted ER Network}
\label{weightERgraph}

To develop and measure the success of our methodology, two network models
with known embedded anomalies are employed. 
The first model guided the  method development stage  which lead to our
pipeline (see  Fig.~\ref{fig:process}). 
This synthetic model is a weighted directed Erd\H{o}s-R\'enyi (ER) random graph model, in which small structures with unusually large weights are planted at random. 
The edge weights of the underlying ER network are drawn independently and uniformly from the interval $(0,1)$, and the distribution of a  non-anomalous weighted edge $ W_{ij}$  is given by
    $$ W_{ij} \sim \text{Bern}(p)\text{Uni}(0,1), $$
where $\text{Bern}(p)$ is a Bernoulli trial with success probability $p$, and
$\text{Uni}(0,1)$ is a uniform random draw from the interval $(0,1)$ which is independent of the network and of all other draws.

\begin{figure}[ht!]
\centering
\includegraphics[width=0.5\textwidth]{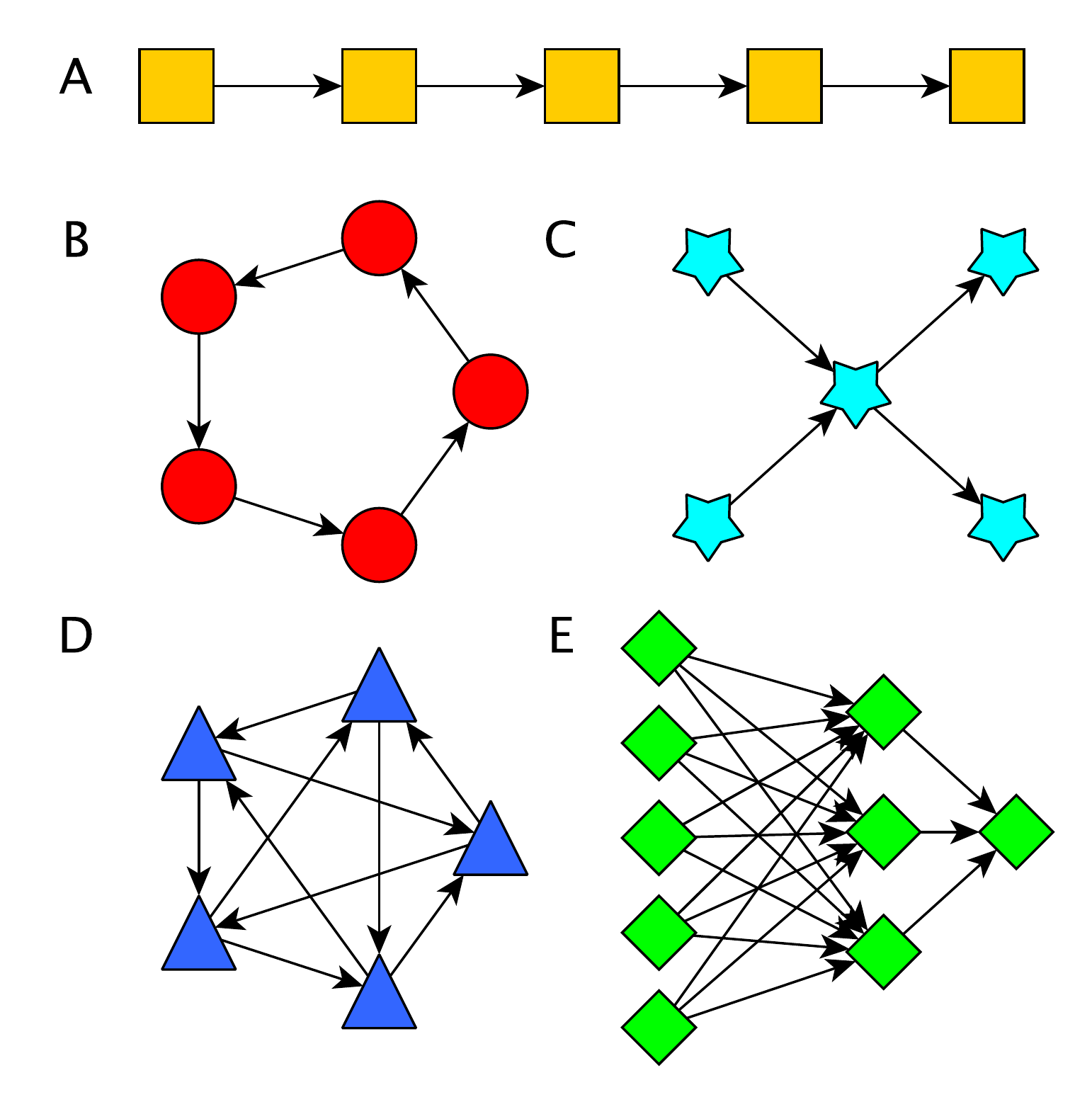}
\caption{
\label{fig:anomaliesFramework} 
    Diagram of the embedded anomalies.  
    {\bf (A) Top yellow squares} - Directed path.
    {\bf (B) Middle left red circles} - Ring structure.
    {\bf (C) Middle right light blue star} - Star structure with a directed flow from left to right. 
    {\bf (D) Bottom left blue triangles} - Clique structure.
    {\bf (E) Bottom right green diamonds} - Directed multipartite structure showing flow of money/information from left to right. }
\end{figure}

A random number of randomly sized anomalous structures of five different types are then embedded at random locations throughout the network. 
The considered five types are paths, rings, stars, cliques and a particular directed multipartite network (which we refer to as {\it trees}, for simplicity), as shown in Fig.~\ref{fig:anomaliesFramework}. 

The embedding procedure of such structures is as follows. For a path of size $k$, a directed series of edges $(\gamma_1,\gamma_2),...,(\gamma_{k-1},\gamma_k)$ is embedded, with an ordered set of distinct nodes ($\gamma_1,..,\gamma_k$) selected uniformly at random. For a ring, we additionally place a directed edge between $\gamma_k$ and $\gamma_1$. For the star,
(Fig.~\ref{fig:anomaliesFramework}C),  $k$ distinct nodes are chosen  uniformly at random, and one of them is randomly selected to be the centre
node. In order to make the network directed, the direction of each edge is selected uniformly at random. 
For the clique, (Fig.~\ref{fig:anomaliesFramework}D), 
we select $k$ distinct nodes uniformly at random, and choose a random direction between each pair of nodes.

The final embedded structure, slightly more complex, is a special case of a directed multipartite network (tree), as shown in  Fig.~\ref{fig:anomaliesFramework}E, and is inspired by the dynamics and flow of information or money in a real-world system. In this setting, we consider a one-dimensional ordered series of independent sets (to which we refer to as shells), with nodes in a given shell passing input to all (or a subset of) nodes in the neighbouring shell upstream (to the right in Fig.~\ref{fig:anomaliesFramework}E). In our experiments, 
we place $5$ nodes in the leftmost shell, which are connected by directed edges to all of the $3$ nodes in middle shell,  which in turn are full connected to the $1$ node in the rightmost shell. For brevity in this document and a slight abuse of notation, we shall refer to this multipartite network as a `tree'. For simplicity, the directional flow of all (existing) edges across shells is from left to right, without exception. Such a structure could be representative, for example, of a hierarchical structure in a criminal organisation, with resources flowing from the lower-ranked members towards the head of the organisation.

To construct a heavy structure on the planted anomalies, rather than drawing the weights on the edges from $\text{Uni}(0,1)$ as for the other edges in the network, we draw them independently from $\text{Uni}(w,1)$, where $w$ is a parameter to the model. 
The construction of potential edges which are not part of the embedded structure (i.e., edges in the clique which are in the direction opposite to that selected for an anomalous edge) are subject to the standard edge construction. 

The detailed steps of our network construction is as follows. 
\begin{enumerate}
    \item Generate a directed ER network with $n$ nodes and probability of connection $p$.
    \item Add weights from $\text{Uni}(0,1)$ to each of the existing edges.
    \item Draw the number of anomalies to embed  
    uniformly at random from $[5,...,20]$.
    \item Select the type of anomaly uniformly at random from $\{ \mbox{rings, paths, cliques, stars, trees} \}$.
    \item For each anomaly, select its size uniformly at random from $[5,...,20]$, except for the tree which has fixed size. 
    \item Select the appropriate number of nodes from the network, uniformly at random.
    \item Insert the chosen anomaly in the network.
    \item Add/replace the edges in the chosen induced copy of the anomaly with weights 
    drawn from $\text{Uni}(w,1)$.
\end{enumerate}

The choice of the size and number of the embedded structures is motivated by the consideration that we would like the structure to be 
large enough such that it can be found,  
but not too large or frequent to overwhelm the network. Our choice of constants in the above procedure  produces on average around $147$ anomalies (assuming no overlaps). 
Therefore, for $n=10,000$, on average 
around $1.47$ percent of nodes in the network are anomalies. This level of imbalance was chosen so that the anomalies do not overwhelm the underlying network.

\subsection{Synthetic Data Set: Accenture model}
\label{sec:accentureModel}

The second network generating model was developed by the Accenture team (MML, PR), following discussions with the Turing team (AE, MC, GR); a network which was generated from this model was shared with the Turing team for performing anomaly detection. They were given access to several samples from the model, and they were informed that the anomalies would include heavy edges. The Turing team was oblivious to the model details; the details were only revealed to the Turing team when generating the final set of 100 test networks for inclusion in this paper. Therefore, we regard this model as an independent test of our proposed scoring methodology.

The synthetic Accenture network is constructed as follows. The number of nodes is fixed to $55000$, and each node is assigned an in-degree and an out-degree. The assignment is made by rounding down random draws from a normal distribution; for the in-degree, the mean is set to $21$ with a standard deviation of $3$, while for the out-degree, the mean is set to $19$ with a standard deviation of $2$. If the sum of the in-degrees does not agree with the sum of the out-degrees, then the larger sum is reduced by decreasing by $1$ the appropriate degree of a random node until the totals match.

As in a standard configuration model, each node is assigned an in-degree and an out-degree, and a corresponding number of stubs are created.  The stubs are then
randomised and matched. In the case of self loops, the algorithm randomly
selects a set of edges of same length as the number of random edges, and swaps
the
`sender' node of a self loop with the sender node of the randomly selected edge.
This procedure can generate multi-edges, which are permitted. This process
repeats until all stubs are matched. Edge weights are then drawn from a normal
distribution with mean $1000$ and standard deviation of $200$, with all edge   weights being positive.

The following anomalies were added to the resulting network: a clique of size $8$ and one of size $12$, a ring of size $4$ and one of size $10$, and a heavy path of size $5$ and one of size $10$. In contrast to the weighted ER case (Sec.~\ref{weightERgraph}), all of the weights on the ring are constant, and the heavy path is sampled from paths that already exist in the network.  Moreover, the edges in the clique of size $k$ are not randomly arranged; instead, there is one edge between each pair of nodes and the out- (in-) degrees of the nodes in the clique are set to $[k-i : i \in [1,...,k]]$. Finally, the weights on the heavy edges are drawn as follows: draw
a number  uniformly at random from the interval $(0.99, 0.999990)$,
and use as weight the value that would be required to obtain this number as
percentile from the normal distribution with mean $1,000$ and variance $200$. 
Due to lack of other information about the network structure,  
the Turing team used a configuration model as the null model for the Monte Carlo tests in the anomaly detection pipeline.

\section{Anomaly Detection Method}
\label{sec:overviewOfPipeline}
In this section, we describe the node-based network summary statistics which are used to construct features. Unless  otherwise stated, a feature is constructed from a node-based statistic as follows. We construct $N=500$ replicas (realisations) from the configuration model which fixes the in- and out-degrees, and randomly shuffles the weights. If the $p$-value $p_{val}$ of this test is at least $0.05$, the node score is set to 0. If $p_{val}$ is less than 0.05, we use the percentiles of the standard normal distribution: the nodes score is $\Phi^{-1}(1 - p_{val})$, where $\Phi$ is the c.d.f. of the standard normal distribution.

\subsection{Basic Detection Module}
\label{sec:basicDetection} \label{sec:nodeLevelStats}
First, we include node-level tests of basic properties of the network in order to check for anomalies,
which mainly investigate if the weights of edges which are adjacent to a node 
appear to deviate from random. 
As a null model for the test,
we randomly shuffle the weights while keeping the network fixed, 
and thus, when testing if the local edge weights of a node deviate 
from random, we simply test if it deviates from a random draw with replacement of the weights.

The first set of test statistics are 
based on the geometric average of weights (GAW).  
To give the statistic more power in the case of an
anomaly with extremely high weights but small number of nodes,  we also
construct a statistic over the largest 10\% and 20\% of weights of the edges attached to a given node. 
Letting
$s^{(i)}_1,s^{(i)}_2,...,s^{(i)}_{d_i}$ denote the
ordered edge weights (both incoming and outgoing) of a given node $i$ with $d_i$ neighbours, we define the following scores 
\begin{flalign*}
&
    \text{GAW}(i) 
\qquad
= 
\qquad
\quad 
    \left( \prod_{j=1}^{d_i} s^{(i)}_j \right)^{d_i^{-1}}
 && \\ &
    \text{GAW10}(i)
\quad
= 
\qquad
\left( \prod_{j=
d_i-
  \left\lceil
  0.1d_i
  \right\rceil
    +1
}^{d_i} 
    s^{(i)}_j \right)^{
\left(  \left\lceil
  0.1 \; d_i
  \right\rceil
\right)^{-1}
}
 && \\ &
    \text{GAW20}(i)  
\quad
= 
\qquad
 \left( \prod_{j=
d_i-
  \left\lceil
  0.2d_i
  \right\rceil
    +1
}^{d_i} s^{(i)}_j \right)^{\left(  \left\lceil 0.2 \; d_i \right\rceil
    \right)^{-1}.
}
\end{flalign*}

To compare against random, we use the configuration model; however, 
as our implementation of the configuration model shuffles the edge weights (see Section~\ref{app:configModelImplementation}), rather than constructing the configuration model networks directly, we save computational time by comparing with random draws from the observed weights. To further save computational time, we draw with replacement, and for the Monte Carlo test we share an empirical null distribution between nodes with the same number of edges. The increased efficiency allows us to consider a larger number of samples, and thus, for the results in this paper we use $10,000$ potentially shared samples in each Monte Carlo test. 

As an additional feature, we include the standardised version (z-score based on empirical mean and empirical standard deviation) of the node degree (total number of connections), in order to capture when the degree is larger or smaller than expected at random.

\subsection{Community Detection}
\label{sec:communityDetection}
The more intricate parts  of our anomaly detection pipeline (NetEMD and Localisation) look for more subtle differences than the basic tests.
However, looking for such more subtle changes is difficult, as it is unlikely that a small anomaly will cause a deviation over the whole network, that is large enough to trigger an alert. To address this, we perform a {community detection} step, in order to divide a network into a number of components, such that the density if higher within communities compared to the density across communities.

As we are looking for structures with heavy flow through the network, we augment paths of heavy edges in order to increase the likelihood of them being placed within the same community, 
as follows.
For each path of size $3$ (with $3$ nodes and $2$ edges)  such that the weight on both edges ($w_1$ and $w_2$) is greater than the 99\textsuperscript{th} percentile of the observed weight distribution:
    \begin{enumerate}
        \item If there is not an edge between the first and last node, then  place one with weight $\text{min}(w_1,w_2)$; 
        \item If there is an edge of weight $w_3$, then  update the weight to $\text{max}(w_3,\text{min}(w_1,w_2))$. 
   \end{enumerate}
This procedure is repeated  until no new edges can be added.
The  intuition behind this  augmentation step is that structures of heavy average weight are transformed into structures that resemble cliques. 
This augmentation step can also be seen as a regularisation of the graph,  similar to what is typically performed in spectral clustering in the sparse regime; 
for examples, see 
\cite{amini2013pseudo},  \cite{jin2015fast},
\cite{chaudhuri2012spectral} 
and  \cite{qin2013regularized}. 
As a next step, we perform community detection on the augmented networks  using 
the Louvain algorithm~\cite{louvain} on the symmetrised graph
($A+A^{T}$), with the standard setting and resolution parameter 1. For this task, we use the {\it python-louvain} package from \cite{louvainpython}.

To test if a given community density or edge weights deviates from random, we
employ five  statistics on each community; three statistics are based on the density of the community and two are based on the strengths within the community. The density-based measurements are defined as follows. 

1. The edge density of the community is divided by the density of the full network, in order to capture communities that are denser than expected.

2. The ratio above, divided by the size of the community, is also used in an attempt to further distinguish small communities (for which the penalty is smaller) from large communities. 

3. We use replicas from a configuration null model  and perform community detection 
on each of those to obtain a distribution for the expected density of a
community. We consider one community from each network randomly, and measure the density in the non-augmented network to obtain a null distribution. As these community detection procedures are computationally expensive, we perform only $20$ of them and then use the resultant null distribution to measure a $p$-value for the density of each of the communities in the resultant network, focusing on the upper tail, i.e. when they are denser than we would expect at random. 
If  the resultant $p$-value is more than $0.5$, we attribute all nodes in the community a score of $0$, and otherwise use the normal percentiles as before. 
To account for very small communities (with less than 4 nodes) we return a separate feature which has value  $1$ for nodes in such a small
community, and $0$ otherwise.
 
The edge weight measures are similar to the first two density statistics, however, rather than the density, we use the geometric average of the weights (GAW) within each community and across the network.
After this community detection step, we construct two sets of networks, the first induced by the communities in the original network, and the second induced by the communities in the augmented network. Crucially, each set of networks is useful for a different part of our pipeline, as we will describe in the following sections.

\subsection{Spectral Localisation}
\label{sec:SpecLocal}

Spectral \textit{localisation} in networks is the phenomenon that a large amount of the mass of an eigenvector is placed on a small number of nodes. In many cases, such nodes are different in some way compared to the remaining ones, and hence they may be good candidates for the anomaly detection task.
\cite{delocalisation} calls an unit eigenvector \textit{delocalised} if its entries are distributed roughly uniformly over the real or complex sphere, and quantifies this in various ways using tools from non-asymptotic random matrix theory. 
While most of the work on eigenvector localisation, including \cite{geod,BZ,jms}, has focused on the high frequency eigenvectors (i.e., those corresponding to the larger eigenvalues), recent work \cite{bob} revealed localised eigenvectors associated to small eigenvalues. Sapoval \cite{RSH} studied localised eigenvectors in
different domains and pointed out their importance for physical applications,
including the design of efficient noise-protective walls.

In this section, we also consider other forms of eigenvector behaviour which are not strictly localisation, such as a small number of nodes that have positive entries, in contrast to the majority of the nodes having negative values. However, for compactness, in this paper we shall broadly refer to each of these behaviours as localisation.

As shown in Figure~\ref{fig:process}, we look for localisation in each of the communities separately, attempting to use the fact that small anomalies are more likely to be identifiable within the communities rather than the whole graph. We consider  eigenvectors of three matrices, namely the symmetrised adjacency matrix, the combinatorial Laplacian of the symmetrised
adjacency matrix, and the random walk Laplacian of the symmetrised adjacency matrix, which are defined as follows. Let $\mathbf{W}=(\mathbf{W}_{ij}) \in {\rm
I\!R}^{n \times n}$ be a weighted (possible directed)
adjacency matrix with non-negative entries,
and let ${\bf \mathbf{W}^{(s)}}={\bf \mathbf{W}}+{\bf \mathbf{W}^{T}}$ be the symmetrised version. We consider the following Laplacians 
\begin{flalign}
\label{eq:comb}
& \text{combinatorial Laplacian:} \qquad \qquad {\bf L}_{Comb} = {\bf D}- 
\mathbf{W^{(s)}}, 
&& \\ &
\label{eq:rand}
\text{random-walk Laplacian:} \qquad \qquad \quad \   {\bf L}_{RW} = {\bf D}^{-1}
\mathbf{W^{(s)}}, 
\end{flalign}
where ${\bf D}$ is a diagonal matrix with the 
weighted degrees (row-sums)
of the symmetrised adjacency matrix ${\bf \mathbf{W}^{(s)}}$ 
on the diagonal, i.e. $ {\bf D}_{ii} = \sum_{j=1}^{n} \mathbf{W}^{\mathbf{(s)}}_{ij}$. Eigenvectors are always normalised so that the sum of the squares of the entries is 1. 

Rather than looking at the complete network, localisation is assessed in each of the communities separately, in order to increase the chance  of finding small-sized anomalies. Furthermore, rather than using the subnetworks corresponding to the communities in the original network,  subnetworks of the augmented network are used as the additional augmented edges around the anomalies (as well as other heavy structures that appear at random) are likely to help the localisation methods.

Localisation is assessed 
in the top $20$ and bottom $20$ eigenvectors of the symmetrised adjacency matrix $ \mathbf{W^{(s)}} $, 
and the top $20$ non-trivial eigenvectors of the combinatorial Laplacian ${\bf L}_{Comb}$ and random-walk Laplacian  
${\bf L}_{RW}$  of the augmented subnetwork corresponding to each community.  Note that, in the case of the Laplacian matrices, the trivial eigenvectors are ignored; 
for ${\bf L}_{Comb}$ the bottom eigenvalue (smallest) is always $0$, and in the case of ${\bf L}_{RW}$, the top eigenvalue (largest) is always $1$. Whenever the network has fewer than $20$ nodes, we consider as many eigenvalues as there are available. Ties (i.e., eigenvalues with multiplicity) are broken at random.

To test if a vector is localised, we compare the vectors to random,   
following the approach used in Sec.~\ref{sec:basicDetection}, we compare to the eigenvectors of configuration model replicas.
If a replica network, which is generated under the null model, has fewer than the required number of eigenvectors,  the replica is resampled to obtain a network with sufficiently many  eigenvectors for the adjacency comparison (note that the Laplacian comparisons need one more eigenvector than the symmetrised adjacency matrix  due to the trivial eigenvector). To avoid running the procedure indefinitely, if 
the $10$\textsuperscript{th} network has more than $2$ nodes, this network is used, else we continue to resample until the network has more than the required number of eigenvectors. Then, the $i$th eigenvector is compared with the networks in the ensemble of networks which are generated under the null model and which contain at least $i$ eigenvectors. For a more in-depth explanation see Appendix~\ref{localisation:configModelNulls}.

To capture the different types of localisation, we use several statistics from three broad classes, namely: norm-based, sign-based, and direct localisation.

\vspace{-3mm}
\paragraph{Norm-Based Statistics}
As in \cite{mihaiLocalisationPaper}, we look for eigenvectors 
such that the sum of the $4$\textsuperscript{th} powers of the eigenvector entries (also known as the inverse participation ratio, IPR) 
is larger than expected at random. As the sum of the squares of an eigenvector 
is $1$, a large sum of fourth powers
implies there are large elements in the vector. 
Extending this idea, for an eigenvector $v$ with elements $v_i$, we also consider  the statistic 
$\sum_i (e^{|v_i|}-|v_i|-1)$.
An eigenvector with all the mass on one entry would have a value
of $e-2\approx 0.71828$, whereas a perfectly delocalised eigenvector ($v_i=n^{-0.5}, \ \
\forall i$) would have a value of $n (e^{ n^{-0.5}} - n^{-0.5}
 -1)$, which for $n=10,000$ amounts to $\approx 0.50167$.
 Note that \cite{miller2010subgraph} also uses localisation, but focuses on the $\ell_1$ norm rather than the inverse participation ratio.

\vspace{-3mm}

\paragraph{Sign-Based Statistics}
Some eigenvectors can be used to identify anomalous structures by the sign of the
observation on each node, even when many entries in the eigenvector are zero. 
To capture this effect, we compute the number of nodes that
have a strictly positive or strictly negative value, and we use the minimum of
these quantities divided by $\text{Min}($20$,\text{Number of Non} \\ \text{Trivial 
Eigenvectors})$ as a test statistic.
There are a number of boundary cases for this
statistic (for example a vector that is all positive), the treatment of which we detail in 
Appendix~\ref{implementation:localise:SignBased}.

\vspace{-3mm}

\paragraph{Direct Localisation Statistics}
To directly capture the localisation phenomenon, we take the minimum number of entries required such that their sum is at least $90$\% of the sum of a chosen  statistic over the eigenvector. The chosen  statistics for this procedure are, the sum of the
fourth power of each of the entries, and the sum of the absolute values of these
statistics. For consistency, whenever the entries are exactly equal, all nodes with the given value are included in the count.

The next step is to  evaluate whether the above test statistics deviate in the appropriate direction, based on a $p$-value test using a set of replicas generated under the null model.
For computational reasons, we share the set of configuration model realisations across different eigenvectors. 
For these experiments
the number of null model repeats is  $N=500$. 
For computational reasons, we fix the number of null model replicas at $500$, and we share  
the replicas between eigenvectors of the same matrix.
Thus, the tests for different eigenvectors of the same matrix are not independent; this procedure is not a rigorous significance test, but rather a procedure to generate a score.

Once  vectors that deviate significantly from what is expected at random are identified, we then select the nodes which caused the deviation. 
For the norm-based approaches, the divide between significantly different nodes turns out to be  difficult to distinguish, and thus we consider four different statistics to capture this divide, which are related to the raw value, the significance of the test statistic that highlighted the eigenvector, and the element-wise behaviour observed at random; see Appendix~\ref{app:localalisationNormBasedDerivation} for  details. For the sign-based approaches, the nodes which possess the rare sign are selected, and following Sec.~\ref{sec:basicDetection}, a score based on the standard normal percentiles is assigned. 
To further highlight small communities, we also
construct a score which is normalised by the number of nodes that have the rare sign.
A small number of special  cases in selecting the nodes, e.g. if the number of positive entries equals the number of negative entries, is treated in Appendix~\ref{implementation:localise:SignBased}.
A small number of additional implementation details are described in Appendix~\ref{app:localisationImplementationDetails}, and a small number of additional features which highlight special  cases can be found in
Appendix~\ref{additionalFeatures}.

For the random forest or other machine learning approaches, it may also be interesting to consider the absolute value of the raw eigenvector, as this would allow our random forest to decide if there is sufficient localisation based on the raw eigenvector rather than relying on handcrafted statistics. We include this feature for completeness, but for simplicity, in all of the experiments in this paper, we set the value to zero.

To reduce the total number of features, the contribution from a given localisation test and scoring method in each eigenvector of a given matrix is summed to give  one feature. 
Thus, rather than obtaining $1200$ features which are composed of all
combinations of each of the statistics and eigenvectors,
the $20$ eigenvectors in each set (${\bf L}_{RW}$, ${\bf L}_{Comb}$, large eigenvalues of ${\bf W}^{\bf(s)}$, and small eigenvalues of ${\bf W}^{\bf (s)}$) are combined to yield $60$ features;
for a complete list see Appendix~\ref{sec:localisationFullList}.

\subsection{The NetEMD Module}
\label{sec:NetEMDMod}

\begin{figure}
\centering
\includegraphics[width=0.8\textwidth]{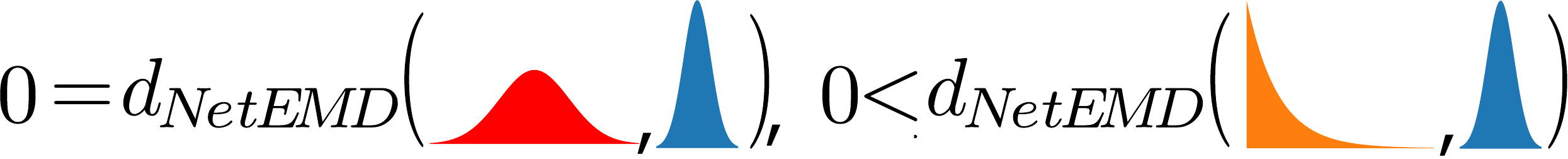}
\caption{\label{fig:combinedNetEMD} 
An example to illustrate that NetEMD only considers the shape of the
distribution rather than the position or the scale: In the left panel, the NetEMD score for comparing two normal distributions with different mean and variances is 0 is $0$, whereas in the right panel the NetEMD score for comparing a normal distribution and an exponential  distribution is greater than $0$.}
\end{figure}

The NetEMD module extends the NetEMD network comparison tool from  \cite{wegner2017identifying} into an anomaly detection method by measuring the difference between the observed network and an expected network. Rather than comparing raw values of a given network statistic, NetEMD compares the shapes of the distributions of a statistic over the nodes of a network.
As an illustration of NetEMD, Fig.~\ref{fig:combinedNetEMD} shows that the NetEMD score for comparing two normal distributions with different mean and variances is 0. In contrast, the NetEMD score for comparing a normal distribution and an exponential distribution is greater than 0.

To describe NetEMD, following the notation in the localisation section, we represent networks as a (possibly) weighted (possibly) directed adjacency matrix ${\bf W} \in {\rm
I\!R}^{n \times n} $. 
Let $T(j,\mathbf{W})$ be a node level statistic for node $j$ on a network $\mathbf{W}$, which takes on real values, and let 
\begin{equation}
    U_T (x, {\bf \mathbf{W}}) = \frac{1}{n}\sum_j \mathbbm{1}(x \le T (j,{\bf \mathbf{W}}))
\end{equation}
denote the empirical distribution function of the statistic $T$ evaluated on node $j$ of network ${\bf \mathbf{W}}$. 
We define the standardised version of $T$ in order to remove the effect of scale
\begin{equation}
    \tilde{T}(j,{\bf W}) =\frac{ T(j, {\bf W})
    } 
    {\sigma(T,\mathbf{W})},  
\end{equation}
where $\sigma^2(T,\mathbf{W}) = \frac1n \sum_{i=1}^n
\left( T(i,\mathbf{W}) - \mu(T,\mathbf{W})\right)^2$ is the empirical variance, and the the empirical mean is given by  $\mu(T,\mathbf{W}) =\frac1n \sum_{i=1}^n T(i,\mathbf{W})$.
For two networks ${\bf W}$ and ${\bf W' }$, and a given statistic $T$, the NetEMD score is given by 
\begin{equation*}
    \text{NetEMD}_{T} ({\bf W},{\bf W'}) = \text{min}_s \int_{\mathrlap{-\infty}}^{\mathrlap{\infty}} | U_{\tilde T} (x+s, {\bf W})-U_{\tilde T} (x, {\bf W'})| dx, 
\end{equation*}
where 
$\tilde{T}$ rather than $T$ is used in $U_{\tilde T}$, in order to
remove the effect of scale. The minimum $s$ of $\int_{-\infty}^{\infty} |
U_{\tilde T} (x+s, {\bf W})-U_{\tilde T} (x, {\bf W'})| dx$ will be achieved as
the empirical distribution functions only have finitely many jumps, see for
example \cite{xureinert}. 

Let $C_{T}=\{T^{(m)}, m=1, \ldots, M \}$ denote a set of node 
statistics such that for each node $j$ and each network $\bf W$, $T^{(m)}(j, {\bf W}) \in {\rm I\!R}$. 
The NetEMD statistic for such a set of node level statistics is the average over the single NetEMD scores 
\begin{equation}
\label{eq:fullNetEMD}
    \text{NetEMD}({\bf W},{\bf W'})
    = \frac{1}{|C_{T}|}\sum_{T \in C_{{T}}} \text{NetEMD}_{T} ({\bf W},{\bf W'}) . 
\end{equation}
In ~\cite{wegner2017identifying}, the statistics often used as $T^{(m)}$ are the so-called orbit counts, i.e., the number of times that a given node is part of a 
given subnetwork in a given position.

\begin{figure}
\centering
\includegraphics[width=0.4\textwidth]{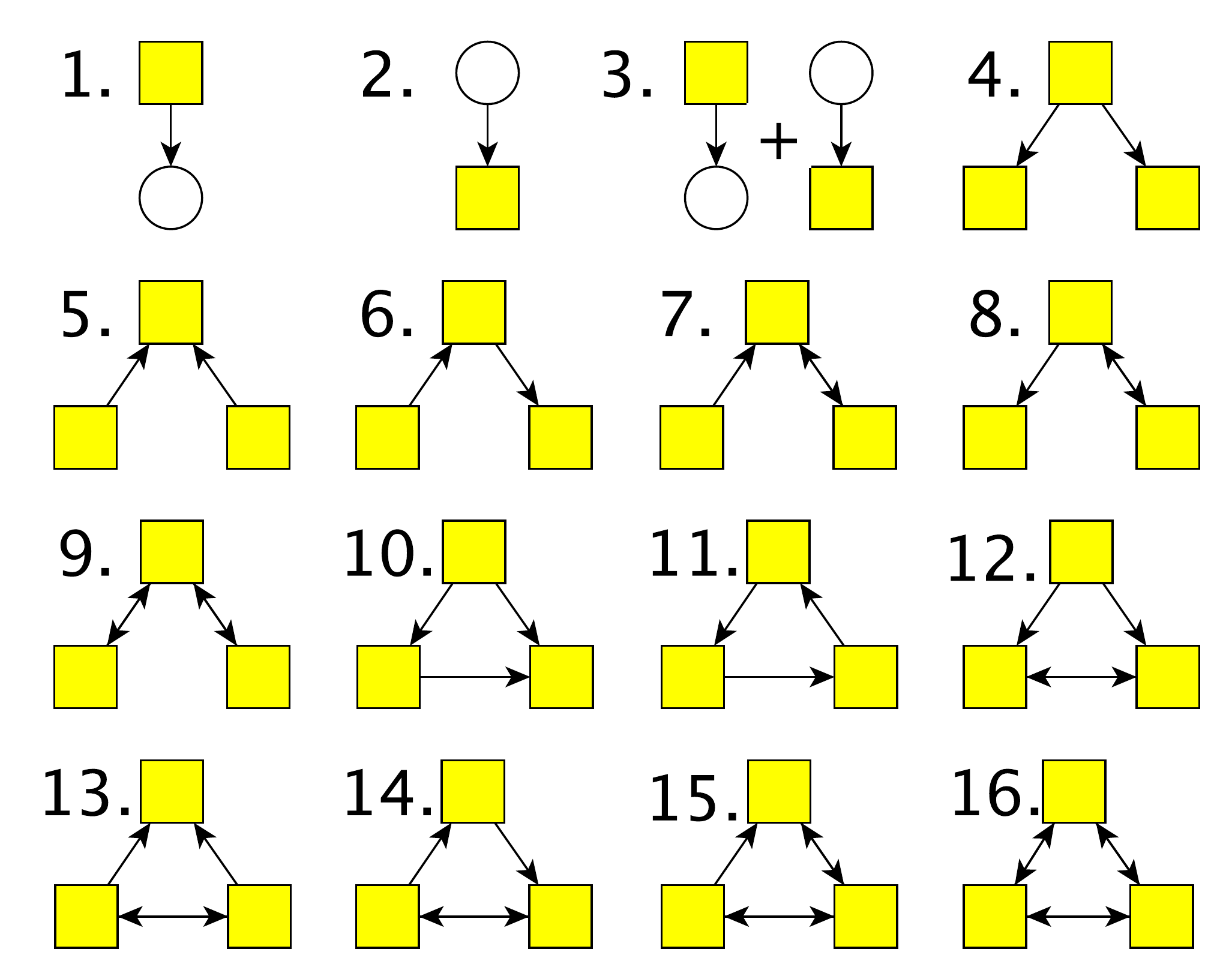}
\caption{\label{fig:motifs} The subnetwork statistics that are used in the
    NetEMD module. Statistic {\bf 1} is the total the out-strength, statistic {\bf 2} is the total the
    in-strength and statistic {\bf 3} is the total of the in-strength and out-strength. The remaining statistics are the directed motifs of size $3$, where the weight is given by the product of the weights on each of the edges. }
\end{figure}

\paragraph{The Set of Statistics}
As a set of NetEMD statistics, 
 we use the in-strength (i.e., the sum of all weights
coming into a node), out-strength (i.e., the sum of all weights going out of a node), the
sum of out-strength and  in-strength, and finally each of the directed
motifs of size $3$; these are shown in  Fig.~\ref{fig:motifs},. Here, a motif is just a small network; no over- or under-representation is implied. 
Following the approach used by \cite{mastersThesisGesine}, we let the weight of each motif
be the product on the weights on the edges. 
Formally, for a given motif ${\bf M}$ on 3 nodes, letting ${\bf W}([i,j,k])$ be the induced subnetwork of ${\bf W}$ with nodes $i$, $j$ and $k$, 
the test statistic is
\begin{flalign*}
    T^{Motif(i)}(j,W) = \sum_{a<b}&  \mathbbm{1}( {\bf W}([j,a,b]) \mbox{ is an induced copy of }M )
    \prod_{c,d \in [j,a,b]} 
    \left(
    ({ W}_{cd}-1)\mathbbm{1}({ W}_{cd}>0)+1
    \right), 
\end{flalign*}
where the indicator function $ \mathbbm{1}$ is 1 if the statement is correct, and
0 otherwise. The indicator in the product ensures that the factor in the product
is $1$ when the edge ${ W}_{cd}$ has zero weight, and it is the weight ${
W}_{cd}$ when this weight is not zero. These motifs are counted in the non-augmented graph.

To the statistics shown in Figure~\ref{fig:motifs} we add the following
spectral measures based on Subsection \ref{sec:SpecLocal},  
which are obtained from the augmented graph. 
For the symmetrised adjacency matrix, we consider the eigenvectors corresponding to the $5$ largest and $5$ smallest eigenvalues, which we name $T^{AdjU(i)}(j,\mathbf{W})$ and $T^{AdjL(i)}(j,\mathbf{W})$ respectively.
For the Laplacians, the top $5$ eigenvalues are used, which corresponds to the smallest eigenvalues in the combinatorial Laplacian and the largest eigenvalues in the random-walk Laplacian. As a remark, we again ignore the trivial first eigenvalue in the Laplacians. 
We refer to these
statistics as $T^{Comb(i)}(j,\mathbf{W})$ and $T^{RW(i)}(j,\mathbf{W})$.
Ties are broken at random.

An issue arises from the sign of the eigenvectors. Let ${\bf B}$ denote one of the
matrices of interest (one of ${\bf W^{(s)}}$, ${\bf L}_{Comb}$, ${\bf L}_{RW}$).
Let $\lambda_1 \ge \lambda_2 \ge ... \ge \lambda_n$ and $v^{(1)},..,v^{(n)}$ such
that ${\bf B} v^{(i)} = \lambda_i v^{(i)}$. 
As the direction of an  eigenvector ${\bf{v}}=(v_1, \ldots, v_n)$ is arbitrary, 
comparing the shapes of the distributions as in Fig.~\ref{fig:combinedNetEMD} has to be adjusted to account for the sign of the eigenvector.
To this end, 
first we check if the distribution of elements in the vector is symmetric. If so, then 
we arbitrarily select the direction as this does not effect the result of the computation. If the distribution of elements in the vector is not symmetric, then we choose the sign which yields the largest number of positive elements. If this procedure results in a tie, we then break ties using 
$g(a)=\sum_j ((v_i)_j)^{2a+1}$; letting  $a^{*}$ be the smallest $a \in \mathbbm{N}\cup \{0\}$ such that $g(a^{*})\neq 0$, and we use the sign of $g(a^{*})$. 
Appendix~\ref{app:directionOfEvectorProof} gives a proof that, for a finite network, the value $a^{*}$ is finite.

\paragraph{Converting NetEMD into an anomaly detection method.}
\label{anomalyDetect}

To convert NetEMD values into node scores, following \cite{ospina2018assessment}, we use a Monte Carlo approach with NetEMD as a statistic. 
First, we generate a reference set of 15 networks, which are generated under the null model, and calculate the NetEMD distance of our target network to the networks in the reference set. 
To produce a robust measure of deviation,  
we use a trimmed mean in which
the largest and smallest distances out of the $15$ reference network distances are excluded.

We then test if the deviation is large enough using a Monte Carlo $p$-value test, with a second set (which we call the null set) of networks which are generated under the null hypothesis and which are independent of the reference set. 
Instead of considering the average NetEMD over a set of NetEMD values for different
statistics as in Eq.~\eqref{eq:fullNetEMD}, we instead compute the trimmed NetEMD mean, and run the Monte Carlo test on each of the statistics in $C_{{T}}^{*}$ individually in each community. We combine the results for a given statistic into a single feature, resulting in one feature per statistic, that we will later combine into an overall ranking. 

To address the case where the number of eigenvectors in the null replica is lower than the required number (6 in this case),  
if a null replica has too few entries, we resample up to $10$ times to construct a network with at least $6$ nodes, and if by the $10$\textsuperscript{th} resample we do not obtain one, then we consider the next network with more than $3$ nodes. Again, following the localisation approach detailed in
Appendix~\ref{localisation:configModelNulls}, we only make comparisons between pairs of networks that have a sufficient number of eigenvectors. 
A few special cases are deferred to Appendix~\ref{app:netemdSpec}.

For statistics for which NetEMD appears to
deviate from the expected behaviour, i.e., with a low Monte Carlo $p$-value, we assign scores to nodes in two ways. As we know we are interested in structures that are likely to be over- or under-represented, we focus on the tails of the empirical distribution of a given statistic over the network. 

First, we select the nodes that deviate more than two empirical standard
deviations from the average of the node contributions, and assign them a score
which is the absolute value of their deviation. Mathematically, for a network $\mathbf{W}$ and a node statistic $T \in C_{{T}}^{*}$, we let 
\begin{equation}
    \label{eq:netemdscoring1}
    score^{NetEMD}_1(j,T,\mathbf{W}) =
    \begin{cases}
        0 & \text{ if }
        |\frac{T(j,\mathbf{W})-\mu(T,\mathbf{W})}{\sigma(T,\mathbf{W})}|<2, \\
        |\frac{T(j,\mathbf{W})-\mu(T,\mathbf{W})}{\sigma(T,\mathbf{W})}| & \text{otherwise}. 
    \end{cases}
\end{equation}
To account for statistics with heavy-tailed distributions, our second method of identifying nodes that contribute most to the deviation, considers 
the top 5\% of node contributions based on the absolute standardised score above, and set 
\begin{equation}
    \label{eq:netemdscoring2}
    score^{NetEMD}_2(j,T,\mathbf{W}) =
    \begin{cases}
        0 & 
        \text{If in bottom $95$\% of } |\frac{T(j,\mathbf{W})-\mu(T,\mathbf{W})}{\sigma(T,\mathbf{W})}| \\
       \Phi^{-1}(1-p_{val}) & 
        \text{else},  \\
    \end{cases}
\end{equation}
where $\Phi^{-1}$ is the inverse c.d.f. of a mean $0$ and variance $1$ normal distribution. 

Finally, in order to reduce the total number of features, we sum the contribution in a given scoring method in each eigenvector of a given matrix into one feature. For example, for a given node $j$, this amounts to 
\begin{equation*}
score^{NetEMD}_X(j,T^{AdjL},\mathbf{W}) = \sum_{i=1}^{5}
score^{NetEMD}_X(j,T^{AdjL(i)},\mathbf{W}),
\end{equation*}
with analogous forms for the different matrix operators we consider.

\subsection{Finding Heavy Paths}
\label{sec:PathFinder}

The separation of the network into communities is computationally advantageous
as it facilitates parallelisation, the disadvantage being that it can split
long paths. To alleviate this risk, we include a separate feature for finding
long paths. As enumerating all paths of a given (long) size is computationally
prohibitive, we take inspiration from \cite{EberleHolder} and
\cite{EberleHolder1}, and use a BeamSearch method in which we retain a small
set of promising examples to extend, with a BeamWidth of 5000. This allows us
to tradeoff the optimality of the solution against the running time, and leads
to the following algorithm.

We start by constructing a set of paths of size $3$ (i.e., with $3$ nodes and $2$ edges). Rather than considering all possible paths of size $3$, 
we consider a more limited set as described in 
Appendix~\ref{app:implementation:path}.
We remove all but the
top BeamWidth paths (for the case of ties, see Appendix~\ref{app:implementation:path}), store the result, and form a new list considering all extensions of the paths which can be
formed by adding an additional edge. We repeat this procedure until the paths in the list are of the required size. The set of paths of different sizes, generated by this procedure, are then used to construct features as usual,  based on a Monte Carlo test and standard normal percentiles.  

Next, we obtain a score for each node by summing the score for each significant path that contains the node in question. For computational reasons, the null replicates are shared between path sizes rather than running the algorithm
for each path size. 
Finally, we test for path sizes. In general, in our method we consider $30$ different path sizes, varying the size from $3$ up to $32$.   
However, for computational efficiency, in the experiments detailed in this paper, we only compute paths of size less than or equal to $21$ (i.e. path length less than or equal to $20$).

\subsection{Combining the scores} 
\label{sec:combineScore}
Each of the network statistics described thus far produces a score (a full list of all procedures can be
found Appendix~\ref{listOfAllMethods}). 
One way to combine the scores is simply to sum them, which yields a raw measure of performance of the pipeline; this is the {\sc Feature sum} method. 
As
the {\sc Feature sum} may not be the most effective combination of the underlying features, 
following \cite{Savage2016DetectionOM}, we also combine the features using a random forest approach which relies on a training set of labelled data. 
Such  training data 
are generated from the model in Sec.~\ref{weightERgraph}, with the parameter sets 
chosen such that under the model without anomalies, each anomaly has expected value less than 1. 
We then perform feature selection, prioritising features that perform well across the ensemble of parameter regimes,  
as will be described in Sec.~\ref{sec:combinationAndFeatureSelection}. Finally, we combine the data from each of the parameter regimes, and fit a random forest on the reduced feature set to produce our final score which we denote by {\sc Random Forest}.

\paragraph{Selecting the parameters of the training set.}
\label{selectingParam}

\begin{figure}
\centering
\includegraphics[width=0.5\textwidth]{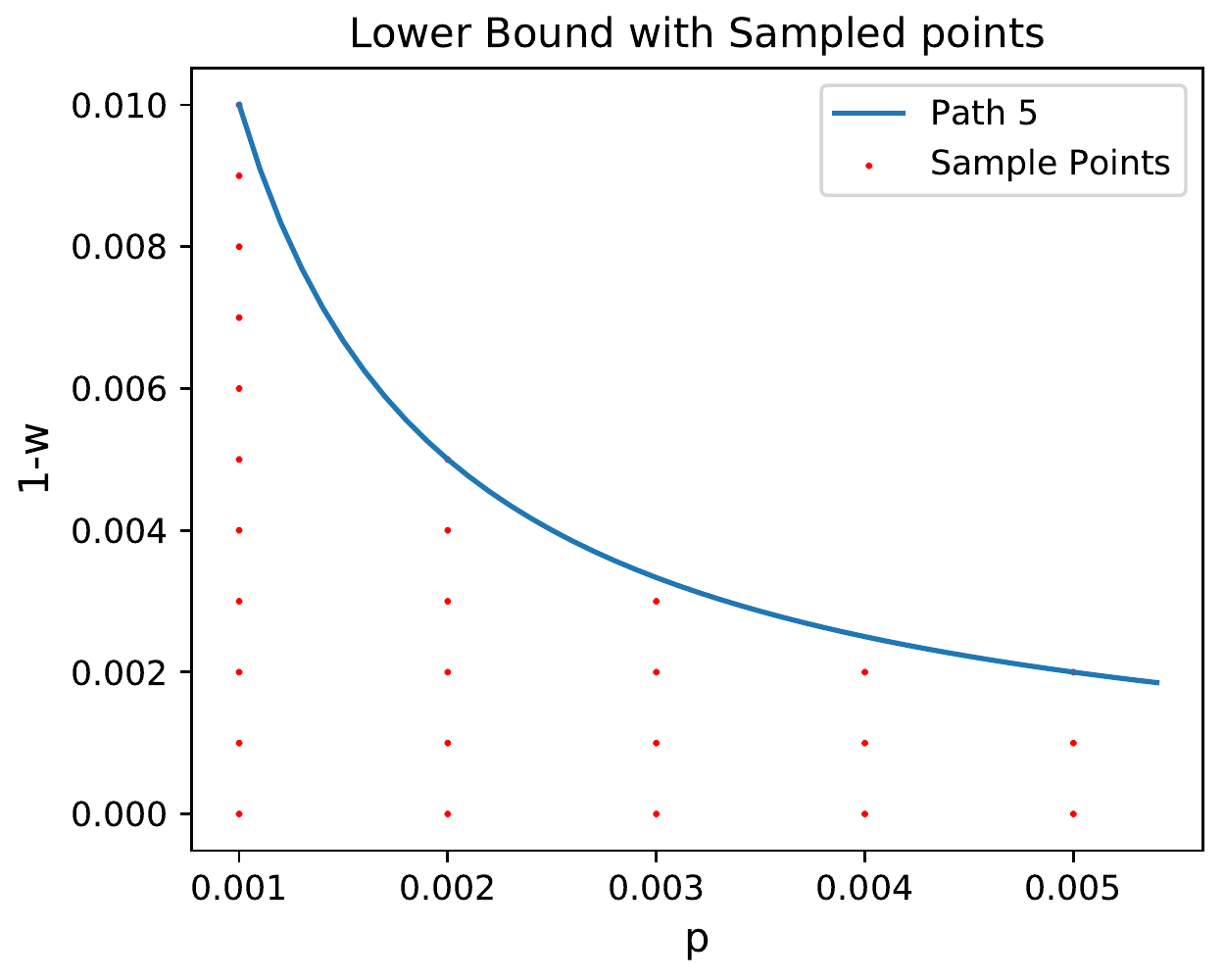}
\caption{\label{fig:limits} Bounds on $(w,p)$ and chosen parameter combinations.
The set of parameter combination  points (red dots) is contrasted with the bounds on $1-w$ for different values of $p$ below which the expected number of paths of size $5$ in a network with $10,000$ nodes is less than  $1$.}
\end{figure}

To ensure that each embedded anomaly is
detectable, the parameter regimes for $(p,w)$ are chosen such that the expected
number of anomalous instances of each of the planted substructures in the
background network is less than $1$. 
In Appendix~\ref{app:inequalities} we derive the bounds for each structure, and it is shown that the path of size 5 is the most restrictive anomaly.  
It requires the expected counts of this anomaly under the null model to be less than 1, leading to the inequality 
\begin{flalign}
    &
    \label{eq:path} 
(1-w)p < \left( { {n \choose 5} 5! }\right)^{-\frac{1}{4}} .
\end{flalign} 
This region is plotted
in Fig.~\ref{fig:limits}. 
We vary $p$ in fixed
increments of size $\frac{1}{1000}$, between $\frac{1}{1000}$
up
to and including
$\frac{1}{200}$, 
and we vary $w$, starting with $0.99$ and
increasing in increments of $\frac{1}{1000}$ up to and including $1.0$. This results in $55$ possible sample points, of which $27$ are within the bounds of detectable points.
The points can be found in
Fig.~\ref{fig:limits}.
For each of the 27 parameter sets, we construct $100$ networks and split them, with the first $70$\% as the training set and the remaining $30$\% as the test set. We run the entire pipeline on each of the $2700$ networks.

\paragraph{Feature selection across the ensemble.}
\label{sec:combinationAndFeatureSelection}

To select a set of features which work well across each of our parameter
regimes, we fit a random forest to the training data for each given parameter set using the scikit-learn \textit{RandomForestRegressor} with default parameters in version 0.19.2, 
and use the feature importance vector 
\cite{randomForestSklearnReference,sklearn,webpage:mediumFeatureImportance} to rank the features. 
Next,  
we compute the average rank of each feature
\vspace{-2mm}
\begin{equation}
\text{FeatureRank}_i^{(\text{All})}
=
\frac{1}{27}\sum_{p,w}
\text{FeatureRank}_i^{(p,w)}, 
\end{equation} 
where $\text{FeatureRank}_i^{(p,w)}$ is the rank of feature $i$ in the parameter set $(p,w)$. 
We select an appropriate cut-off by finding a large jump in a plot of the resulting average ranks.

\subsection{Oddball}
\label{oddball}

The original formulation of Oddball from \cite{akoglu2010oddball} is built 
to detect $4$ different types of anomalies in weighted networks which can be described as {\it ``almost'' star,  ``almost'' clique,} the so-called {\it ``heavy vicinities''} based on weighted degree, and heavy edges.
\cite{akoglu2010oddball} observes an empirical relationships between measured statistics related to each type of anomaly. The relationships are power-laws which are identified between statistics of the $1$-hop snowball sample of each node, namely the number of edges against the number of nodes, the total weight of edges against the number of edges, and the largest eigenvalues (of the adjacency matrix on the sample) against the total weight. For each relationship, Oddball fits the resultant power-law and then computes the following outlier score for each node 
\begin{equation}
\label{eq:oddball}
\frac{\max(observed,predicted)}
{\min(observed,predicted)}
(log(|observed-predicted|+1)).
\end{equation}
The authors augment this score
by summing the scores~\eqref{eq:oddball} with
the local outlier factor (see \cite{breunig2000lof}) 
to capture points which may be close to prediction, but far from other observed points. 
The authors' website, available at {http://www.andrew.cmu.edu/user/lakoglu/} provides a version of Oddball, called {\it Oddball-lite}, which does not use Python and which we use in this paper; {\it Oddball-lite} appears not to carry out the augmentation by the local outlier factor. 
Furthermore, while Oddball was originally formulated as an anomaly detection scheme for weighted non-directed networks, the {\it Oddball-lite} implementation provided by the authors of
\cite{oddballTechnicalReport} 
and
\cite{akoglu2010oddball} 
also provides a directed version, which we use. 
For completeness, we include in Appendix~\ref{app:oddballComparisons} the relationships used in the directed case, as well as small modifications to the code we performed in order to run the experiments in this paper.

Finally, in the original Oddball paper \cite{akoglu2010oddball}, deviations from each observed relationships are considered separately with a ranking constructed for each relationship (e.g., number of nodes versus number of edges), and each relationship then related to a different type of anomaly. However, we construct one overall ranking from  our main features. For a meaningful overall comparison with Oddball, we consider the sum over each of the relationships to construct an overall ranking. In Appendix~\ref{app:Oddball}, we consider the individual performance of each of the features  
and discuss any well performing underlying relationships in the main paper.

\subsection{Performance Measures}
\label{sec:assess}

To measure the performance of our methods we consider three different statistics. Our primary use case is in identifying financial crime, and as we know that each of our hits will have to be validated by a human, only a relatively small number of candidates could be validated. To that end, we consider the performance of the
top $\{1,2,4,8,16,32,64,128,256,512,1024\}$ nodes in a ranking. We measure performance and compare graphically, 
using precision and recall which are 
defined as follows~\cite{sklearn}
\begin{flalign*}
\text{Recall} = \frac{\text{Number of Anomalies in sample}}{\text{Total Number of Anomalies}}, 
\\
\vspace{5mm}
\text{Precision} = \frac{\text{Number of Anomalies in sample}}{\text{Size of Sample}}. 
\end{flalign*}
Furthermore, we construct our ranking and thus our fixed-size sample using a score. In the case of ties, we use the average number of anomalies if we selected randomly between the tied nodes. 
Further, if a node is involved in two anomalous structures we only count the node once, and the total number of anomalous nodes is reduced accordingly.

Comparing the precision and recall of two different networks with different
numbers of anomalies can be misleading. As the number of anomalies in the
weighted ER graph is unknown, ranging from $25$ to $400$, we interpret the
precision and recall results as advisory. In the Accenture model case, the
number of anomalous nodes is fixed (with the exception of a small number of overlapping anomalous structures), and hence the results are easier to interpret. 

The third performance measure is the average precision, which is
defined as follows. Let $R_{ec}(i)$ denote the recall if we threshold at the top
$i$-th nodes in a ranking, and $P_{re}(i)$ denote the precision for the same threshold.
Following \cite{sklearn}, the average precision is defined as 
\vspace{-2mm} 
\begin{equation*}
\sum_{i=1}^{n-1} (R_{ec}(i+1)-R_{ec}(i))P_{re}(i), 
\end{equation*}
and has been shown to give additional weight to the
highest ranked items~\cite{averagePrecision}, thus making it a good measurement
for this work. 
Other  assessment methods for classification performance are discussed, for example, in  
\cite{precisionRecallPlot}.

Finally, we also compare the performance of our measures to that of a random
classification. As shown in \cite{averagePrecision}, under a continuous
approximation, a random ordering will have an average precision equal to the
percentage of anomalies in the network. Therefore, in order to compare our performance to random, we compare against the percentage of anomalies in the network.

\section{Results}\label{results}

The performance of the {\sc Feature sum} and the {\sc Random forest} methods is first tested on the weighted ER networks, and then on the Accenture model. 

\subsection{Results on the Weighted ER Network}

\subsubsection{\sc Feature sum}
\label{sec:linearModel}

The {\sc Feature sum} method simply sums the scores of all features. 
To avoid biasing the testing data, we compute the performance using only the training data. 

\begin{figure}[!ht]
\centering 
\includegraphics[width=\textwidth]{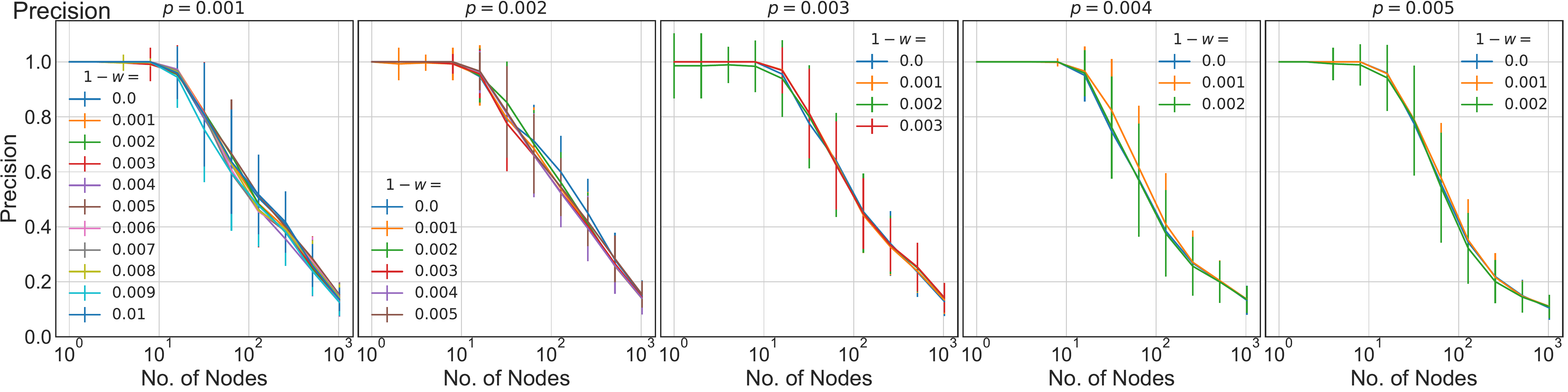}
\includegraphics[width=\textwidth]{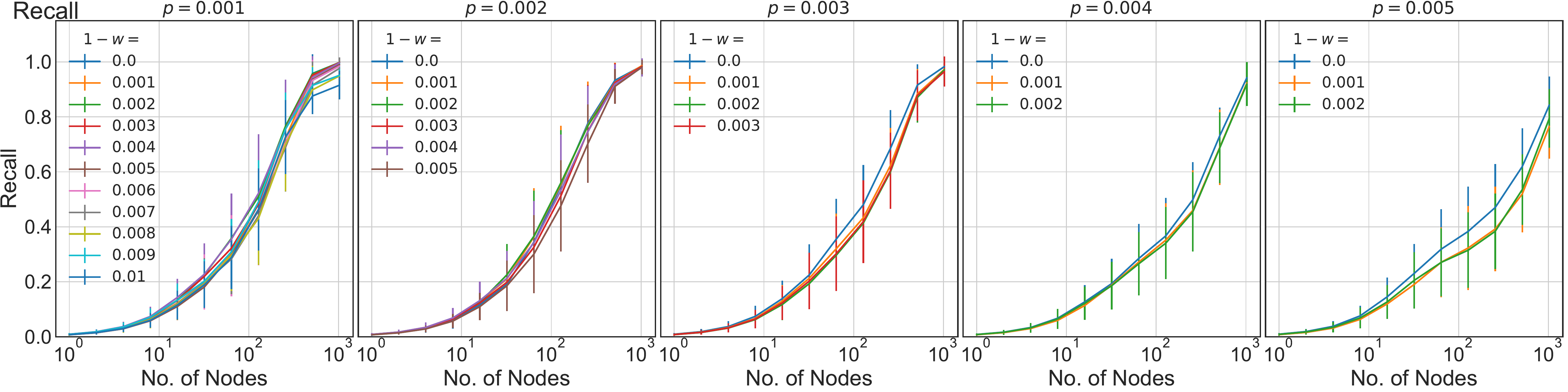}
\caption{The mean precision (upper plots) and recall (lower plots) for
the {\sc Feature sum} method across the parameter regimes over the training networks. In the set of panels on the left, the mean precision and recall over the training networks is plotted over the parameter regimes $(p,w)$ which have $p=0.001$, considering the $2^{i}$ top scoring nodes, where $i=0,...,10$. The error bars are given as one estimated standard error 
over the relevant networks. The remaining panels show the same plots for different densities.
}
\label{fig:linearRecallPrecisionPlot} 
\end{figure}

The results can be seen in Fig.~\ref{fig:linearRecallPrecisionPlot}. Regarding
the precision, the {\sc Feature sum} method performs well in all networks, with
the top scoring $8$ nodes 
often all being true anomalies. Regarding the recall, for low values of $p$, 
the {\sc Feature sum} method recovers most anomalies within the $1024$ top scoring  nodes, with a reasonable fraction being recovered within the first $512$ top scoring nodes (approximately $5\%$ of nodes). As the number of anomalies varies between $50$ and $400$, obtaining a good proportion of the nodes within the first $5$\% is a reasonably good performance. For higher densities, the performance is similarly high for the top nodes, 
but only capturing around $75$\% of the anomalous nodes. The standard deviation tends to be larger for the medium node score range, indicating that the random network generation appears to affect the ordering more in the medium node score range than it does in top and bottom ranges.

\begin{figure*}[h!]
\centering
\includegraphics[width=\textwidth]{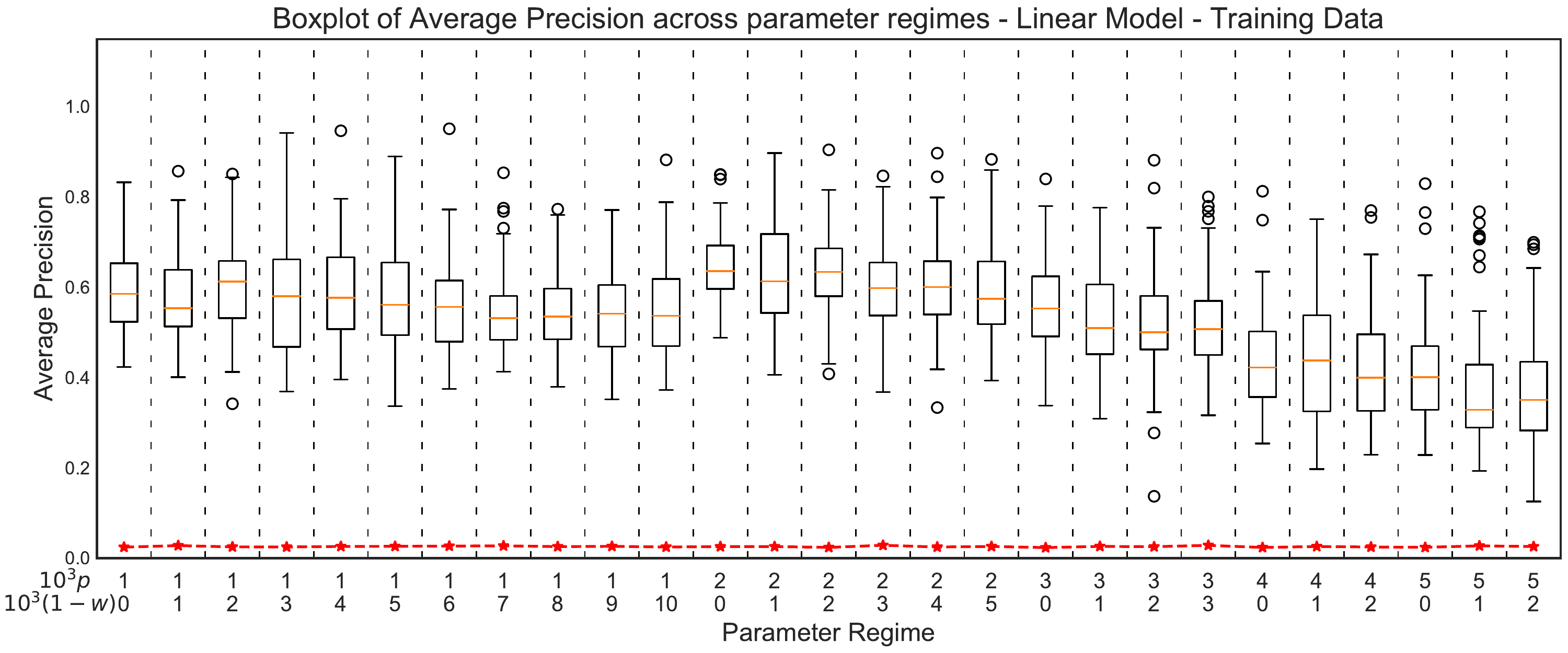}
\caption{\label{fig:linearModelBoxPlot} 
The average precision (computed using scikit-learn~\cite{sklearn}), for each of the parameter regimes using the {\sc Feature sum} method. 
With a random ordering, the average precision should be equal to the percentage of anomalous nodes~\cite{sklearn} (see Sec.~\ref{sec:assess} for details).
The red star line shows  the largest percentage of anomalies over the networks which are  observed in each parameter regime. 
}
\end{figure*}

Finally, we compare the performance of the {\sc Feature sum} in different parameter
regions using the average precision; the results can be seen in
Fig.~\ref{fig:linearModelBoxPlot}. The {\sc Feature sum} scoring system outperforms
random classification (red dashed line) in all parameter regimes considered.
The median average precision for the low density networks ($p\le 0.003$)
is reasonably consistent, with values around $0.5-0.6$. For larger densities,
we observe lower values, albeit still considerably above the random regime.

\subsubsection{{\sc Random forest}}
\paragraph{{\sc Random forest} feature selection.}
\label{sec:result:featureSelection}

\begin{figure*}[h!] 
\centering
\includegraphics[width=\textwidth]{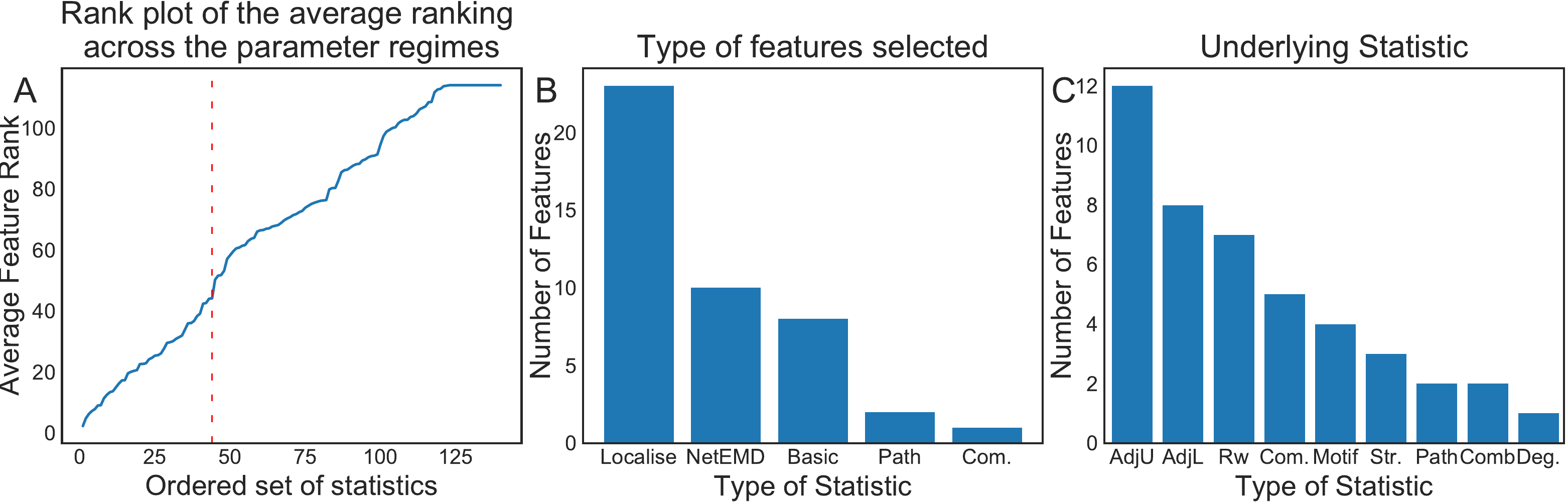}
\caption{\label{fig:averageRankPlot} 
    {\bf (A)} Rank plot of the average rank of each of the statistics over the parameter regimes. The red dashed line details a small jump at the 44\textsuperscript{th} feature which is  used as a threshold for feature selection.
    {\bf (B) } Histogram of the underlying type of method behind each feature. 
    {\bf (C)} Histogram of the underlying matrix or statistic used in each feature. 
    }
\end{figure*}

As described in Sec.~\ref{sec:combinationAndFeatureSelection},  a separate random forest is fitted to each of the training sets and  the ranking in each feature importance vector is obtained. We then plot the average importance rank of each feature across the parameter range; features with the same importance score are  each assigned the lowest possible rank. The resultant curve can be seen in Fig.~\ref{fig:averageRankPlot}A. Although there is not a strong cut off candidate,  there is a reasonable jump in average rank at $44$, and hence  $44$ is used as the cut off. We also tested a similar procedure using the average score rather than the
average rank (results in Appendix~\ref{app:featureSelection2}), which produces a much
clearer jump with only $6$ features. However, concerns about correlated features reducing the  
importance score and thus removing useful signal, led to favouring the rank-based threshold. 
The rankings are relatively stable with $42$ of the most informative $44$ features in
the ranking ordering also in the average score based ordering. A list of the selected features can
be found in Appendix~\ref{app:featureSelectionListOfFeatures}.

With $44$ as the cut off,  Fig.~\ref{fig:averageRankPlot}B shows the underlying modules which dominate the selected features. Localisation measures appear in $\approx 52\%$ of the selected features. 
Furthermore, all of the basic methods are present in the $44$ selected features.
Fig.~\ref{fig:averageRankPlot}C shows the underlying statistic which drives each of the
selected features. Around half of the statistics are driven by a version (non-augmented or augmented) of the adjacency matrix.

\begin{figure*}
\centering
\includegraphics[width=\textwidth]{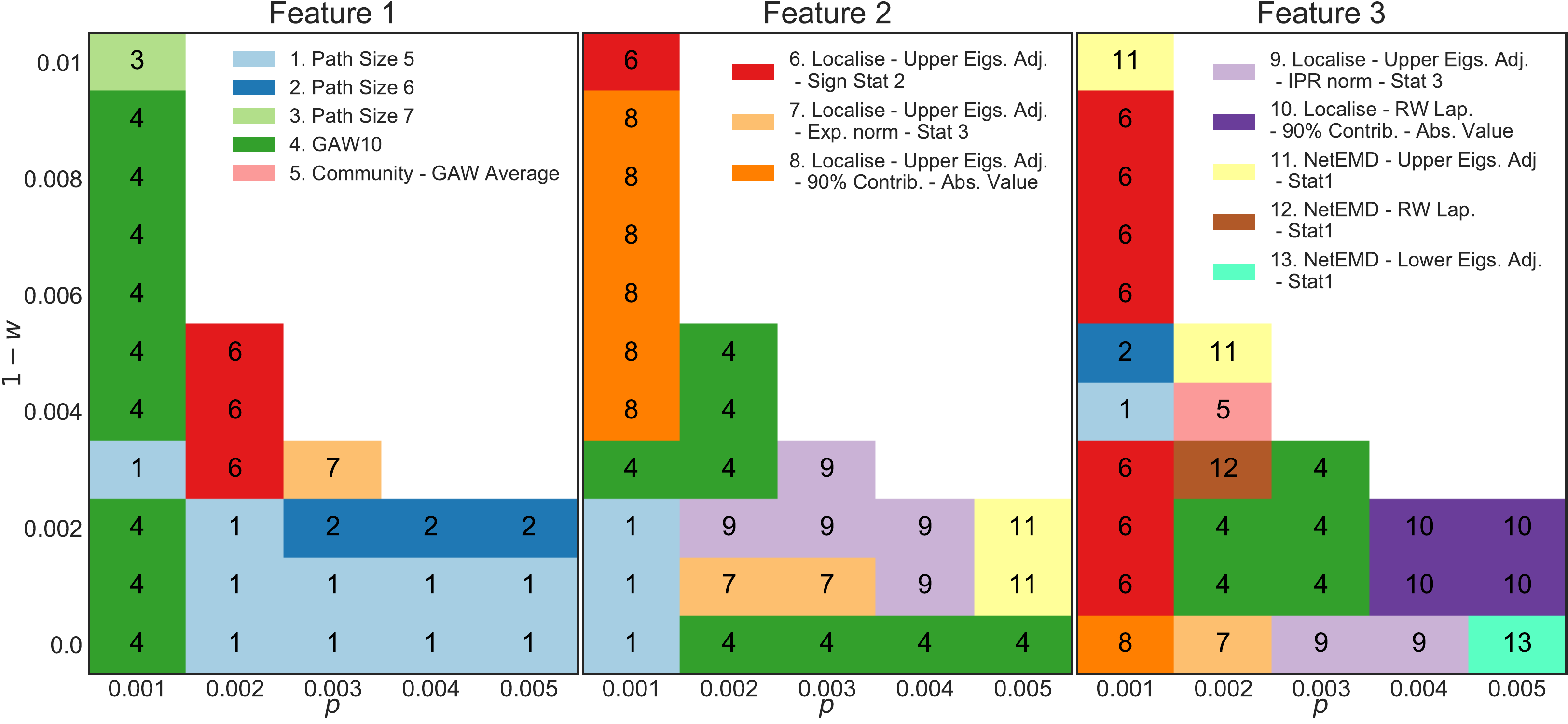}
\caption{\label{fig:regionsOfStatistics} 
The top three most important statistics in each parameter regime, according to the
 feature importance calculation performed by Sklearn and detailed in
 Sec.~\ref{sec:combinationAndFeatureSelection}. Feature 1 is the most important feature, followed by Feature 2, then Feature 3.
    To aid comprehension when printed in black and white, we 
    include a numerical index of each feature.
    }
\end{figure*}

To explore the variation of the most important features between the
parameters regimes, Fig.~\ref{fig:regionsOfStatistics} displays the top $3$
features of each of the parameter regimes. Local statistics dominate the
figure, with test statistics based on the top $10$\% of neighbour edges by
weight, and paths of size $5$, appearing to be important. 
The high performance of these paths could be related to the fact that the size
of the smallest embedded structure is $5$ and therefore paths of this size are present in all embedded paths and all embedded rings. Slightly longer path-based methods also are well represented, with paths of size $6$ taking up a reasonably large amount of the region where $p$ is large but $1-w$ is small. While paths of size $7$ appears to be important for $(p,1-w)=(0.001,0.01)$, this statistic does
not appear in the list of $44$ statistics that were selected earlier in this section, as it is not important enough in other parameter regimes. 

As the second most important feature, a larger fraction of the more complex measures appear, with features relying on both localisation and NetEMD. The underlying statistic appears to always be the adjacency matrix, perhaps indicating that the adjacency matrix has a stronger signal than the other matrices for the anomalies we are aiming to detect. The third most important feature shows a similar
pattern, with many of the NetEMD/Localisation based statistics directly using the adjacency matrix. 

Finally, the most important and second most important features appear to 
contain several regions in which the selected statistic is stable. We take this 
level of stability as an indication that the number of training examples is reasonably large. In the case of the third most important statistic, we observe a
little more variation, which could either indicate that different statistics specialise in different parameter regimes, or it could simply be a consequence of the number of repeats being relatively small.

Using  the selected $44$ features and the standard parameters in the RandomForestRegressor from Sklearn, we fit a random forest over the full training data set, resulting in $1,890,000$ observations with $278,056$  anomalies ($1.47\%$).  We tested in-sample performance using the same procedure we used in the {\sc Feature sum} model, achieving near perfect scores. The relevant plots can be seen in Appendix~\ref{sec:withinSamplePerformance}.

\paragraph{Performance of  {\sc Random forest} on the test set.}
\label{sec:testingSet}

\begin{figure*}[h!]
\centering
\includegraphics[width=\textwidth]{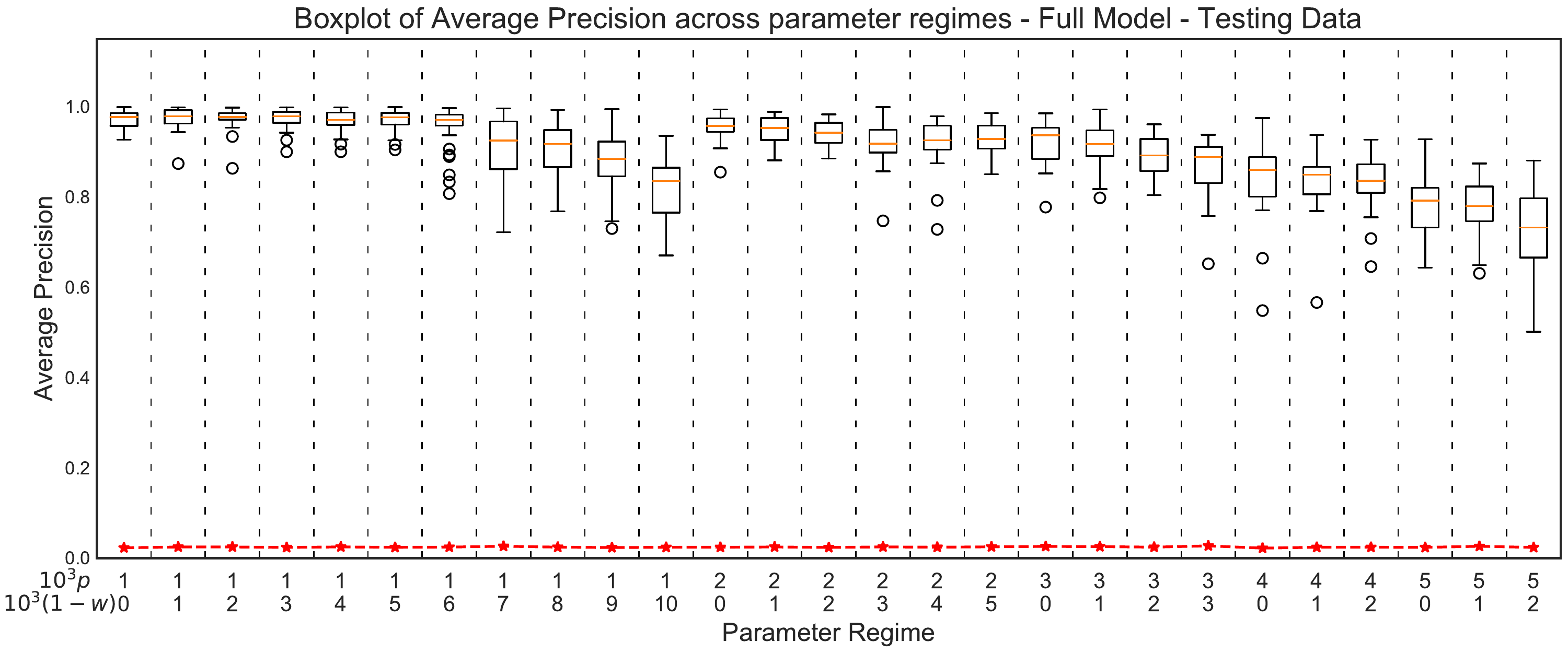}
\caption{\label{fig:fullTestModelBoxPlot} 
    The average precision for each of the
    parameter regimes using the {\sc Random forest} Model after feature selection and training on the test data. 
With a random ordering, the average precision should be equal to the percentage of anomalous nodes~\cite{sklearn} (see Sec.~\ref{sec:assess} for details).
The red star line  shows
the largest percentage of anomalies over the networks which are observed in each parameter regime. 
    }
\end{figure*}

To measure out-of-sample performance, we applied the fitted {\sc Random forest} to the test set. When interpreting these results, it is crucial to understand that the default RandomForestRegressor has $10$ trees, which means that each node will be given a score from $\{0,0.1,...,1.0\}$.

The average precision, as seen in Fig.~\ref{fig:fullTestModelBoxPlot}, shows a substantially better performance than was achieved using the {\sc Feature sum} model on the training data. For low $p$ and low $1-w$ (left-hand side of the plot), almost perfect performance is achieved, and while larger values of $1-w$ appear to reduce the level of performance, the {\sc Random forest} still outperforms the {\sc Feature sum} method. For low $p$, the region which shows the different behaviour is also the region in which the feature selection step proposed different features which were not included in the model (namely Paths of size $7$), possibly suggesting that including these features could improve performance in this domain.
The mid-range densities $p=\{0.002,0.003\}$ also show reasonable performance, with performance in the higher densities namely $p=\{0.004,0.005\}$ performing a little worse but still above the {\sc Feature sum} method.

\begin{figure*}[h!]
\centering
\includegraphics[width=\textwidth]{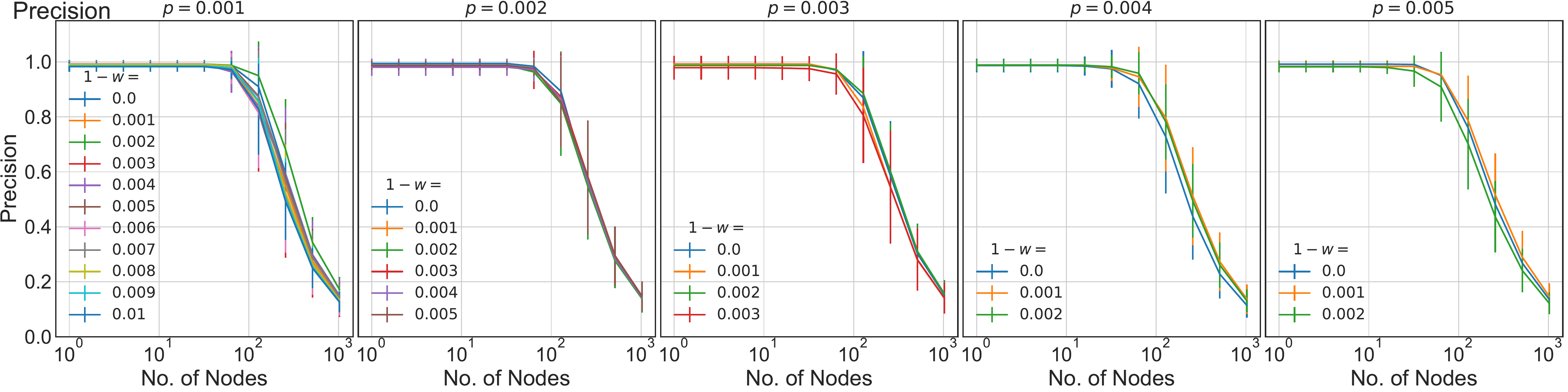}
\includegraphics[width=\textwidth]{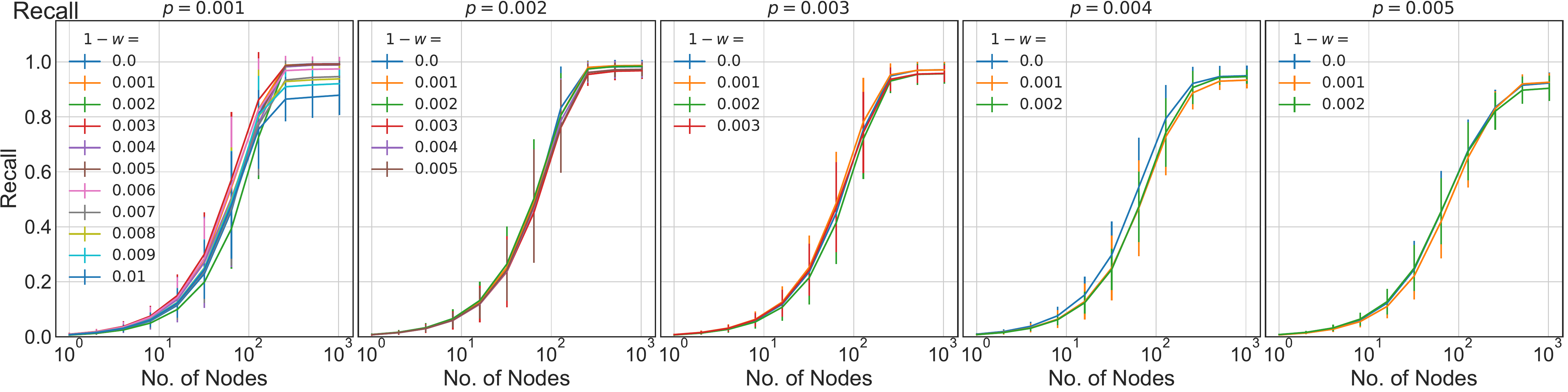}
\caption{\label{fig:fullTestRecallPrecisionPlot} 
The mean precision (upper plot) and recall (lower plot) across each of the parameter regimes. The set of panels on the left shows the mean precision and recall over the training networks of the parameter regimes which have $p=0.001$, using the {\sc Random forest} model to compute the scores considering the $2^{i}$ top scoring nodes where $i=0,...,10$. The error bars are given as one estimated standard error 
over the relevant networks. The remaining panels show the same plots for different densities. 
    }
\end{figure*}

In the recall plots of Fig.~\ref{fig:fullTestRecallPrecisionPlot}, 
the recall performance appears to plateau, indicating that there is a subset of
nodes which is difficult to classify correctly by the {\sc Random
forest}; a similar pattern appears for the {\sc Feature sum} method, see
Fig.~\ref{fig:linearRecallPrecisionPlot}. 
The precision plots in Fig.~\ref{fig:fullTestRecallPrecisionPlot} indicate that, in most networks across all parameter regimes, the first $32$ nodes are all anomalous nodes in all networks in the ensemble, with the first $64$ being almost all anomalies for densities $0.001-0.003$.

\paragraph{Performance of Oddball on the test set.} 

To further test the performance of our method, 
we also compared the performance of {\sc Feature sum} and {\sc Random forest}  to that of Oddball \cite{akoglu2010oddball}, see Sec.~\ref{oddball}. 
\begin{figure*}[h!]
\centering
\includegraphics[width=\textwidth]{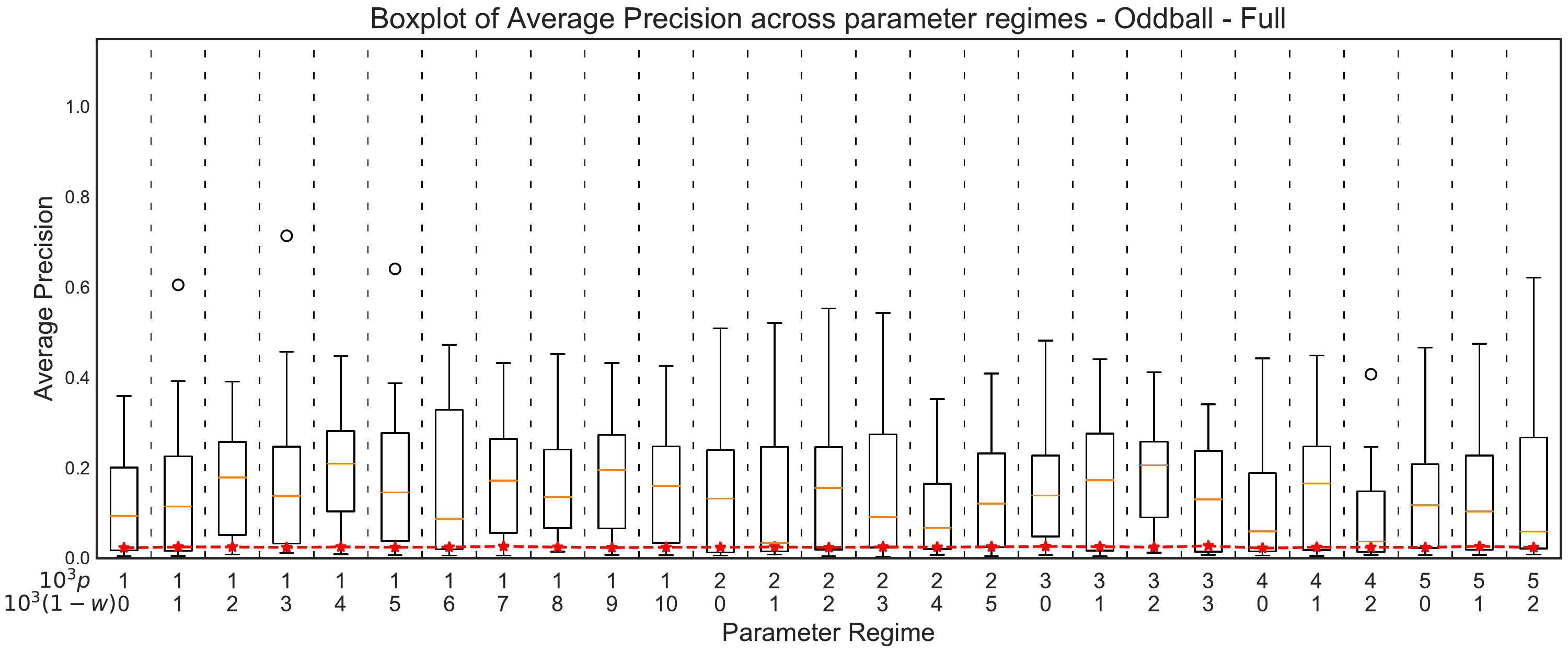}
\caption{\label{fig:oddball1} 
    The average precision for each of the
    parameter regimes, using the sum of Oddball over each of the features. 
With a random ordering, the average precision should be equal to the percentage of anomalous nodes~\cite{sklearn} (see Sec.~\ref{sec:assess} for details).
As our networks have different numbers of anomalies to compare our performance against random, we plot (red star line)
the largest percentage of anomalies over the networks we observe in each parameter regime.}
\end{figure*}
The performance of the combined Oddball statistics can be found in Fig.~\ref{fig:oddball1}, 
and the performance on each of the underlying statistics can be seen in
Appendix~\ref{app:oddballsplit}. The performance of Oddball
appears to be driven by 
summaries that are related to the local density (nodes vs edges) and to the deviations in
average local weight,
namely (and loosely quoting
from the {\it Oddball-lite} code) ``egonet nodes vs egonet edges'',
``egonet edges vs egonet weight'', ``ego out-degree vs ego out-weight'' and
``ego in-degree vs ego in-weight''.

\begin{figure*}[h!]
\centering
\includegraphics[width=\textwidth]{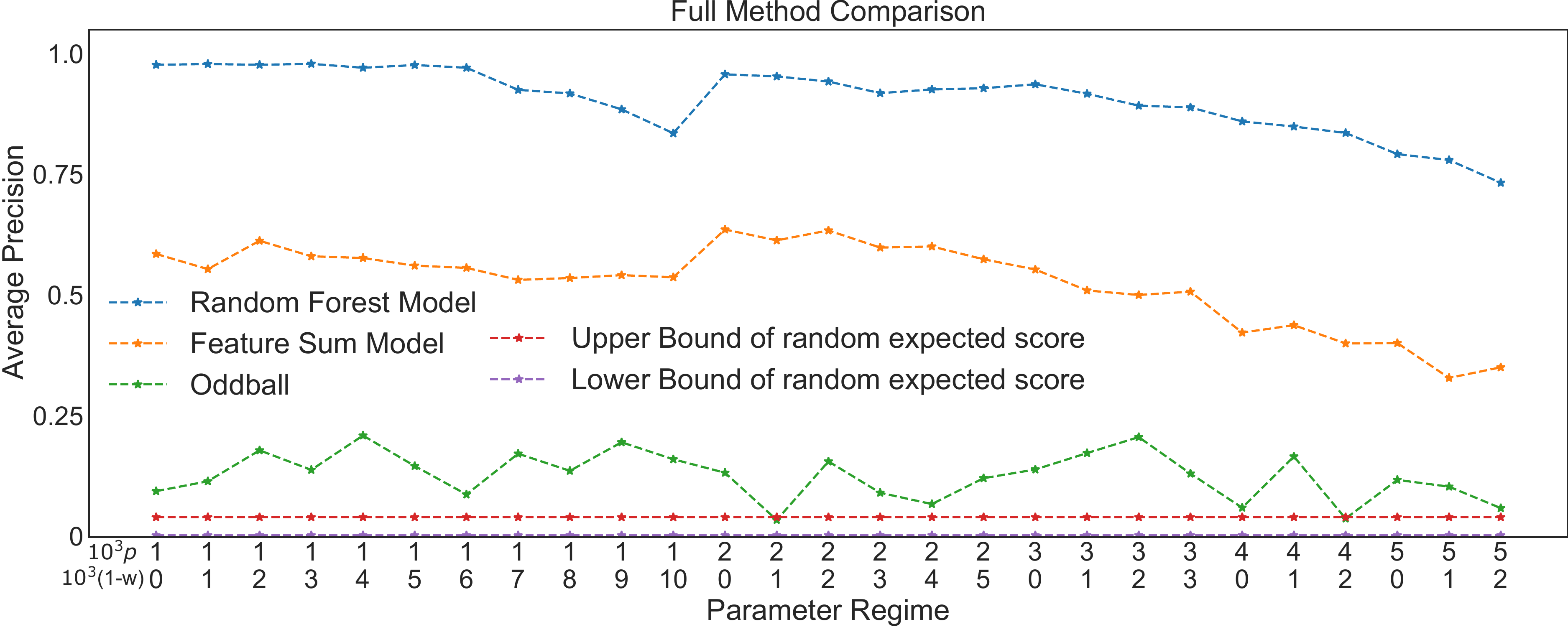}
\caption{\label{fig:combinedPerformance} 
   The median of the average precision from scikit-learn~\cite{sklearn}, for each of the
    parameter regimes. Oddball and the {\sc Random forest} are applied to the test data, whereas the {\sc Feature sum} method is applied to the training data. 
    As the average precision under a random ranking is equal to the percentage of anomalies (see Sec.~\ref{sec:assess} for details), 
    and in this model there are 
    between $25$ and $400$ anomalies embedded (under the assumption of no overlapping structures), we plot the lower bound ($\frac{25}{10000}$) and upper bound ($\frac{400}{10000}$) of the performance of a random ranking. 
    }
\end{figure*}

Fig.~\ref{fig:combinedPerformance} shows a plot of the median of the average
precision; in all parameter regimes the {\sc Random forest} outperforms
Oddball. The {\sc Feature sum} method, ran on the training dataset, also performs
favourably to Oddball, indicating that the underlying features already have
sufficient information to capture to anomalies. 

\vspace{-1mm}
\subsection{Performance on the Accenture Model}
\label{sec:accentureModelRes}
\begin{figure*}[h!]
\centering
\includegraphics[width=\textwidth]{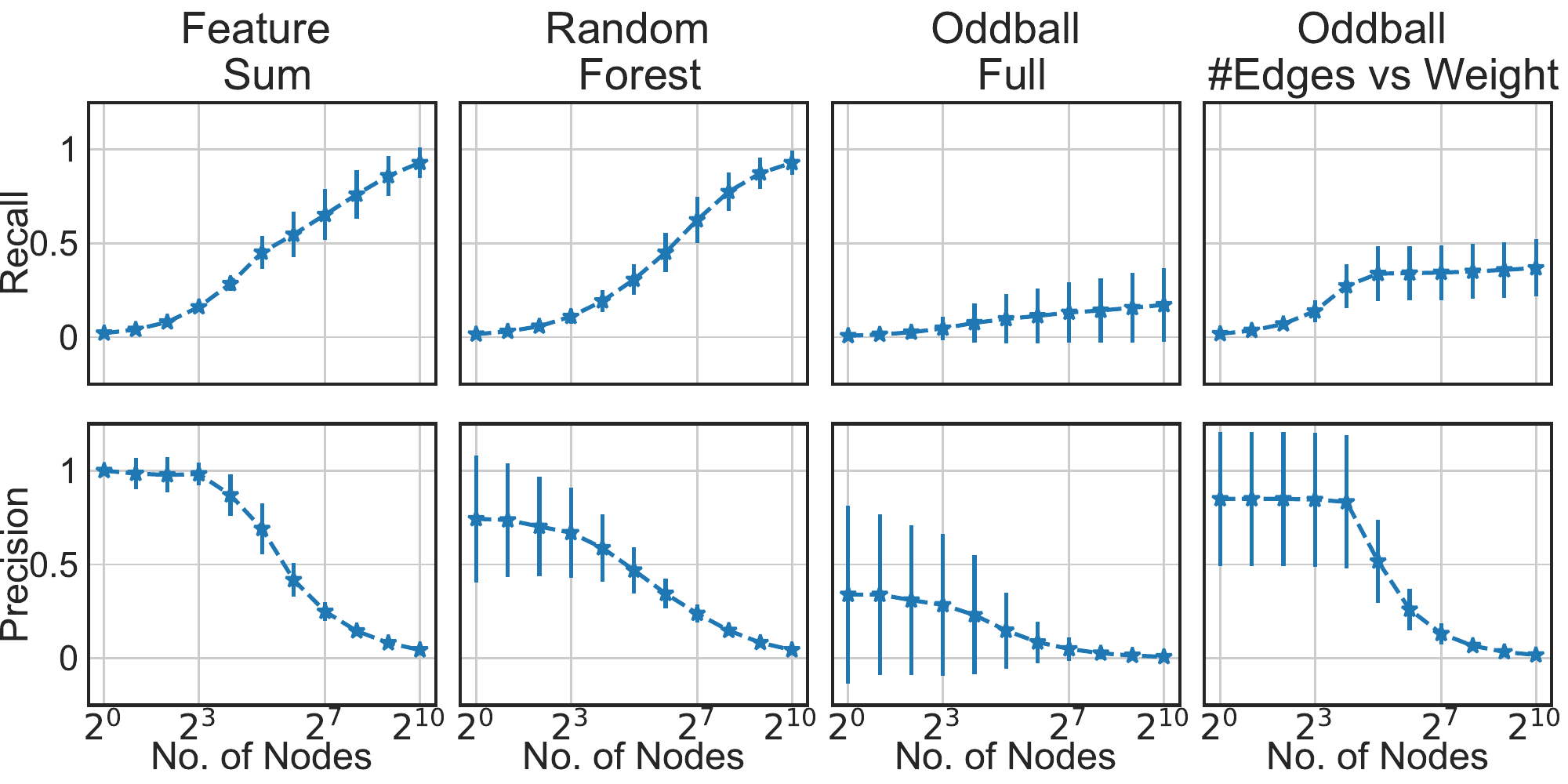}
\caption{\label{fig:accentureRecallPrec} 
   Precision and recall 
   on $100$ networks from the Accenture data generator, using the {\sc Feature sum} and the {\sc Random forest}; averages with error bars (one estimated standard error). 
    }
\end{figure*}

For the Accenture generating model described in
Sec.~\ref{sec:accentureModel},
all of the 
structures from Fig.\ref{fig:anomaliesFramework}, except the path of size $5$, are detectable in the Accenture model under the approximation
that the network is a directed weighted ER network with connection probability $p=19/55000$ and heavy weights exceeding $w=0.99$ (see
Appendix~\ref{sec:accentureModelProof} for details).
We generated $100$ networks from the Accenture model and measured the performance
of the {\sc Random forest} which is obtained using the training data from the weighted ER network in Sec.~\ref{sec:combinationAndFeatureSelection},
and of the {\sc Feature sum}, on this data set, see
Fig.~\ref{fig:accentureRecallPrec}. Unlike in the weighted ER case, the number
of anomalies is fixed in the Accenture model, so that the average of the recall and the
precision is well defined.

\begin{figure*}[h!]
\centering
\includegraphics[width=0.9\textwidth]{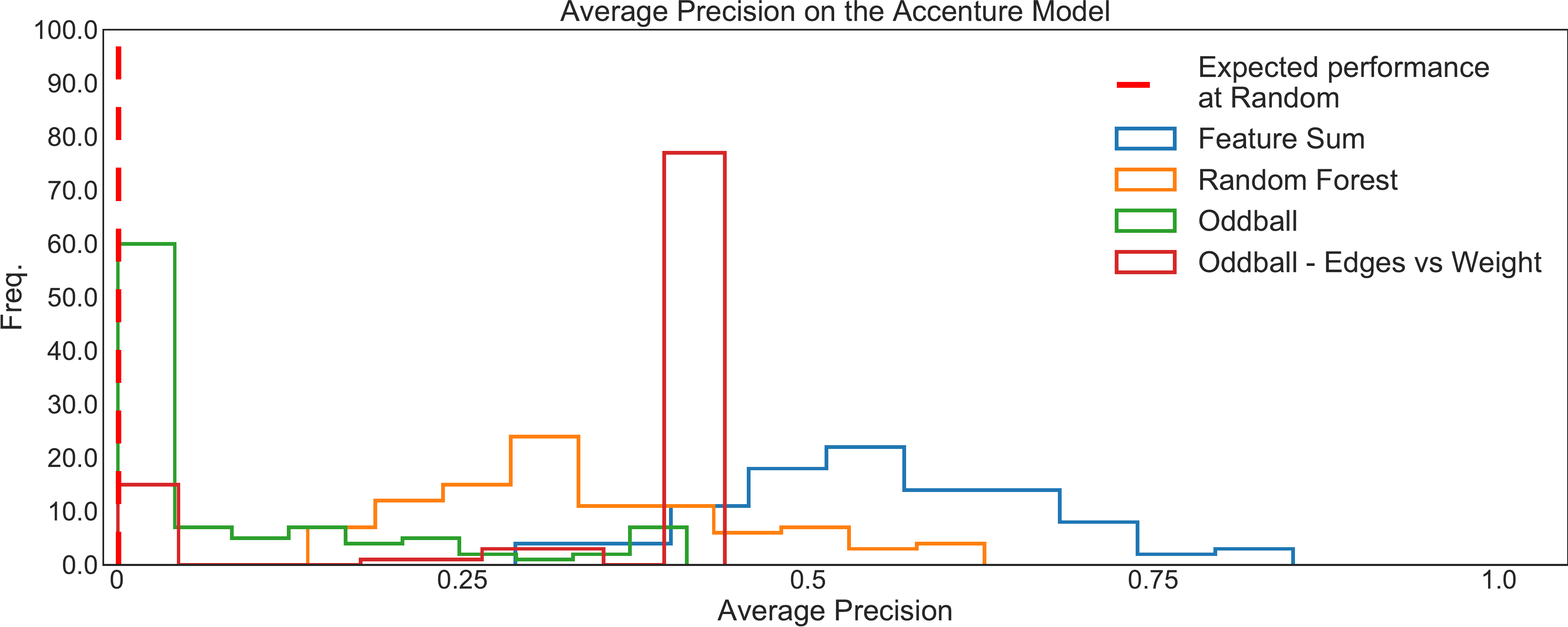}
\caption{\label{fig:accentureHist} 
The average precision on $100$ networks from the Accenture data generator,
using {\sc Feature sum}, {\sc Random forest}, Oddball, and Oddball - Edges vs
Weight. The dotted line shows the performance of the random classifier, which is
indistinguishable from the $x$-axis. 
    }
\end{figure*}

First focusing on the average precision, when analysing Oddball,
one of the
underlying statistics, namely comparing the number of edges and
the weight, outperforms the {\sc Random forest} method,
although it is still outperformed by the {\sc Feature sum}; the results can be seen in Fig.~\ref{fig:accentureHist}. 
An inspection of the recall and precision plots of Fig.~\ref{fig:accentureRecallPrec} indicate that
Oddball Edges vs Weight does very well on the first few nodes, finding the
clique nodes, 
but then, in contrast to  {\sc Feature sum} and {\sc Random forest}, it struggles to find any of the additional planted anomalies.
One additional underlying Oddball statistic also highlighted cliques but similarly struggled to uncover other planted anomalies, see~\ref{app:oddballsplitAccenture}) for full results of the underlying statistics.

Concentrating on the $1024$ nodes with highest scores, as shown in Fig.~\ref{fig:accentureRecallPrec},
both the {\sc Feature sum} and the {\sc Random forest} methods have an
average recall of $0.93$. Both of these methods are, on average, able to recover the vast majority of the embedded anomalies within the top $1.8\%$ of the nodes, thus indicating good performance. With regards to precision, on this data set the {\sc Feature sum} performs better at smaller thresholds, while the {\sc Random forest} does not always achieve a precision close to 1, even for the very top scoring nodes. 

Finally, we consider the average precision of all of the methods as shown in Fig.~\ref{fig:accentureHist}. 
The average precision of the {\sc Random forest} is lower in absolute value than on the weighted ER model, but still considerably higher than the performance of the random classifier.
The {\sc Feature sum} significantly outperforms the {\sc Random forest} on this data set, which could be an indication that the {\sc Random forest} is 
overfitted.
Based on average precision, both the {\sc Feature sum} and the {\sc Random forest}  outperform Oddball, echoing the results from the weighted ER model.

\subsection{Finding Non-planted Structures}
\begin{figure}[ht]
\center
\includegraphics[width=0.9\linewidth]{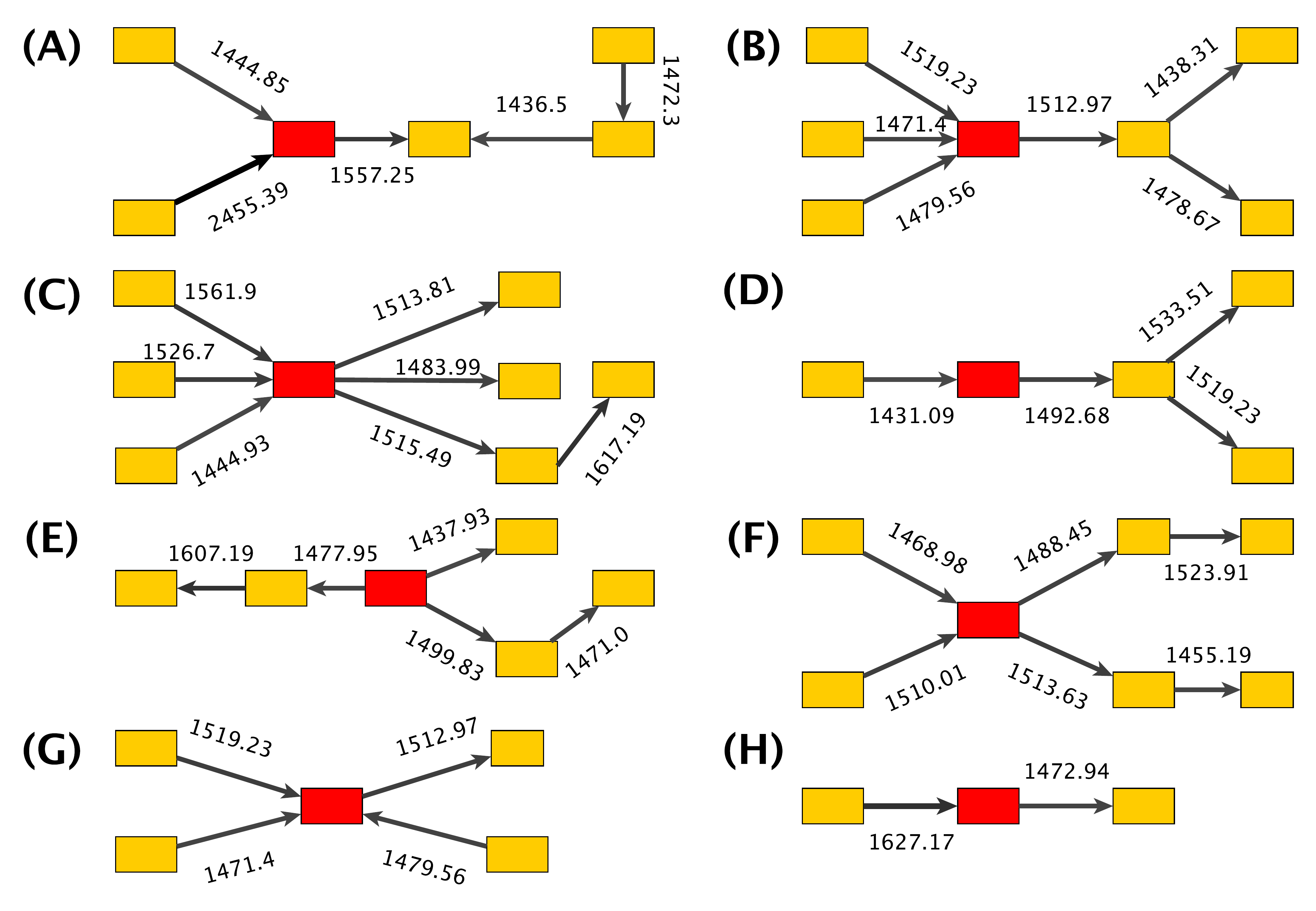}	 \caption{Figures shows the heavy structures around nodes are identified as potential anomalies using {\sc Random Forest} on the Accenture data generator. The identified node is in red with the remaining nodes in yellow, the edges show their weight both by their colour and the annotated weight next to each edge.}
\label{fig:heavySubGraph}
\end{figure}
While we measure the performance based on an embedded set of structures, 
we now illustrate that 
the method can uncover structures that we have not embedded.
To this end, we inspect 
the pattern of connections around those nodes which are ranked highly, yet are not part of an embedded anomaly. Keeping in mind that the random forest scores, based on an average over 10 decision trees, take values in $\{0, 0.1, 0.2, \cdots, 0.8, 0.9, 1\}$, we performed this analysis in the first of our networks from the Accenture benchmark. We considered the top $8$ nodes in the ranking constructed from {\sc Random Forest},  corresponding to those nodes with a score of $0.8$ or higher. To visualise, we plot the two-hop snowball sampled network, around each of these anomalous nodes. For brevity, in the paper, we only display a subnetwork of the 2 hop snowball sample, namely all edges  with weight greater than $1400$ (two standard deviations of the null weight distribution), for which there is a path (ignoring direction) of heavy edges to the seed node. The subnetworks are displayed in Fig.~\ref{fig:heavySubGraph}, whereas the whole snowball samples can be seen in Appendix~\ref{app:nonSelectedNodes}.

We observe that each node has a heavy structure around it.
In many cases,  substantially higher than expected edge weights  are present and 
several heavy flow-like structures, for example in (B), (C), (D), (F), (G) and (H), which
were not embedded anomalies.


\vspace{-1mm}
\section{Discussion}
\label{sec:discussion}
\vspace{-1mm}

\paragraph{Summary} 
This paper presents a novel anomaly detection method which combines
spectral approaches with recent advances in network comparison, to uncover  unknown anomalies in networks. The method is developed with applications to money laundering in mind. Using such spectral and network comparison approaches, as well as exploring paths in the network, we construct $140$ main features. 
We consider two methods to classify nodes based on the resulting features. The first method, denoted by {\sc Feature sum}, simply sums over the 140 main features. 

To improve on the {\sc Feature sum} method, we 
fitted a random forest to the underlying set of $140$ main features. The random forest was trained using a
synthetic network generator which plants anomalies in weighted directed
Erd\H{o}s-R\'{e}nyi networks. Using this network generator, we constructed
$1,890$ training networks with $10,000$ nodes each, sampled evenly across the
parameter range of interest, 
which is determined by requiring that some known anomalies have expected counts less than 1 under the weighted directed Erd\H{o}s-R\'{e}nyi model.
We select $44$ features, focusing on features which have high scores
across the parameter regimes of interest, and we fit a random forest using
these features; we denote the resulting methodology as {\sc Random forest}. It performs favourably on a test set in comparison to {\sc Feature sum}, as well as when compared to random selection and OddBall, a popular anomaly detection method introduced in \cite{akoglu2010oddball}. 

We have also explored in more detail the features which appear to be driving our performance in the weighted ER network, and observed that many features relate to localisation and basic statistics.
Around half of the selected features are driven by eigenvectors of the symmetrised augmented adjacency matrix. 
This fact could be related to the augmentation step, which repeatedly adds edges around heavy paths of size $3$, as described in Sec.~\ref{sec:communityDetection}, and can be thought of as a form of regularisation of a sparse graph. 
Thus, in augmented matrices, heavy structures are turned into cliques.
In contrast, the most important features, as computed by the feature importance score across the parameter ensemble, are dominated by non-augmented statistics related to small structures around nodes, such as the strength of the top $10$\% of edges, or short but heavy weighted paths. 

The {\sc Feature sum} and the {\sc Random forest} 
were employed to detect anomalies in an unseen network model developed by co-authors PR and ML at Accenture. On this network, both the {\sc Feature sum} and the {\sc Random forest } outperform random classification, and more importantly, the Oddball algorithm. Both {\sc Feature sum} and {\sc Random forest} extract, on average, $92.3\%$ of anomalous nodes within the top $1.86$\% ranking nodes.
On the Accenture generating model, using the average precision measure, 
{\sc Feature sum} outperforms {\sc Random forest}. 

We further validated our method by looking at the local neighbourhood of nodes that were not embedded in the graph, yet were highlighted by our method. A visual inspection of the heavy structures (edge weights more than  2 standard deviations above the mean) showed heavy star-like structures, with several examples having a flow-like structure, giving evidence that these structures are indeed anomalous.

The method used in this paper includes a number of parameter choices. Usually,  $p$-values of tests are used when they are below $5$\%; other choices are possible. In the generating models, the weights of the anomalies were chosen such that the anomalies have a fairly strong signal, with other choices being also possible. As a guideline, the edge probability $p$ should be chosen such that the resulting density of the network matches that of the network of interest, and the weight parameter $w$ should be chosen such that anomalies have low expected value (taken as less than 1 in this paper) under the model without added anomalies.

\paragraph{Potential improvements}
There are a few drawbacks of the pipeline we proposed in this paper. First, the limited number of trees in a default \textit{RandomForestRegressor} reduces the number of possible states, which further limits our ability to distinguish between nodes. A solution to this problem would be to simply increase the number of trees, which we will pursue in future work. It should also be noted that a future version of  \textit{RandomForestRegressor} will have $100$ trees by default. Second, the network augmentation used to perform community detection,  starts to break down at higher densities; at such densities, the expected number of heavy paths under the null model is fairly high. Community detection on these dense networks produces a small number of very large communities, along with some very small communities. 
While a high edge density is rarely encountered in large real networks of financial transactions, an exploration of 
replacing the fixed augmentation threshold with a threshold which is adjusted to the network density could be of general interest. 

It remains to note that our implementation of the routines described in this paper is not fast. This is partly a consequence of our choice of programming language (Python), but also of the assumptions about an unknown null model, and the aim to detect general anomalies, which prevented us from specialising either on the anomalies or the null model. There are several future directions one could pursue, in order to make this method substantially faster and thus scalable to very large networks. Firstly, as the run-time of this algorithm is heavily driven by the number of null replicates used, with the complexity of the algorithm scaling multiplicatively with the number of replicates, reducing the number of replicates is a first option. 
Secondly, throughout this paper,  the so-called resolution parameter in modularity-based community detection has been fixed at $1$. To reduce the number of communities, and thus computational complexity, one could modify this parameter. Thirdly, there are clear computational advances from exploiting the embarrassingly parallel nature of the multiple communities, from moving the development language from Python to C, and finally, from taking advantage of new speed-ups in NetEMD~\cite{elliottnetemd}.

\paragraph{Future work}
There are also some promising avenues for further research. Firstly, many of the statistics in our method are expensive to compute. It would be interesting to assess the impact on performance of removing some features, such as long path counts, which appear as important in only a few parameter settings and which take a long time to compute.  Moreover, in future work we shall explore training sets which are less homogeneous than the set we used. We shall also investigate refined weightings of the summands in {\sc Feature Sum}. A more systematic analysis of the non-planted anomalies which the methods detect is also of interest.

Thus far, the method developed in this paper assumes that the underlying network is static. It 
would be interesting to expand these approaches for temporal networks (e.g., as treated in \cite{knight2016modellingpaper}), and to incorporate node level features to uncover anomalies which otherwise cannot be extracted from static topological data alone.
Yet another interesting direction is to no longer consider the symmetrised matrix ${\bf W}^{\bf (s)} = {\bf W}+{\bf W}^T$ (where ${\bf W}$ is the adjacency matrix of the given directed graph), but to rely on recent developments in the literature for clustering directed networks. The algorithms proposed in \cite{DirectedClustImbCuts} directly rely on a (potentially normalised) complex-valued Hermitian matrix representation of the asymmetric adjacency matrix ${\bf W}$, an approach which could also be adapted to the setting considered in our current anomaly detection pipeline.

Finally, in real networks, the nodes which have a high anomaly score should be investigated further with respect to their properties. 
This question falls under the area of behavioural analysis, and hence falls outside the scope of this paper. 

\paragraph{Acknowledgements}
We thank Nicolas Guernion, Stephen Law, Luis Ospina Forero, Martin O'Reilly, Anastasia Shteyn, Kieran Towey, and Mahlet Zimeta for useful discussions and support. This work was funded by EPSRC grant EP/N510129/1 at The Alan Turing Institute and Accenture Plc.

\bibliographystyle{plainnat}
\bibliography{mainbibfile}

\newpage 
\renewcommand\thesection{\Alph{section}}
\setcounter{section}{0}

\section*{Appendix} 
For convenience here is Table of Contents which includes the main body of the paper. 

\tableofcontents

\section{Implementation Details and Special  Cases}

We present additional details on the underlying network construction and the implementation of the methods, paying particular attention to special cases (boundary cases), such as graphs which do not have a sufficient number of eigenvectors. 

\subsection{Configuration Model Implementation}
\label{app:configModelImplementation}
The null model
for the underlying network  is a
slightly modified configuration model. 
The implementation is heavily inspired by the following  procedure from the directed NetworkX implementation~\cite{networkx} of the configuration model on a fixed finite node set with fixed in-degrees and out-degrees. In this setting, each node is given as many outgoing stubs as its out-degree, and as many incoming stubs as its in-degree. Nodes with degree 0 are excluded. The next steps are as follows.  
\begin{enumerate}
    \item Construct a list of outgoing stubs and incoming stubs.  
    \item Construct a list of weights.
    \item Shuffle the list of incoming stubs.
    \item Shuffle the list of weights.
    \item Construct a network from the resultant set of outgoing stubs, incoming stubs and weights, by connecting outgoing stubs with incoming stubs at random.
    \item Remove all self-loops.
    \item For multiple edges, randomly select an edge at random, and remove all others.
    \item Remove all resulting degree 0 nodes.
\end{enumerate}
Note that for multi-edges, an edge is selected uniformly at random  
between the multi-edges regardless of the edge weight. 

\subsection{Localisation}
\label{app:localisationImplementationDetails}

In the computation of localisation, we made a small number of implementation choices which are not obvious from the description of the method in
the main body. In this section, we describe these implementation details.

\subsubsection{Eigenvector Computation}
\label{app:localisation:eigenvectorRoutines}
For computational reasons, in cases where the number of eigenvectors required is less than the
size of the matrix minus one,  SciPy's (Ref.~\cite{scipy}) sparse eigenvector routines are used, which only
computes the top $k$ eigenvectors.

However, during the experiments, the sparse routines were occasionally hanging, which may indicate that the underlying routines were
unable to converge, possibly due to badly conditioned matrices. In contrast, 
the non-sparse eigenvector routines had
comparatively little problem with these matrices. Keeping the same random seed, we used the following procedure.
\begin{enumerate}
    \item Attempt to compute the eigenvectors using the fast sparse methods.
    \item If the fast method has completed within $30$ seconds, then return
        the resultant eigenvectors. 
    \item If the method has not completed within $30$ seconds, then stop the
        fast sparse approach and use the slower eigenvector routine from NumPy
        which computes all of the eigenvectors.
\end{enumerate}

\subsubsection{Node Selection Derivation for Norm-based Localisation Statistics}
\label{app:localalisationNormBasedDerivation}
In many of the approaches in localisation statistics, selecting the nodes that cause the localisation is comparatively simple; for the norm-based approaches (i.e.,  the exponential norm and the inverse participation coefficient), it is a little more arbitrary. A main issue is that the eigenvectors themselves do not include information about how likely they are. To address this, we construct $4$ different statistics using the following procedure. 
\begin{enumerate}
\item
The first statistics are the absolute values of entries of  the eigenvector, divided by  
the absolute value of the constant vector, scaled to have Euclidean norm $1$. 
\item
To identify  nodes for which the eigenvector has an entry larger than expected at
random, we carry out a corresponding Monte Carlo test.
For $p$-values that are below $0.5$, we perform the normal c.d.f.
transformation to obtain a feature vector, for those above $0.5$ we assign it a
value of $0$.
\item For the third statistic, 
the inverse normal c.d.f. of the $p$-value of the
norm-based test statistic which highlighted the localisation in this vector
is assigned to all nodes whose element in the relevant eigenvector
is greater than the average of the largest element in absolute value over the
null distribution. 
\item  Finally, combining the ideas behind the first and third statistic, we follow the procedure in the third statistic, but rather than giving the
relevant nodes the value of the transformed eigenvector level inverse normal
c.d.f. $p$-value, we assign them their element-wise absolute value. 
\end{enumerate}

Mathematically, this is equivalent to the following.
Let $\Psi$ be the set of $m$ eigenvectors generated from the null distribution (shared with the Monte Carlo $p$-value), 
and let $v^{*}$ be the observed eigenvector. Let us represent the value of feature $a$ for node $u$ by $g_a(u)$, $a=1, 2, 3, 4$. Denoting by $n$ the number of nodes, the first feature is given by 
\begin{flalign}
\label{eq:normBasedF1}
g_1(u)= \sqrt{n}|
v^{*}_u|.  
\end{flalign}

For the remaining features, let 
$S=\{ \text{max}_j |x_j| \ : \ x \in \Psi\}$.
Furthermore, let $t(a)$ be the Monte Carlo $p$-value of an observation $a$
from over the empirical null distribution from $S$, 
i.e.  $t(u)=\frac{1}{m+1} +\frac{1}{m+1}\sum_{x \in S}  \mathbbm{1} ( v^{*}_u
\le x)$.
Let $\tilde{s}$ be the average over $S$, and let $p_{val}$ be the $p$-value of the norm-based test statistic which highlighted localisation in this vector. The remaining three features are  
\begin{flalign} 
\label{eq:normBasedF2}
 g_2(u) & = 
\begin{cases}
\Phi^{-1}(1-t(u)) & \text{if } \ \ t(u)<0.5 \\
0 & \text{else,}
\end{cases}
&& \\ 
\label{eq:normBasedF3}
g_3(u) & = 
\begin{cases}
\text{Max}(\Phi^{-1}(1-p_{val}),0) & \text{if } \ \ |v^{*}_u| \ge \tilde{s} \\
0 & \text{else,}
\end{cases}
&& \\ 
\label{eq:normBasedF4}
g_4(u) & = 
\begin{cases}
 |v^{*}_u| & \text{if } \ \ |v^{*}_u| \ge \tilde{s} \\
0 & \text{else}, 
\end{cases}
\end{flalign}
where $\Phi^{-1}$ is the inverse c.d.f. of a mean $0$ variance $1$ normal
distribution.

\subsubsection{Sign-based Localisation Statistics}
\label{implementation:localise:SignBased}

As discussed in the main text, there are a small number of special cases for the
sign-based statistic, but for brevity we did not include them in the main body. 
Here, we adjust the test statistic to address these cases, and then adjust the procedure which constructs features from vectors that appear to deviate from their expectation.

\paragraph{Test Statistic}
First, we let $\theta(G)=\text{Min}(20,\text{Number of Non-Trivial
Eigenvectors})$, and we recall that the sign-based test statistic for
eigenvector $v$ from $G$ is 
$
    \frac{\text{min}(
    N_{pos}(v)
    ,
    N_{neg}(v)
    )}
    {\theta(G)}, 
$
where $N_{pos}(v)= \sum_i \mathbbm{1}(v_i>0)$ and $ N_{neg}(v) =    \sum_i
\mathbbm{1}(v_i<0)$.

Many of the special cases relate to the setting where all entries have the same
sign, resulting in a test statistic of $0$. A test statistic of $0$ is
problematic, as we are interested in cases where a small non-zero number of
entries have the same sign (and thus those entries are anomalous).
Thus,rather than assigning $0$ to the test statistic, we assign it $\theta(G)$, alleviating the problem. The resulting test statistic is
\begin{equation*}
  \frac{1}{\theta(G)}
\text{Min}\Big(
    N_{pos}(v)
    +
    \mathbbm{1}(N_{pos}(v) = 0)
\theta(G)
\ , \
    N_{neg}(v)
    +
    \mathbbm{1}(N_{neg}(v)=0)
\theta(G)
    \Big)
, 
\end{equation*}
which is equivalent to the original test statistic when $N_{pos}\neq 0$ and
$N_{neg}\neq 0$. 

\paragraph{Feature Construction}
As with the other localisation features, in vectors where the test statistic
appears to deviate from expectation, we attempt to highlight the nodes causing
the deviation. In doing so, we consider three special cases, namely
the presence of vectors where $N_{neg}$ or $N_{pos}$ equals $0$, if $N_{neg}
=N_{pos}$, and finally the extremely rare case where more than $50\%$ of the
nodes are considered anomalous. 

The original features, as described in the paper, are given by
\begin{eqnarray*}
    SignStat^{(1)}_i&=& 
    \mathbbm{1}(N_{pos}(v)<N_{neg}(v)) 
    \mathbbm{1}(v_i>0)\Phi^{-1}(1-p_{val})\\
    \nonumber
    &&
+
    \mathbbm{1}(N_{pos}(v)>N_{neg}(v)) 
    \mathbbm{1}(v_i<0)\Phi^{-1}(1-p_{val}), 
    \\
    SignStat^{(2)}_i&=& 
    \frac{\mathbbm{1}(N_{pos}(v)<N_{neg}(v)) 
    \mathbbm{1}(v_i>0)\Phi^{-1}(1-p_{val})
}{N_{pos}(v)}\\
    \nonumber
&& +
    \frac{\mathbbm{1}(N_{pos}(v)>N_{neg}(v)) 
    \mathbbm{1}(v_i<0)\Phi^{-1}(1-p_{val}), 
}{N_{neg}(v)}
\end{eqnarray*}
where $\Phi^{-1}$ is the inverse of the c.d.f. of a normal distribution with mean $0$
and variance $1$, and $\mathbbm{1}$ is the indicator function.

When $N_{pos}$ or $N_{neg}$ equals zero, we
replace $N_{pos}$ with $\widehat{N_{pos}}=
N_{pos}+\mathbbm{1}(N_{pos}=0)|v|$ 
and we replace $N_{neg}$ with 
( $\widehat{N_{neg}}=
N_{neg}+\mathbbm{1}(N_{neg}=0)|v|$), leading to the following features

\begin{eqnarray}
\label{eq:signStat1}
    SignStat^{(1)}_i&=& 
    \mathbbm{1}(\widehat{N_{pos}}(v)<\widehat{N_{neg}}(v)) 
    \mathbbm{1}(v_i>0)\Phi^{-1}(1-p_{val})\\
    \nonumber
    &&
+
    \mathbbm{1}(\widehat{N_{pos}(v)}>\widehat{N_{neg}}(v)) 
    \mathbbm{1}(v_i<0)\Phi^{-1}(1-p_{val} , 
    \\
\label{eq:signStat2}
    SignStat^{(2)}_i&=& 
    \frac{\mathbbm{1}(\widehat{N_{pos}}(v)<\widehat{N_{neg}}(v)) 
    \mathbbm{1}(v_i>0)\Phi^{-1}(1-p_{val})
}{\widehat{N_{pos}}(v)}\\
    \nonumber
&& +
    \frac{\mathbbm{1}(\widehat{N_{pos}}(v)>\widehat{N_{neg}}(v)) 
    \mathbbm{1}(v_i<0)\Phi^{-1}(1-p_{val})
}{\widehat{N_{neg}}(v)} ,
\end{eqnarray}
which equal  their original formulation in the cases where $N_{pos}\neq 0$ and  $N_{neg}\neq 0$.

The second special case occurs if $N_{neg}=N_{pos}$; in this case, we note that the
features as defined in~\eqref{eq:signStat1} and~\eqref{eq:signStat2}
are always $0$. This case can happen if the vector has a large number of $0$'s
and a small number of non-zero entries, thus allowing the test statistic to be
significant, while $N_{pos}=N_{neg}$.  To address this, we introduce 
\begin{flalign}
\label{eq:signstatequal1}
    SignStatEqual^{(1)}_i= 
    \mathbbm{1}(\widehat{N_{pos}}(v)=\widehat{N_{neg}}(v)) 
    (\mathbbm{1}(v_i>0) + \mathbbm{1}(v_i<0)) 
    \Phi^{-1}(1-p_{val})
    \\
\label{eq:signstatequal2}
    SignStatEqual^{(2)}_i= 
    \frac{\mathbbm{1}(\widehat{N_{pos}}(v)=\widehat{N_{neg}}(v)) 
    \mathbbm{1}(v_i>0)
+
    \mathbbm{1}(v_i<0))
    \Phi^{-1}(1-p_{val})
    }{\widehat{N_{pos}}+\widehat{N_{neg}}}.
\end{flalign}
There is one additional extremely rare edge case for this statistic. If many elements of the null distribution have the same sign, then any observed vectors with more than $50$\% of their entries with one sign but the remaining entries $0$ can appear significant. To address this issue and to avoid adding noise to the feature set, we exclude these vectors from the features, and we simply create a new set of features for the cases where the entries are large.
Finally, as the extremely rare case does not appear in our data sets, we exclude the related features from our main set of $140$ features; we list them in the additional feature section in Sec.~\ref{additionalFeatures}.

\subsubsection{Configuration Null Model Replicates which have Insufficient Size}
\label{localisation:configModelNulls}
Sometimes, a null model replica is too small to have sufficient eigenvectors to
run a comparison to the real data. The underlying reason for this is that in
the stub-based configuration model described in  Appendix~\ref{app:configModelImplementation}, the removal of degree $0$ nodes can affect the number of eigenvectors. When the networks are large,  the
impact of these missing nodes is small. However, an issue arises in Monte Carlo
tests when some simulated networks are so small that the size of the network
drops below the number of eigenvectors required for the null comparison. 

To overcome this issue, as described in the main body, we re-sample any networks that is too small until a network is generated that has sufficient size; 
we stop 
at the $10$\textsuperscript{th} generated network. As long as the $10$\textsuperscript{th} generated network has at least $2$ nodes, we use that network; otherwise, we re-sample again, repeating the loop for as long as necessary. 
As our method requires fewer than 25 eigenvectors, and the communities uncovered in the generating model are often much larger than this, 
the resampling only required in a small percentage of communities.

To assess statistical significance in our Monte Carlo tests when the null replicas have insufficient size, 
we  compare the $i^{th}$ real eigenvector only with networks that have at least $i$ eigenvectors, thus giving a well defined comparison.

\subsection{NetEMD Implementation Details}
\label{app:netemdSpec}
NetEMD 
standardises empirical 
distributions and then computes the minimum EMD over translations. 
Distributions in which all the mass is concentrated on a single point do not
undergo variance scaling, as the variance is $0$.  
However, when using floating point features, such as eigenvectors, 
variance scaling will inflate 
the small floating point difference.
To address this issue, if the range of a
statistic of the network is less than $10^{-10}$, we replace the statistic with a point
distribution.

\subsection{PathFinder}
\label{app:implementation:path}
 
Here we discuss how we construct our original paths of size $3$ from which to expand.
\paragraph{Selecting the Starting Paths of Size $3$}
To avoid considering all paths of size $3$, which can be very large, we start
with a subset of paths of size $3$, and then
we 
expand on the {\it BeamWidth} paths with the highest
fitness,
where fitness is smallest edge weight in a path.  The procedure to construct our paths of size
$3$ is as follows. 
\begin{enumerate}
\item Make a List ({\it
PathsToConsider}) with {\it BeamWidth} dummy paths of weight $0$: 
\item For each node ({\it CurrentNode}) in the network: 
\begin{enumerate}
\item Consider the node ({\it CurrentOut}) with highest-weighted outgoing
edge to {\it CurrentNode}. 
\item Consider the node ({\it CurrentIn}) with highest-weighted incoming
edge from {\it CurrentNode}. 
\item If the minimum fitness of the paths in {\it PathsToConsider} is less
than the fitness of the path {\it CurrentIn}  -> {\it CurrentNode} -> {\it
CurrentOut}. 
\begin{enumerate}
\item Add path to {\it PathsToConsider}, removing the path with the lowest fitness.
\item If there are no more incoming and outgoing edges to consider, go to {\bf 2}
and consider the next {\it CurrentNode}.
\item If the largest unconsidered incoming edge to {\it CurrentNode} has a higher weight than
the largest unconsidered outgoing edge from {\it CurrentNode}, then update {\it CurrentIn} to the
relevant node and go to {\bf (c)}. 
\item Update {\it CurrentOut} to the node corresponding to the largest
unconsidered outgoing edge and go to {\bf (c)}. 
\end{enumerate}
\item Go to {\bf 2} and consider the next {\it CurrentNode}
\end{enumerate}
\end{enumerate}

The order in which the nodes are considered does play a role in the
selection of the paths of size $3$, as the if statement in {\bf (c)} will
only add paths if they have higher fitness than those previously observed.

\paragraph{Underlying Data structure, and Breaking Ties}

Our underlying implementation of the method uses a fixed size heap 
(using the Python module heapq) to which we are adding and removing elements. The computational complexity of adding an element to a heap and removing the least fit element is
$O(\log(\text{number of elements in the heap}))$. We set the size of the heap
to the BeamWidth. Adding the $\rho$ path extensions has
complexity $O(\rho \log(\text{number of}$ $\text{elements in the heap/ beam width}))$, 
with $O(\text{number of}\ \allowbreak\text{elements in the heap/ beam width})$ storage.

As we are using the inbuilt heapq module, we break ties in fitness using
the inherent Python 2 order of lists, which compares the first element, and then the second and so on. The lists in question are the lists of node indices involved in a path arranged in the order
they are visited, in which larger indexes are considered fitter.  

\section{Regions of Detectability}

In the main body of the paper, we discussed limiting the parameter regimes to
those where the expected number of anomalous networks that would occur at random
is less than $1$. In this section of the appendix, we derive the expression
used in the paper (Sec.~\ref{app:inequalities}) and we show that the case of
path $5$ is the most restrictive case (Sec.~\ref{app:fiveMostRestrict}).

\subsection{Derivation of Regions of Detectability}
\label{app:inequalities}
Anomalies will only be detectable if they are unusual in some way, in the underlying network. In this paper, we require that the expected counts of the anomaly in a network without planted anomalies is less than 1. The parameter region which ensures that the expected counts are less than 1 is what we call the {\it{region of detectability}} for that anomaly. 
To compute the bounds on the expected number of anomalies, let $f^{(A(k,w))}(G,X)$ be the number of  anomalies of type $A$ of size $k$ and minimum weight $w$, observed on the subnetwork of $G$  induced by the nodes in $X$. Let the total number of such anomalies over the network be
$F^{(A(k,w))}(G) = \sum_{\substack{X \subseteq V \\ |X|=k}} f^{(A(k,w))}(G,V)$. We assume that the function $f$ preserves the exchangeability in the sense that the second equality in \eqref{eq:linearExpectation} holds
\begin{flalign}
\label{eq:linearExpectation}
    E[F^{(A(k,w))}(G)] 
    = 
    \sum_{
        \substack{X \subseteq V \\ |X|=k}
    }
    E[f^{(A(k,w))}(G,X)] 
    = 
    {n \choose k}
    E[f^{(A(k,w))}(G,X_{arb})],  
\end{flalign}
where the final equality comes from the fact that all nodes in an ER network are exchangeable, and $X_{arb}$ is an arbitrary set of nodes of size $k$. For example, subnetwork counts preserve the exchangeability. 

Here, we calculate the expectations under a directed ER model with edge probability $p$ and edge weight $W_{ij} \sim Uniform (0,1)$. An edge is {\it{heavy}} if its edge weight $W_{ij}$ is at least $w$, where $w$ is a fixed number. Thus the probability of seeing a heavy edge is $p(1-w)$. 
The structures for which we find the expected counts are 
\begin{enumerate}
\item a clique on $k$ nodes in which there is at least one heavy edge between any two nodes;
\item a star on $k$ nodes in which the centre has in-degree at least $k_1$ and out-degree at least $k-k_1 - 1$, with the additional requirement that the $k_1$ in-degree edges come from different nodes than the nodes that the $k-k_1 - 1$ out-degree edges connect to;
\item a path on $k$ nodes $v_1, \ldots, v_k$ starting in $v_1$, with a directed heavy edge from $v_i$ to $v_{i+1}$, for $i=1, \ldots, k-1$; 
\item a ring of $k$ nodes - a directed completion of a heavy path; 
\item a particular tree with 1 root (the inner shell) connecting to 3 nodes (the middle shell) connecting to 5 nodes each (the outmost shell), so that there are 9 nodes in the tree and $ 5 \times 3 + 3 \times 1 = 18$ edges in total, with the specified connecting edges again being heavy.  
\end{enumerate}
Note that, in all cases, we do not condition on the edges which are not specified by the anomaly. 
The calculations are as follows. 

\begin{enumerate}
\item 
The derivation for the clique is as follows.  The probability of observing a
clique of size $k$ on a fixed set of $k$ nodes in an undirected ER
network is given by $p^{{k \choose 2}}$. Further, the probability of
observing an edge in either direction in a directed ER network is given
by $1-(1-p)^2$ (one minus the probability of there being no edge in
either direction). Observing that the probability of a heavy edge in our
network model is given by $p(1-w)$,  we obtain the following
inequality
\begin{flalign*}
   1 & >  E[F^{(A^{(clique)}(k,w))}(G)] 
    \\
   & =  {n \choose k}
    E[f^{(A^{(clique)}(k,w))}(G,X_{arb})] \\
    & =
    {n \choose k}
    (1-(1-(1-w)p)^2)^{{k \choose 2}}
    \end{flalign*}
    which is satisfied when 
 \begin{eqnarray} \label{eq:clique}
    (1-w)p
    & < 
    1-\left(1-\left({{n \choose k}}^{-1}\right)^{\frac{1}{{k \choose 2}}}\right)^{\frac{1}{2}}. 
\end{eqnarray}
\item The derivation for the star is as follows. We calculate the expected counts for stars on $k$ nodes. To make the computation easier, we parameterise the number of nodes that have incoming heavy edges into the star as $k_1$, where $k_1<k$. For a star with $k$ nodes, the probability that $k-1$ heavy edges occur is $((1-w)p)^{k-1}$.
Starting with a set of $k$ nodes, there  are ${k \choose k_1}$ ways of
selecting $k_1$ nodes with edges into the centre node. There are then $k-k_1$
ways of choosing the central node, which also fixes the number of nodes with
edges from the centre node. Putting this all together in an expression for
$f^{(A^{(Star)}(k,w))}(G,X_{arb})$, we arrive at 
\begin{flalign*}
E[f^{(A^{(Star)}(k,w,k_1))}(G,X_{arb})] & =  \sum_{ \substack{X_1,X_2,X_3 \subset X_{arb} 
\\
X_1 \cup X_2 \cup X_3 =X_{arb}\\ 
|X_1 \cap X_2|= |X_2 \cap X_3|= 
|X_1 \cap X_3| =0 \\ 
|X_1|=k_1,\ |X_2|=1,
}}
((1-w)p)^{k-1} 
\\
& =
{k \choose k_1} 
{k-k_1 \choose 1} 
((1-w)p)^{k-1} .
\end{flalign*}
Substituting this into equation~\eqref{eq:linearExpectation} and rearranging we obtain
\begin{flalign*}
 1 & >    E[F^{(A^{(Star)}(k,w,k_1))}(G)] 
    = 
    {n \choose k}
    {k \choose k_1}{k-k_1 \choose 1}
    ((1-w)p)^{k-1}
    \end{flalign*} 
    giving that 
    \begin{eqnarray} \label{eq:star} 
    (1-w)p 
    < 
    \left(
    \frac{1}
    {
    {n \choose k}
    {k \choose k_1}{k-k_1 \choose 1}
    }
    \right)^{\frac{1}{k-1}}.
\end{eqnarray}

\item The derivation for the heavy path is similar, as it also has $k-1$ heavy edges. In this case, there are $k$ unique positions and therefore $k!$ different ways to assign the nodes in $X_{arb}$ to the positions. Substituting this into equation~\eqref{eq:linearExpectation} and rearranging leads us to 
\begin{flalign*}
  1 & >  E[F^{(A^{(Path)}(k,w))}(G)] 
    \\
    &= 
    {n \choose k}
    (k)!
    ((1-w)p)^{k-1}
\end{flalign*}
which is satisfied if 
\begin{flalign*}
   (1-w)p
    & < 
    \left(
    \frac{1}{
    {n \choose k}
    (k)!
}\right)^{\frac{1}{k-1}},
\end{flalign*}
giving the expression \eqref{eq:path} in the main text. 
\item
In the case of the ring, due to its rotational symmetry  we  fix the position of one node and then note that there are $(k-1)!$ ways of placing the remaining nodes. 
Substituting this into equation~\eqref{eq:linearExpectation} and rearranging we obtain
\begin{flalign*}
  1 & >  E[F^{(A^{(Ring)}(k,w))}(G)] 
    \\
    &= 
    {n \choose k}
    (k-1)!
    ((1-w)p)^{k}
\end{flalign*}
giving that 
\begin{eqnarray}\label{eq:ring} 
    (1-w)p 
    & < 
    \left(
    \frac{1}{
    {n \choose k}
    (k-1)!
}\right)^{\frac{1}{k}}.
\end{eqnarray}
\item 
The final structure is the tree with $5$ nodes in the outmost shell, $3$ nodes in
the middle shell and $1$ node in the inner shell, so that there are  $5\times 3+ 3\times 1 = 18$ edges. Fixing $k$ as $9$, as this structure is only defined on $9$ nodes, there are ${9 \choose 5}$ ways of selecting the first shell nodes, and ${4 \choose 1}$ ways of choosing the third shell. Substituting this into equation~\eqref{eq:linearExpectation} and rearranging we obtain
\begin{flalign*}
  1 & >   E[F^{(A^{(Tree)}(k,w))}(G)] 
    \\
    &= 
    {n \choose 9}
    {9 \choose 5}{4 \choose 1}
    ((1-w)p)^{18}
  \end{flalign*}
which amounts to  
\begin{eqnarray} \label{eq:tree}
    (1-w)p
    & < 
    \left(
    \frac{1}{
    4{n \choose 9}
    {9 \choose 5}
}\right)^{\frac{1}{18}}.  
\end{eqnarray}
\end{enumerate}

\subsection{Proof that Path $5$ is the most restrictive case for $k \ge 5$}
\label{app:fiveMostRestrict}
To validate our use of Path $5$ as the most restrictive structure among all structures on at least $k=5 $ nodes (not taking the special case of the tree into account) we compare the respective expectations. 
First we show that a path of size $5$ is more restrictive than any other path size, and then we  demonstrate that a path of size $k$ is more restrictive than another structure of size $k$.

Let us first consider the path of size $k$ and a path of size $k+1$ where $k \ge 5$. We first note that
$$
\left(
\frac{1}{{n \choose k}
(k)!}\right)^{\frac{1}{k-1}}
=
\left(
\prod_{i=0}^{k-1}\frac{1}{n-i}
\right)^{\frac{1}{k-1}}.
$$
Now let us consider the ratio between the bound for a path of size $k$ and the bound for a path of size $k+1$.
We first note that for $\alpha<0$ and $\beta < k <n$, then $(n-\beta)^{\alpha}< (n-k)^{\alpha}$, then 
\begin{flalign*}
\frac
{
\left(
\prod_{i=0}^{k-1}\frac{1}{n-i}
\right)^{\frac{1}{k-1}}
}
{
\left(
\prod_{i=0}^{k}\frac{1}{n-i}
\right)^{\frac{1}{k}}
}
&
=
\frac
{
\left(
\prod_{i=0}^{k-1}\frac{1}{n-i}
\right)^{\frac{1}{k-1}}
}
{
\left(
\frac{1}{n-k}
\right)^{\frac{1}{k}}
\left(
\prod_{i=0}^{k-1}
\frac{1}{n-i}
\right)^{\frac{1}{k}}
}
=
{
\left(
n-k
\right)^{\frac{1}{k}}
}
\prod_{i=0}^{k-1}
\left(
n-i
\right)^{\frac{1}{k}-\frac{1}{k-1}}
\\ &
<
{
\left(
n-k
\right)^{\frac{1}{k}}
}
\prod_{i=0}^{k-1}
\left(
n-k
\right)^{\frac{1}{k}-\frac{1}{k-1}}
=
\left(
n-k
\right)^{
1-\frac{k}{k-1}+\frac{1}{k}}
=
\left(
n-k
\right)^{
\frac{
-1
}{k(k-1)}} < 1.
\end{flalign*}
We note that the first equality comes from observing that $\frac{1}{k} - \frac{1}{k-1}<0$, and applying the inequality  
$(n-\beta)^{\alpha}< (n-k)^{\alpha}$.
As the power of the penultimate statement is negative, the ratio is less than $1$. Hence  the sequence is monotonically increasing, and therefore path $5$ is the most restrictive case of the path-based options. 

Let us now consider the remaining structures. First consider the ring, and consider the ratio between a ring of size $k$ and a path of size $k$: 
\begin{flalign*}
\frac{
    \left(
    \frac{1}{
    {n \choose k}
    (k-1)!
}\right)^{\frac{1}{k}}
}
{
\left(
\prod_{i=0}^{k-1}\frac{1}{n-i}
\right)^{\frac{1}{k-1}}
}
=
\frac{
    \left(
k\prod_{i=0}^{k-1}\frac{1}{n-i}
\right)^{\frac{1}{k}}
}
{
\left(
\prod_{i=0}^{k-1}\frac{1}{n-i}
\right)^{\frac{1}{k-1}}
}
=
k^{\frac{1}{k}}
\prod_{i=0}^{k-1}
    \left(
{n-i}
\right)^{\frac{1}{k-1}-\frac{1}{k}} \ge 1.
\end{flalign*}
For $k<n$, each term in the product is greater than or equal to $1$, and therefore the product is also, implying that the ratio is greater than $1$ and thus the bound for the ring is larger. 

Next,  let us consider an analogous calculation for the star graph  
\begin{flalign*}
\frac{
    \left(
    \frac{1}
    {
    {n \choose k}
    {k \choose k_1}{k-k_1 \choose 1}
    }
    \right)^{\frac{1}{k-1}} 
    }
    {
    \left(
\frac{1}
{
{n \choose k}(k)!
}
\right)^{
\frac{1}{k-1}
}
}
=
    \left(
    \frac{k!}
    {
    {k \choose k_1}(k-k_1)
    }
\right)^{
\frac{1}{k-1}
}
=
    \left(
    \frac{k_1!(k-k_1)!}
    {
    k-k_1
    }
\right)^{
\frac{1}{k-1}
}
=
    \left(
    k_1!(k-k_1-1)!
\right)^{
\frac{1}{k-1}
}
\ge 1, 
\end{flalign*} 
where the penultimate equality holds as we know that $k_1<k$ - the case $k_1 = k$ is not possible because one node must be the centre node. Clearly, this ratio is greater than $1$, therefore the path case is more restrictive.
 
For the clique, the bound is given by the following calculation 
\begin{flalign}
    1-\left(1-\left({{n \choose k}}^{-1}\right)^{\frac{1}{{k \choose 2}}}\right)^{\frac{1}{2}}
   \ge 
    1-\left(1-\left({{n \choose k}}^{-1}\right)^{\frac{1}{{k \choose 2}}}\right)
    =
    \left({{n \choose k}}^{-1}\right)^{\frac{1}{{k \choose 2}}}
\end{flalign}
As ${n \choose k}^{-1}<1$,  the quantity in the main bracket is between $0$ and $1$ and thus as $x\ge x^2$ for $0\le x \le 1$, the inequality holds. 
Therefore let us consider the ratio of the lower bound and the path bound on $k$ nodes:
\begin{flalign*}
\frac{
    \left({{n \choose k}}^{-1}\right)^{\frac{1}{{k \choose 2}}}
    }
    {
  \left(
  \frac{1}
  {
  {n \choose k}(k)!
  }
  \right)^{
  \frac{1}{k-1}
  }
}
=
\frac{
    \left({{n \choose k}}^{-1}\right)^{\frac{1}{\frac{1}{2}k(k-1) }}
    }
    {
  \left(
  \frac{1}
  {
(  {n \choose k}(k)!)^{\frac{1}{2}k}
  }
  \right)^{
  \frac{1}{\frac{1}{2}k(k-1)}
  }
}
=
\left(
\frac{
({n \choose k}(k)!)^{\frac{k}{2}}
    }
    {
    {n \choose k}
  }
\right)^{\frac{1}{\frac{1}{2}k(k-1) }}
=
\left(
{n \choose k}^{\frac{k}{2}-1}
((k)!)^{\frac{k}{2}}
\right)^{\frac{1}{\frac{1}{2}k(k-1) }}
\ge 1.
\end{flalign*}
The final term is greater than $1$ as by assumption $k\ge 5$, thus $\frac{k}{2}-1$ is greater than $0$, and thus all terms in the product are greater than $1$. Thus the ratio is greater than one, and therefore the path case is more restrictive. 

Finally, as  the tree  is of fixed size, we can simply compare the values directly. The bound for the tree is $\approx 0.0144154$, whereas the bound for a path of size $5$ is around $10^{-5}$, indicating that the path bound is more restrictive.
Hence the smallest overall bound for the subnetwork probabilities under
consideration is the bound for paths of size 5.

\begin{figure}
\centering
\includegraphics[width=1.0\textwidth]{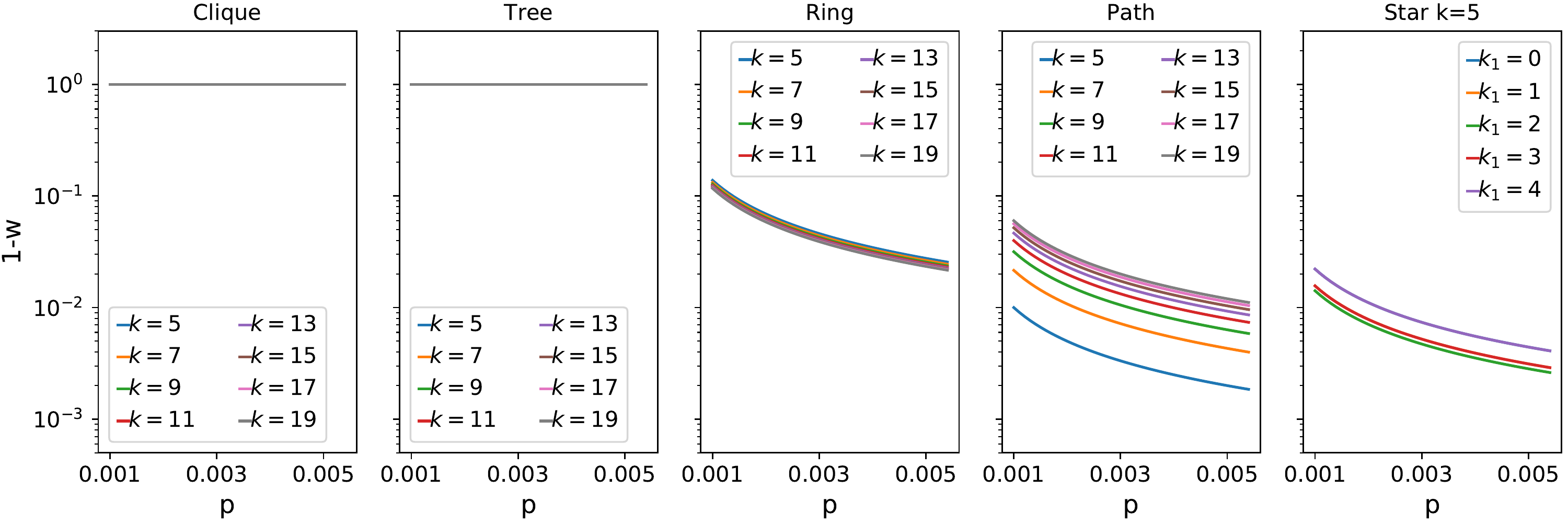}
    \caption{\label{fig:fulllimits} The bounds on the values of $p$ and $1-w$ such that the expected number of anomalous structures are less than one. Each plot shows the bounds of a given structure (described in the title) of different sizes, with the exception of the final panel, which shows a star of size 5 with differing numbers of incoming and outgoing edges.}
\end{figure}

\begin{figure}
\centering
\includegraphics[width=1.0\textwidth]{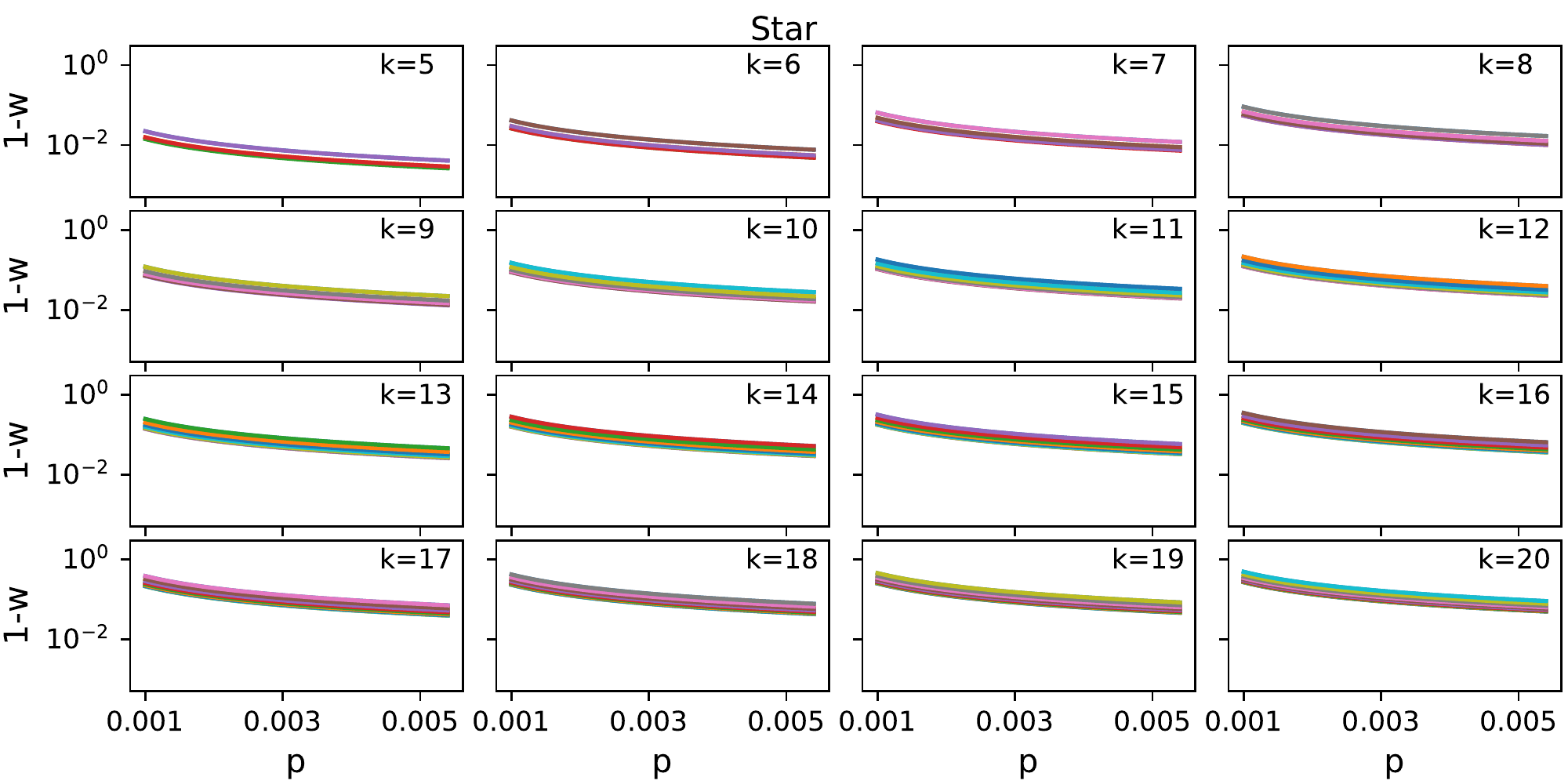}
    \caption{\label{fig:fulllimitsStar}
   The bounds on the values of $p$ and $1-w$ such that the expected number of stars of a given size and a given number of incoming connections are less than one. The label in each plot shows the size of the structure in each panel; each line in the panel represents a different number of incoming edges.} 
\end{figure}

Fig.~\ref{fig:fulllimits} shows the bound for different sized versions of
each structure. In the case of the clique, tree, ring and path we plot
structures of size $k=5,7,...,19$, whereas for the star we fix $k=5$ and vary the number of incoming edge $k_1=0,2,...,4$. We observe that the curve for the Path of size $5$ is the most restrictive of all ($4^{th}$-panel blue line). To confirm that there is not a star with a combination of $k$ and $k_1$ which would outperform Path $5$,  
we  plot each combination in Fig.~\ref{fig:fulllimitsStar}. The restrictions for $k=5$ are stronger than those of all other sizes, thus illustrating the analytical result. 

\subsection{Approximate Detectability of  the Anomalies in the Accenture Model}
\label{sec:accentureModelProof}
We claimed in the paper that the planted structures (with the exception of the path of size $5$)
in a simplified version of the model
constructed by Accenture are detectable. In this section, we provide justification for this
assertion. 

The Accenture model fixes the in- and out-degree sequence of the network using draws from different
normal distributions (mean $21$ and standard deviation $3$ for the incoming
edges, and mean $19$ with standard deviation of $2$ for the outgoing edges). To guarantee a feasible in- and out-degree sequence, the model
 reduces the in-degree or the out-degree of randomly selected nodes until
the sum of the degree sequences match. 

While analysing this procedure exactly is difficult, the number of connections is likely to
be driven by the out-degree as it has a smaller mean. 
Then, the relevant bounds for a directed weighted ER network with $p=19/55000$ are a reasonable approximation to the detectability of the planted anomalies under this model. 

Applying Eq.~\eqref{eq:clique},  Eq.~\eqref{eq:ring} and Eq.~\eqref{eq:path} to this
case with $n=55,000$, with cliques of sizes $8$ and $12$; rings of sizes $4$ and $10$, and heavy paths of sizes $5$ and $10$ we observe the following 
\vspace{-1mm}
\begin{flalign*}
&    \text{\bf Clique $8$} \quad  \     & (1-w)p & < 1-\left(1-{{55000 \choose 8}}^{-
\frac{1}{{8 \choose 2}}
}\right)^{\frac{1}{2}} \approx 0.0328
    &&    \\ &
   &  (1-w)\phantom{p} & \lessapprox 95.0316 
&&    \\
&    \text{\bf Clique $12$} \quad  \    & (1-w)p & < 1-\left(1-{{55000 \choose 12}}^{-
\frac{1}{{12 \choose 2}}
}\right)^{\frac{1}{2}} \approx 0.0978
    &&    \\ &
    & (1-w)\phantom{p} & \lessapprox 2707.79
&&    \\
    &
    \text{\bf Ring $4$} \quad \quad     & (1-w)p  & < \left( { {55000 \choose 4} (4-1)! }\right)^{-\frac{1}{4}} 
    \approx 2.57 \time 10^{-5}
    &&    \\ &
    & (1-w)\phantom{p}  & \lessapprox 0.0744343
&&    \\ & 
    \text{\bf Ring $10$} \quad \quad     & (1-w)p  & < \left( { {55000 \choose 10} (10-1)! }\right)^{-\frac{1}{10}} 
    \approx 2.29 \time 10^{-5}
    &&    \\ &
   & (1-w)\phantom{p} & \lessapprox 0.0662647
&&    \\ 
    &
    \text{\bf Path $5$}  \quad \quad  & (1-w)p & < \left( { {55000 \choose 5} (5)! }\right)^{-\frac{1}{5-1}}
    \approx 
    1.19 \times 10^{-6}
    &&    \\ &
   & (1-w)\phantom{p} & \lessapprox 0.00118732
&&    \\
    &
    \text{\bf Path $10$}  \quad \quad  & (1-w)p & < \left( { {55000 \choose 10} (10)! }\right)^{-\frac{1}{10-1}}
    \approx 
    5.41 \times 10^{-6}
    &&    \\ &
&    (1-w)\phantom{p} & \lessapprox 0.0156523. 
\end{flalign*}

As $1-w$ is equal to the range of the c.d.f. covered by the weights, we can
compare the bound directly on the range of the c.d.f. considered in the Accenture
model. This range is from $0.99$ to $0.99999$, spanning $ \approx 1\%$.
Therefore, all planted anomalies other than the path of size $5$ are
detectable in the Accenture model, in the sense that the number expected at
random is approximately less than $1$.

\section{Proof that the Selection of the Direction in the Eigenvector Terminates}
\label{app:directionOfEvectorProof}

In NetEMD,  the distribution of the entries of the eigenvectors are used as a
feature in the comparison. However, as for an eigenvector $v$, both $v$ and $-v$ are solutions to the eigenvector equation, and we need to select between them. To this end, we use the following procedure. 
\begin{enumerate}
    \item Compute the number of entries greater than $0$ ({\it numGreater}).
    \item Compute the number of entries less than $0$ ({\it numLesser}).
    \item If the vector is symmetric then use $v$ (as in this case the choice of sign does not matter).
    \item If {\it numGreater } $>$ {\it numLesser} then use $v$.
    \item If {\it numGreater } $<$ {\it numLesser} then use $-v$.
    \item If {\it numGreater } $=$ {\it numLesser} then:
    \begin{enumerate}[label=(\roman*)]
        \item Set {\it a} to $0$
        \item If $\sum_j (v_j)^{2a+1}>0$ use $v$
        \item If $\sum_j (v_j)^{2a+1}<0$ use $-v$
        \item If $\sum_j (v_j)^{2a+1}=0$ use increase $a$ by $1$ and go back to (ii) 
    \end{enumerate}
\end{enumerate}

As, potentially, such an algorithm may result in an infinite loop, we show that the procedure will always result in a selection between $v$ and $-v$ in finite time. More precisely, we shall show that there always exists an exponent $a \in \mathbb{N} \cup \{0\}$ such that $\sum_j (v_j)^{2a+1} \ne 0$. For this proof, we borrow ideas from~\cite{stackoverflowcite}.

Let $w_1 \le w_2 \le ... \le w_n$ be the elements of the eigenvector $v$ which are not equal to $0$. Assume that the procedure has reached Step 6. 
Then the number of positive entries must be equal to the number of
negative entries. Let the negative entries be $q_i$ such that 
$q_1 \le q_2 \le q_3 \le ... \le q_{\frac{n}{2}} < 0$
and $r_i$ be the positive entries
$0 < r_1 \le r_2 \le r_3 \le ... \le r_{\frac{n}{2}}$.

As $\sum_j (v_j)^{2a+1} $ is a sum of odd powers (power sums), we can remove 
any pair of elements $(q_i,r_j)$ such that $|q_i|=|r_j|$, as their odd powers
cancel each other in the sum. We first consider
$(q_1,r_1)$, and the pair remove if they are equal. We then consider each pair in a fixed ordering
assuming that either of the elements 
have not already been removed. 
This leaves  
$\tilde{q}_1 \le \tilde{q}_2 \le \tilde{q}_3 \le , ... \le \tilde{q}_{
\frac{
\tilde{n}
}
{2}
} \le 0$
and 
$0 \le \tilde{r}_1 \le \tilde{r}_2 \le \tilde{r}_3 \le ... \le \tilde{r}_{\frac{\tilde{n}}{2}}$.
By assumption, the sequence is not symmetric and therefore $\tilde{n} \neq 0$. 

To show that the procedure terminates in finitely many steps, we  show that for every  set of values which is not symmetric around 0,  there is an odd power such that the sum of the values to this odd power sums is not equal to $0$. 
Clearly, if  
$
\sum_i v_i^{2a+1}
=
\sum_i \tilde{q}_i^{2a+1}
+
\sum_i \tilde{r}_i^{2a+1}
=0
$ 
then 
$
\sum_i (b v_i)^{2a+1}
=
\sum_i (b\tilde{q}_i)^{2a+1}
+
\sum_i (b\tilde{r}_i)^{2a+1}
=0$. 

Further, as all pairs of values that cancel each other out in the sum of odd
powers have been removed, 
$
|\tilde{q}_{1}|
\neq 
|\tilde{r}_{\frac{\tilde{n}}{2}}|$. 
Without loss of generality, let
$
|\tilde{q}_{1}|
\le
|\tilde{r}_{\frac{\tilde{n}}{2}}|$. Let 
$b=\frac{1}{|\tilde{r}_{\frac{\tilde{n}}{2}}|}$, let  
$ \tilde{q}_{i}' = \frac{ \tilde{q}_{i} }{b}$ and let $ \tilde{r}_{i}' = \frac{
    \tilde{r}_{i} }{b}$. Then
 \begin{flalign*}
0 =     \sum_{i=1}^{\frac{\tilde{n}}{2}}
    (b\tilde{q}_i)^{2a+1}
+
    \sum_{i=1}^{\frac{\tilde{n}}{2}}
    (b\tilde{r}_i)^{2a+1}
=
    \sum_{i=1}^{\frac{\tilde{n}}{2}}
    (\tilde{q}_i')^{2a+1}
+
    \sum_{i=1}^{\frac{\tilde{n}}{2}-1}
    (\tilde{r}_i')^{2a+1}
    +
    1.
 \end{flalign*}   
Using that the $r_i > 0$ and that $q_i<0$ , it follows that 
 \begin{flalign*}
\frac{\tilde{n}}{2}
    (\tilde{q}_1')^{2a+1}
+
    1
\le 0 .
\end{flalign*}   
 Re-arranging and taking logarithms, 
 \begin{flalign*}
    log(1)
\le 
    log(\frac{\tilde{n}}{2})
   + 
   (2a+1)
   log(-(\tilde{q}'_1)). 
   \end{flalign*}   
Thus, 
 \begin{flalign*}
    -\frac{ log(\frac{\tilde{n}}{2}) }
        { log(-(\tilde{q}_1')) }
\ge 
    (2a+1), 
\end{flalign*}
where the direction of the inequality changes as by assumption 
$
-1< \tilde{q}_1'<0
$, and therefore
$
log(-(\tilde{q}_1'))<0
$.

As
$-(\tilde{q}_1') \neq 1$, the left-hand side of the last inequality is a finite
positive quantity which does not depend on $a$. Therefore, there exists a positive integer $a$ such that the inequality does not hold. Thus, there is a power sum which is non-zero, and hence the algorithm  terminates after finitely many steps, which concludes the proof.

\section{List of All Anomaly Detection Features}
\label{listOfAllMethods}

\subsection{Basic Node Statistics} 
All of the statistics below are node statistics in that they give a different
value to each node - but some of them give the same value to each node in a
community.  Here is the list of basic node statistics.

\begin{enumerate}
\item  Standardised Degree
\item  Community Density Full, i.e. the density of the community divided by overall density  
\item  Community Density Average, i.e. the density of the community divided by overall density and also divided by the size of the community
\item  Community GAW Full, i.e. the Geometric Average Weight of the community divided by overall Geometric Average Weight
\item  Community GAW Average, i.e. the Geometric Average Weight of the community divided by overall Geometric Average Weight and also divided by the size of the community
\item  Geometric Average Weight (GAW)
\item  Geometric Average Weight Top 10\% (GAW Top 10\%)
\item  Geometric Average Weight Top 20\% (GAW Top 20\%)
\item  Community Density Configuration Model - $\text{Max}(\Phi(1-p_{val}),0)$,  where $\Phi^{-1}$ is the inverse c.d.f. of a standard normal distribution and $p_{val}$ is the $p$-value of the Monte Carlo test  which compares the  
density of the community with respect to a null distribution of densities of communities of configuration model graphs
\item Small Community Category - As discussed in Sec.~\ref{sec:communityDetection}
\end{enumerate}

\subsection{Path Statistics }
The statistics labelled {\bf 11-40} are obtained by using the PathFinder
method.  To maintain consistency with the embedded paths, we denote the size of a
path is the number of nodes in the path. in general 
we include path statistics of size $3$-$32$.

\begin{enumerate}
\setcounter{enumi}{10}
\item  Path of size 3 
\item  Path of size 4 
\newline
$\vdots$
\setcounter{enumi}{38}
\item  Path of size 31 
\item  Path of size 32 
\end{enumerate}

For the main experiments we only compute path of size up to $21$ (which corresponds to $20$ edges), with the remaining features left blank. 

\subsection{NetEMD Measures}
The statistics labelled {\bf 41-72} are features related to $16$ motifs with NetEMD statistics $1$ and $2$. The 
statistics {\bf 73-80} are features related to the 
upper adjacency matrix, the lower adjacency matrix, the combinatorial Laplacian and the random-walk Laplacian with  
NetEMD scores $1$ and $2$ from Eq.~\eqref{eq:netemdscoring1} and Eq.~\eqref{eq:netemdscoring2}.

\begin{enumerate}
\setcounter{enumi}{40}
\item  Motif 1 - NetEMD Score $1$ (Eq.~\eqref{eq:netemdscoring1})             
\item  Motif 1 - NetEMD Score $2$ (Eq.~\eqref{eq:netemdscoring2})            
\item  Motif 2 - NetEMD Score $1$ (Eq.~\eqref{eq:netemdscoring1})             
\item  Motif 2 - NetEMD Score $2$ (Eq.~\eqref{eq:netemdscoring2})             
\newline
$\vdots$
\setcounter{enumi}{68}
\item  Motif 15 - NetEMD Score $1$ (Eq.~\eqref{eq:netemdscoring1})            
\item  Motif 15 - NetEMD Score $2$ (Eq.~\eqref{eq:netemdscoring2})            
\item  Motif 16 - NetEMD Score $1$ (Eq.~\eqref{eq:netemdscoring1})            
\item  Motif 16 - NetEMD Score $2$ (Eq.~\eqref{eq:netemdscoring2})            
\item  Adj. Upper Eigenvectors - NetEMD Score $1$ (Eq.~\eqref{eq:netemdscoring1})            
\item  Adj. Upper Eigenvectors - NetEMD Score $2$ (Eq.~\eqref{eq:netemdscoring2})                        
\item  Adj. Lower Eigenvectors - NetEMD Score $1$ (Eq.~\eqref{eq:netemdscoring1})
\item  Adj. Lower Eigenvectors - NetEMD Score $2$ (Eq.~\eqref{eq:netemdscoring2})
\item  Combinatorial Laplacian Eigenvectors - NetEMD Score $1$ (Eq.~\eqref{eq:netemdscoring1})         
\item  Combinatorial Laplacian Eigenvectors - NetEMD Score $2$ (Eq.~\eqref{eq:netemdscoring2})         
\item  Random Walk Laplacian Eigenvectors - NetEMD Score $1$ (Eq.~\eqref{eq:netemdscoring1})            
\item  Random Walk Laplacian Eigenvectors - NetEMD Score $2$ (Eq.~\eqref{eq:netemdscoring2})            
\end{enumerate}

\subsection{Localisation Statistics}
\label{sec:localisationFullList}
The statistics labelled {\bf 81 - 140} are the localisation statistics on the 
upper adjacency matrix, the lower adjacency matrix, the combinatorial Laplacian and the random-walk Laplacian with  
the following localisation measures:

\begin{enumerate}
\setcounter{enumi}{80}
\item Adj. Upper Eigenvectors - Inverse Participation Ratio - Norm Node Statistic 1 (Eq.~\eqref{eq:normBasedF1})
\item Adj. Upper Eigenvectors - Inverse Participation Ratio - Norm Node Statistic 2 (Eq.~\eqref{eq:normBasedF2})
\item Adj. Upper Eigenvectors - Inverse Participation Ratio - Norm Node Statistic 3 (Eq.~\eqref{eq:normBasedF3})
\item Adj. Upper Eigenvectors - Inverse Participation Ratio - Norm Node Statistic 4 (Eq.~\eqref{eq:normBasedF4})
\item Adj. Upper Eigenvectors - Exponential - Norm Node Statistic 1 (Eq.~\eqref{eq:normBasedF1})
\item Adj. Upper Eigenvectors - Exponential - Norm Node Statistic 2 (Eq.~\eqref{eq:normBasedF2})
\item Adj. Upper Eigenvectors - Exponential - Norm Node Statistic 3 (Eq.~\eqref{eq:normBasedF3})
\item Adj. Upper Eigenvectors - Exponential - Norm Node Statistic 4 (Eq.~\eqref{eq:normBasedF4})
\item Adj. Upper Eigenvectors - 90 percent Contribution in Inverse Participation Ratio
\item Adj. Upper Eigenvectors - 90 percent Contribution in absolute value 
\item Adj. Upper Eigenvectors - Sign Statistic 1 (Eq.~\eqref{eq:signStat1})  
\item Adj. Upper Eigenvectors - Sign Statistic 2 (Eq.~\eqref{eq:signStat2})  
\item Adj. Upper Eigenvectors - Sign Statistic Equal 1 (Eq.~\eqref{eq:signstatequal1})  
\item Adj. Upper Eigenvectors - Sign Statistic Equal 2 (Eq.~\eqref{eq:signstatequal2})  
\item Adj. Upper Eigenvectors - Absolute value of the eigenvector 
\item Adj. Lower Eigenvectors - Inverse Participation Ratio - Norm Node Statistic 1 (Eq.~\eqref{eq:normBasedF1})
\item Adj. Lower Eigenvectors - Inverse Participation Ratio - Norm Node Statistic 2 (Eq.~\eqref{eq:normBasedF2})
\item Adj. Lower Eigenvectors - Inverse Participation Ratio - Norm Node Statistic 3 (Eq.~\eqref{eq:normBasedF3})
\item Adj. Lower Eigenvectors - Inverse Participation Ratio - Norm Node Statistic 4 (Eq.~\eqref{eq:normBasedF4})
\item Adj. Lower Eigenvectors - Exponential - Norm Node Statistic 1 (Eq.~\eqref{eq:normBasedF1})
\item Adj. Lower Eigenvectors - Exponential - Norm Node Statistic 2 (Eq.~\eqref{eq:normBasedF2})
\item Adj. Lower Eigenvectors - Exponential - Norm Node Statistic 3 (Eq.~\eqref{eq:normBasedF3})
\item Adj. Lower Eigenvectors - Exponential - Norm Node Statistic 4 (Eq.~\eqref{eq:normBasedF4})
\item Adj. Lower Eigenvectors - 90 percent Contribution in Inverse Participation Ratio
\item Adj. Lower Eigenvectors - 90 percent Contribution in absolute value 
\item Adj. Lower Eigenvectors -  Sign Statistic 1 (Eq.~\eqref{eq:signStat1})  
\item Adj. Lower Eigenvectors -  Sign Statistic 2 (Eq.~\eqref{eq:signStat2})  
\item Adj. Lower Eigenvectors -  Sign Statistic Equal 1 (Eq.~\eqref{eq:signstatequal1})  
\item Adj. Lower Eigenvectors -  Sign Statistic Equal 2 (Eq.~\eqref{eq:signstatequal2})  
\item Adj. Lower Eigenvectors - Absolute value of the eigenvector 
\item Combinatorial Laplacian - Inverse Participation Ratio - Norm Node Statistic 1 (Eq.~\eqref{eq:normBasedF1})
\item Combinatorial Laplacian - Inverse Participation Ratio - Norm Node Statistic 2 (Eq.~\eqref{eq:normBasedF2})
\item Combinatorial Laplacian - Inverse Participation Ratio - Norm Node Statistic 3 (Eq.~\eqref{eq:normBasedF3})
\item Combinatorial Laplacian - Inverse Participation Ratio - Norm Node Statistic 4 (Eq.~\eqref{eq:normBasedF4})
\item Combinatorial Laplacian - Exponential - Norm Node Statistic 1 (Eq.~\eqref{eq:normBasedF1})
\item Combinatorial Laplacian - Exponential - Norm Node Statistic 2 (Eq.~\eqref{eq:normBasedF2})
\item Combinatorial Laplacian - Exponential - Norm Node Statistic 3 (Eq.~\eqref{eq:normBasedF3})
\item Combinatorial Laplacian - Exponential - Norm Node Statistic 4 (Eq.~\eqref{eq:normBasedF4})
\item Combinatorial Laplacian - 90 percent Contribution in Inverse Participation Ratio
\item Combinatorial Laplacian - 90 percent Contribution in absolute value 
\item Combinatorial Laplacian - Sign Statistic 1 (Eq.~\eqref{eq:signStat1})  
\item Combinatorial Laplacian - Sign Statistic 2 (Eq.~\eqref{eq:signStat2})  
\item Combinatorial Laplacian - Sign Statistic Equal 1 (Eq.~\eqref{eq:signstatequal1})  
\item Combinatorial Laplacian - Sign Statistic Equal 2 (Eq.~\eqref{eq:signstatequal2})  
\item Combinatorial Laplacian - Absolute value of the eigenvector 
\item Random Walk Laplacian - Inverse Participation Ratio - Norm Node Statistic 1 (Eq.~\eqref{eq:normBasedF1})
\item Random Walk Laplacian - Inverse Participation Ratio - Norm Node Statistic 2 (Eq.~\eqref{eq:normBasedF2})
\item Random Walk Laplacian - Inverse Participation Ratio - Norm Node Statistic 3 (Eq.~\eqref{eq:normBasedF3})
\item Random Walk Laplacian - Inverse Participation Ratio - Norm Node Statistic 4 (Eq.~\eqref{eq:normBasedF4})
\item Random Walk Laplacian - Exponential - Norm Node Statistic 1 (Eq.~\eqref{eq:normBasedF1})
\item Random Walk Laplacian - Exponential - Norm Node Statistic 2 (Eq.~\eqref{eq:normBasedF2})
\item Random Walk Laplacian - Exponential - Norm Node Statistic 3 (Eq.~\eqref{eq:normBasedF3})
\item Random Walk Laplacian - Exponential - Norm Node Statistic 4 (Eq.~\eqref{eq:normBasedF4})
\item Random Walk Laplacian - 90 percent Contribution in Inverse Participation Ratio
\item Random Walk Laplacian - 90 percent Contribution in absolute value 
\item Random Walk Laplacian - Sign Statistic 1 (Eq.~\eqref{eq:signStat1})  
\item Random Walk Laplacian - Sign Statistic 2 (Eq.~\eqref{eq:signStat2})  
\item Random Walk Laplacian - Sign Statistic Equal 1 (Eq.~\eqref{eq:signstatequal1})  
\item Random Walk Laplacian - Sign Statistic Equal 2 (Eq.~\eqref{eq:signstatequal2})  
\item Random Walk Laplacian - Absolute value of the eigenvector 
\end{enumerate}

\subsection{Additional Features}
\label{additionalFeatures}
Finally, we include a small number of additional features which are employed primarily 
to highlight special cases.
However, these features were not added
to the set of features considered in the main body of the text, as these special cases did not
appear in the data analysis. 

For brevity, as this does not appear in the list of our $140$ statistics, we do not list
each combination of our statistic and matrix. We simply list the set of statistics.

\paragraph{Localisation}
We flag a special cases by creating the following additional features:
\begin{enumerate}
    \item Sign based statistic - Statistic 1 Large Number 
    (Eq.~\eqref{eq:signStat1}) 
    \item Sign based statistic - Statistic 2 Large Number
    (Eq.~\eqref{eq:signStat2}) 
    \item Sign based statistic - Statistic Equal 1 Large Number
    (Eq.~\eqref{eq:signstatequal1}) 
    \item Sign based statistic - Statistic Equal 2 Large Number
    (Eq.~\eqref{eq:signstatequal2}) 
\end{enumerate}
This series of features flags
cases where the minimum of the number of positive and negative entries
becomes greater
than or equal to $\frac{n}{2}$. As this statistic captures the smaller of the number of
positive entries and the number of negative entries, it seems surprising that
we can get entries that are greater than $\frac{n}{2}$ and are still
significant.  A statistic that is greater than or equal to $\frac{n}{2}$ is possible if
the remaining entries are zero, and thus if a sufficient number of the null
replicas are in the same position. However, as
this is outside of the original motivation of the feature, we split it out into
a set of separate features. We perform a similar action for the case where the
number of positive and negative terms are equal, which is only possible if there are exactly $\frac{n}{2}$ positive entries, and $\frac{n}{2}$ negative entries.

\section{Feature Selection}
\subsection{List of Selected Features} 
\label{app:featureSelectionListOfFeatures}

\begin{enumerate}
%
\item Geometric Average Weight Top 10\% (GAW Top 10\%)
%
\item Localisation - Adjacency Upper Eigenvectors - Sign based statistic - Average score 
%
\item NetEMD -  Adj. Lower Eigenvectors - NetEMD Score $1$ (Eq.~\eqref{eq:netemdscoring1})            
%
\item NetEMD -  Adj. Upper Eigenvectors - NetEMD Score $1$ (Eq.~\eqref{eq:netemdscoring1})            
%
\item Localisation - Adj. Upper Eigenvectors - Exponential - Norm Node Statistic 1 (Eq.~\eqref{eq:normBasedF1})
%
\item Localisation - Adj. Upper Eigenvectors - Inverse Participation Ratio - Norm Node Statistic 1 (Eq.~\eqref{eq:normBasedF1})
%
\item Localisation - Adj. Lower Eigenvectors - Exponential - Norm Node Statistic 1 (Eq.~\eqref{eq:normBasedF1})
%
\item Localisation - Adj. Upper Eigenvectors - 90 percent Contribution in absolute value 
%
\item Localisation - Adj. Lower Eigenvectors - 90 percent Contribution in absolute value 
%
\item Localisation - Random Walk Laplacian - 90 percent Contribution in absolute value 
%
\item Basic Statistics - Community GAW Full i.e the Geometric Average Weight of the community divided by overall Geometric Average Weight.
%
\item NetEMD - Adj. Lower Eigenvectors - NetEMD Score $2$ (Eq.~\eqref{eq:netemdscoring2})
%
\item Basic Statistics - Standardised Degree
%
\item Localisation - Adj. Upper Eigenvectors - Inverse Participation Ratio - Norm Node Statistic 1 (Eq.~\eqref{eq:normBasedF1})
%
\item Localisation - Adj. Upper Eigenvectors - 90 percent Contribution in Inverse Participation Ratio
%
\item  NetEMD - Random Walk Laplacian - NetEMD Score $1$ (Eq.~\eqref{eq:netemdscoring1})            
%
\item  Basic Statistics - Community GAW Average i.e the Geometric Average Weight of the community divided by overall Geometric Average Weight and also divided by the size of the community.
%
\item  Community Density Full i.e the density of the community divided by overall density.
%
\item Localisation - Random Walk Laplacian - Exponential - Norm Node Statistic 1 (Eq.~\eqref{eq:normBasedF1})
%
\item Path Methods - Path of Size 6 
%
\item Path Methods - Path of Size 5 
%
\item Localisation -  Adj. Upper Eigenvectors - Sign Statistic 1 (Eq.~\eqref{eq:signStat1})  
%
\item Localisation - Adj. Upper Eigenvectors - Inverse Participation Ratio - Norm Node Statistic 2 (Eq.~\eqref{eq:normBasedF2})
%
\item Basic Statistics - Community Density Average i.e the density of the community divided by overall density and also divided by the size of the community.
%
\item NetEMD - Motif 5 - NetEMD Score $2$ (Eq.~\eqref{eq:netemdscoring2})             
%
\item Localisation - Random Walk Laplacian - Sign Statistic 2 (Eq.~\eqref{eq:signStat2})  
%
\item Localisation - Adj. Upper Eigenvectors - Inverse Participation Ratio - Norm Node Statistic 4 (Eq.~\eqref{eq:normBasedF4})
%
\item NetEMD - Adj. Upper Eigenvectors - NetEMD Score $2$ (Eq.~\eqref{eq:netemdscoring2})                        
%
\item NetEMD - Motif 5 - NetEMD Score $1$ (Eq.~\eqref{eq:netemdscoring1})             
%
\item Localisation - Random Walk Laplacian - Sign Statistic 1 (Eq.~\eqref{eq:signStat1})  
%
\item Localisation - Adj. Upper Eigenvectors - Exponential - Norm Node Statistic 2 (Eq.~\eqref{eq:normBasedF2})
%
\item Localisation - Adj. Upper Eigenvectors - Exponential - Norm Node Statistic 4 (Eq.~\eqref{eq:normBasedF4})
%
\item Localisation - Random Walk Laplacian - Inverse Participation Ratio - Norm Node Statistic 1 (Eq.~\eqref{eq:normBasedF1})
%
\item Basic Statistic - Geometric Average Weight Top 20\% (GAW Top 20\%)
%
\item Localisation - Combinatorial Laplacian - 90 percent Contribution in absolute value 
%
\item Localisation - Adj. Lower Eigenvectors - Sign Statistic 2 (Eq.~\eqref{eq:signStat2})  
%
\item Localisation - Adj. Lower Eigenvectors - Sign Statistic 1 (Eq.~\eqref{eq:signStat1})  
%
\item Basic Statistics - Community Density Configuration Model 
%
\item NetEMD - Random Walk Laplacian Eigenvectors - NetEMD Score $2$ (Eq.~\eqref{eq:netemdscoring2})            
%
\item NetEMD - Motif 4 - NetEMD Score $2$ (Eq.~\eqref{eq:netemdscoring2})             
%
%
\item Localisation - Adj. Lower Eigenvectors - 90 percent Contribution in Inverse Participation Ratio
%
\item Localisation - Combinatorial Laplacian - Exponential - Norm Node Statistic 1 (Eq.~\eqref{eq:normBasedF1})
%
\item Basic Statistics - Geometric Average Weight  (GAW)
%
\item NetEMD - Motif 6 - NetEMD Score $2$ (Eq.~\eqref{eq:netemdscoring2}) 
\end{enumerate}

\subsection{Alternative Feature Selection}
\label{app:featureSelection2}
To select the feature set for our random forest, we used the average rank over
the $27$ parameter regimes. Another approach would have been to use the average
feature importance score.

\begin{figure*}
\centering
\includegraphics[width=0.5\textwidth]{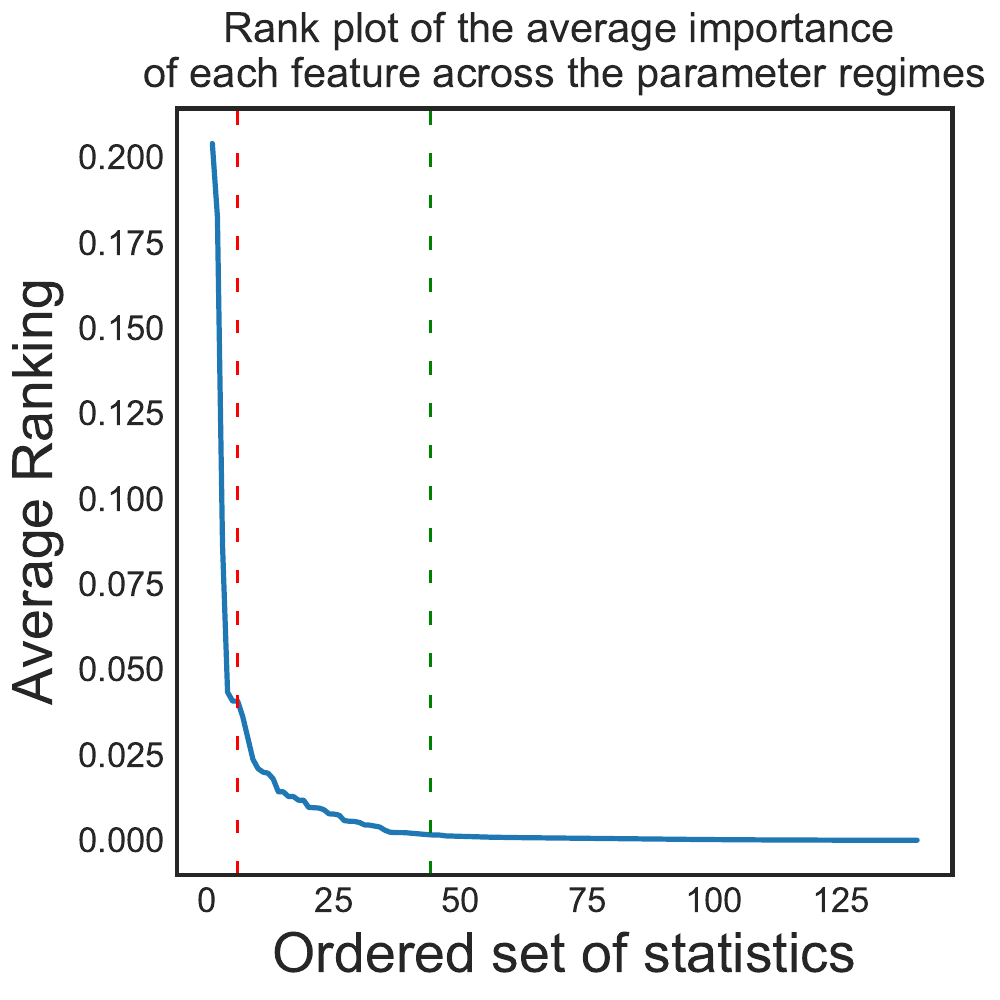}
    \label{app:averageScoreFeature}
    \caption{Feature selection using the average score over the $27$ parameter
    regimes. The red dashed line represents a possible feature cut-off point. The green
    line represents the cut-off point  which we selected when using the average rank, as
    can be seen in Fig.~\ref{fig:averageRankPlot}A}
\end{figure*}

The resultant feature plot, which is a direct comparison to the ranking version
from Fig.~\ref{fig:averageRankPlot}A, can be seen in
Fig.~\ref{app:averageScoreFeature}. This approach would yield
substantially fewer features. We decided to use the feature rank
approach, because  the feature importance scores could be confounded 
by correlated features; a set of features with high correlation may be seen
as less important than a set of features that are not correlated, see for example \cite{correlatedFeatures} or  \cite{rfcors} for details. 

\subsection{Full Feature Selection Plots}
\begin{figure}
\begin{center}
\includegraphics[width=0.9\textwidth]{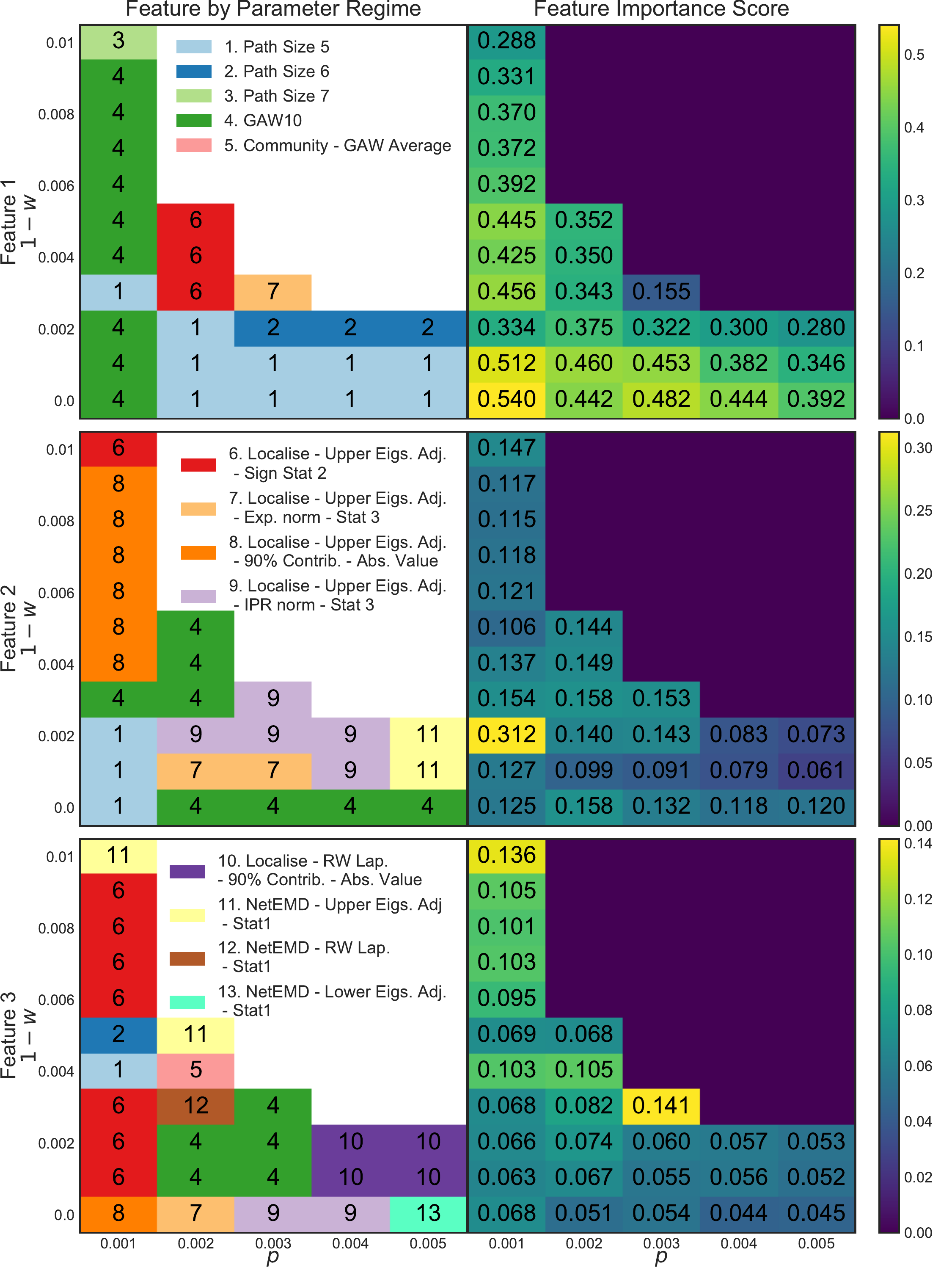}
\end{center}
\label{app:splitoutFeatureImportanceScores}
\caption{ Feature importance plots, as seen in the main body Fig.~\ref{fig:regionsOfStatistics}, combined with the corresponding score
displayed on the right. To aid understanding, we also include the index of the feature in the left set of plots, and the raw value of the statistic in the right set of plots.}
\end{figure}

In the main body, we showed the feature importance detailing the top
three features for each parameter
regime. In a similar manner to Sec.~\ref{app:featureSelection2}, we also show the relative feature importance for each of these features.
These can be seen in Fig.~\ref{app:splitoutFeatureImportanceScores}. The weight on the first few features varies considerably across parameter regimes. 
Notably, when $p=0.001$ and $1-w=0$ or $1-w=0.001$, over $50$\% of the feature importance
is given to  only $1$ feature. At the other extreme,
for example in the region $p=0.003$ and $1-w=0.003$, the feature importance is much less concentrated, with the first three features having feature importances in broadly similar ranges. 

\section{Within-Sample Scores for the {\sc Random Forest} Method}\label{sec:withinSamplePerformance}

\begin{figure*}[h!]
\centering
\includegraphics[width=\textwidth]{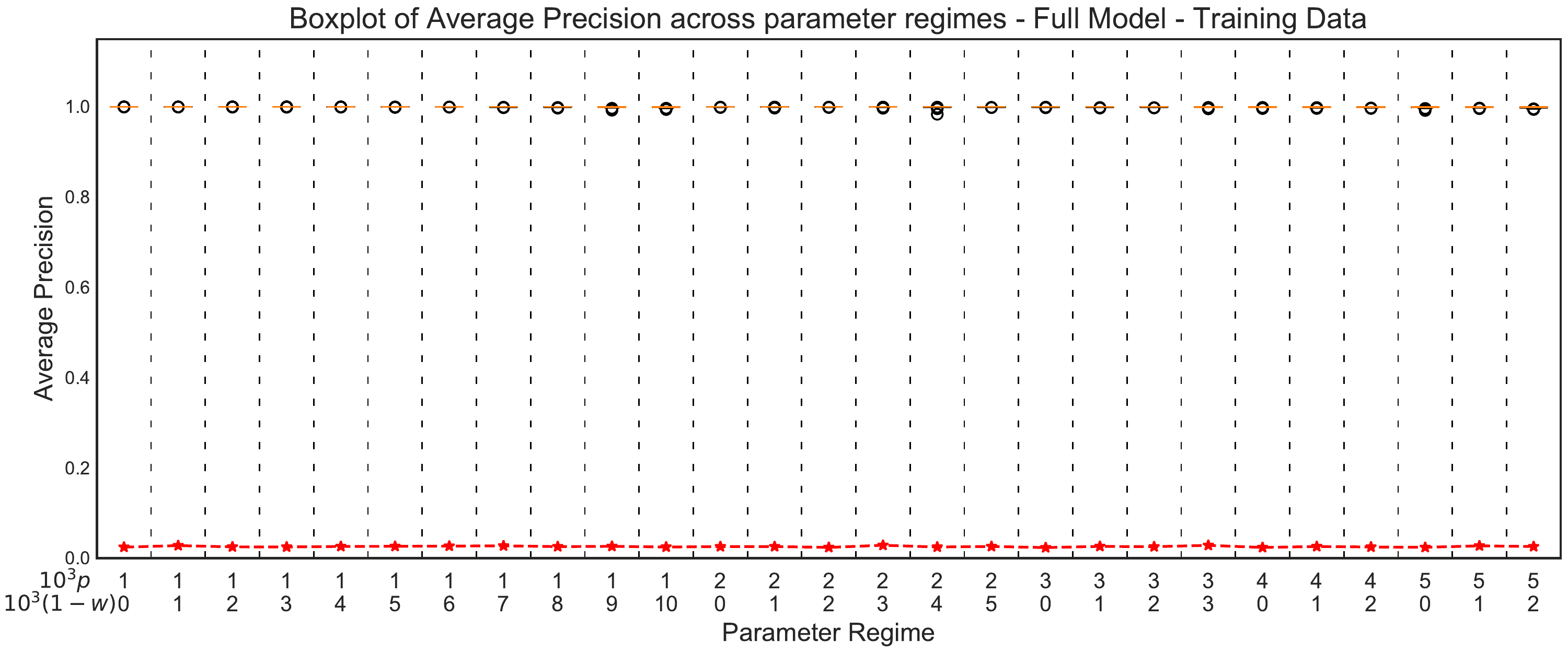}
\caption{\label{fig:fullModelBoxPlot} 
    The average precision for the random forest, for
    each of the parameter regimes using the full model after feature selection
    and training on the training data (in-sample performance).
    The red starred line indicates the largest
    percentage of anomalies expected  at random using this
    measure~\cite{sklearn}. 
    }
\end{figure*}

To test the performance of the  random forest model, we only reported the out-of-sample performance in the main body of the text. For completeness, in this section we report the in-sample performance of the random forest, which can be seen in Fig.~\ref{fig:fullModelBoxPlot}.  We observe almost perfect performance across all of the regimes. This behaviour is not unexpected for a random forest, but it confirms that in the sample there is enough variation in the  features to correctly identify all of the anomalies. 

\section{Oddball}
\label{app:Oddball}

In this section, we present more details on Oddball, with
Sec.~\ref{app:oddballComparisons} addressing the adaptation of Oddball for
directed networks, Sec.~\ref{app:oddballsplit} addressing the performance of
each of the Oddball statistics on the Weighted ER Model, and
Sec.~\ref{app:oddballsplitAccenture} addressing the performance of each of the
Oddball statistics on the Accenture model.

\subsection{Oddball for Directed networks}
\label{app:oddballComparisons}

The original formulation of Oddball in~\cite{akoglu2010oddball} was
primarily concerned with weighted undirected networks, although stating that it is easily generalisable to directed networks. The software which accompanies the paper provides the generalisation. While in the undirected case, many of the statistics are based on the ego-networks of  individual nodes, in the directed generalisation, the ego-networks (egonets) are computed on a symmetrised matrix and thus consider both incoming and outgoing edges. We use this extension in the analysis in the main paper and accompanying appendices. 

As the package was developed on an older version of Matlab, we made the following adjustments to run the package on an updated Matlab version. First, we removed plotting commands, some of which were creating errors on our networks, as the plots were not required for our analysis. Second, the directed version of Oddball uses the number of degree $1$ neighbours as a statistic, which we removed as several of our networks do not possess degree $1$ nodes (the expected number is given by $n(n-1)p(1-p)^{n-2}$ which is $4.52$ when $p=0.001$, but $0.0004$ when $p=0.002$). As this method raised an error when nodes did not exist, we removed the component.

For completeness, we include a list of the comparisons used in the directed version of Oddball.
\begin{enumerate}
    \item Number of nodes vs number of edges
    \item Number of edges vs the total weight in the egonet 
    \item Egonet out-degree vs egonet out-weight (edges from the ego network from
        to the remaining network)
    \item Egonet in-degree vs egonet in-weight (edges to the ego network from
        the remaining network)
    \item Ego out-degree vs ego out-weight
    \item Ego in-degree vs ego in-Weight 
    \item Egonet total weight vs egonet maximum weight
    \item Egonet total in-weight vs egonet maximum in-weight
    \item Egonet total out-weight vs egonet maximum out-weight.
\end{enumerate}
Here, ``ego network'' refers to properties of the ego, while ``ego net'' refers to properties of the 1 hop snowball sample. As above, the software also provides one additional comparison, namely the number of nodes vs number of degree-$1$ nodes, which we do not use as our networks contain very few such nodes.

\subsection{Oddball Individual Statistics - Weighted ER Network}
\label{app:oddballsplit}

\begin{figure*}[h!]
\centering
\includegraphics[width=\textwidth]{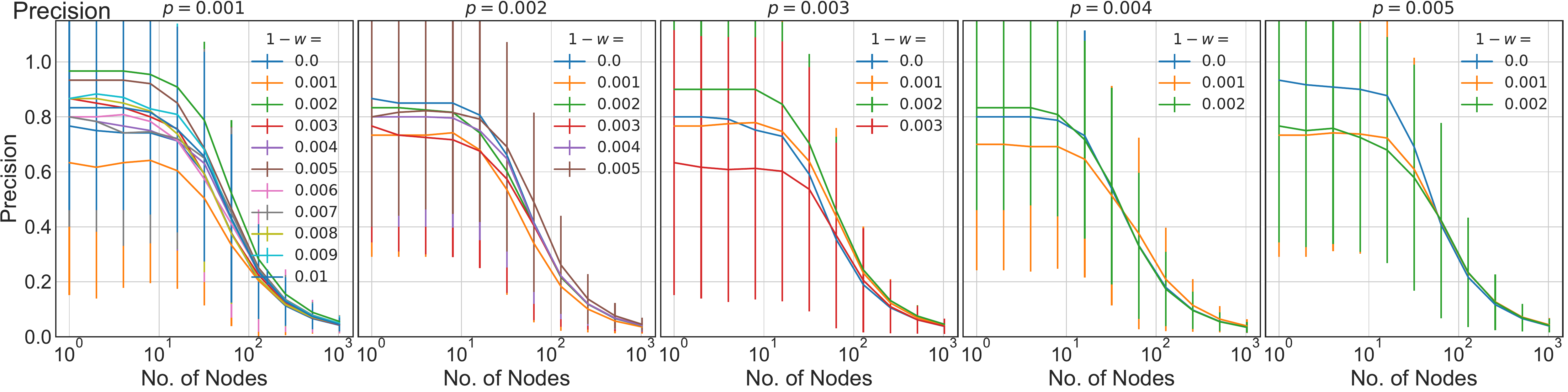}
\includegraphics[width=\textwidth]{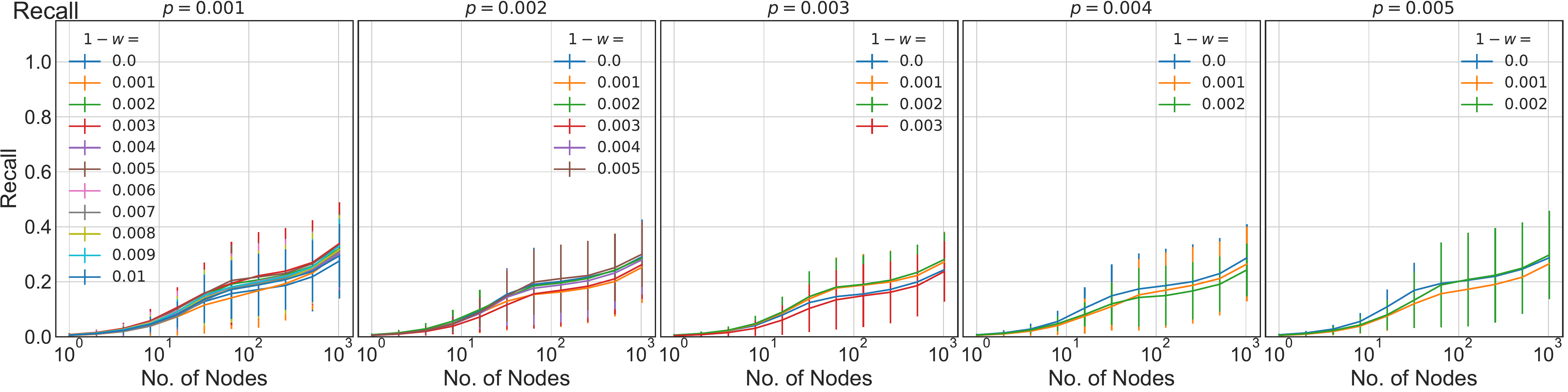}
\includegraphics[width=\textwidth]{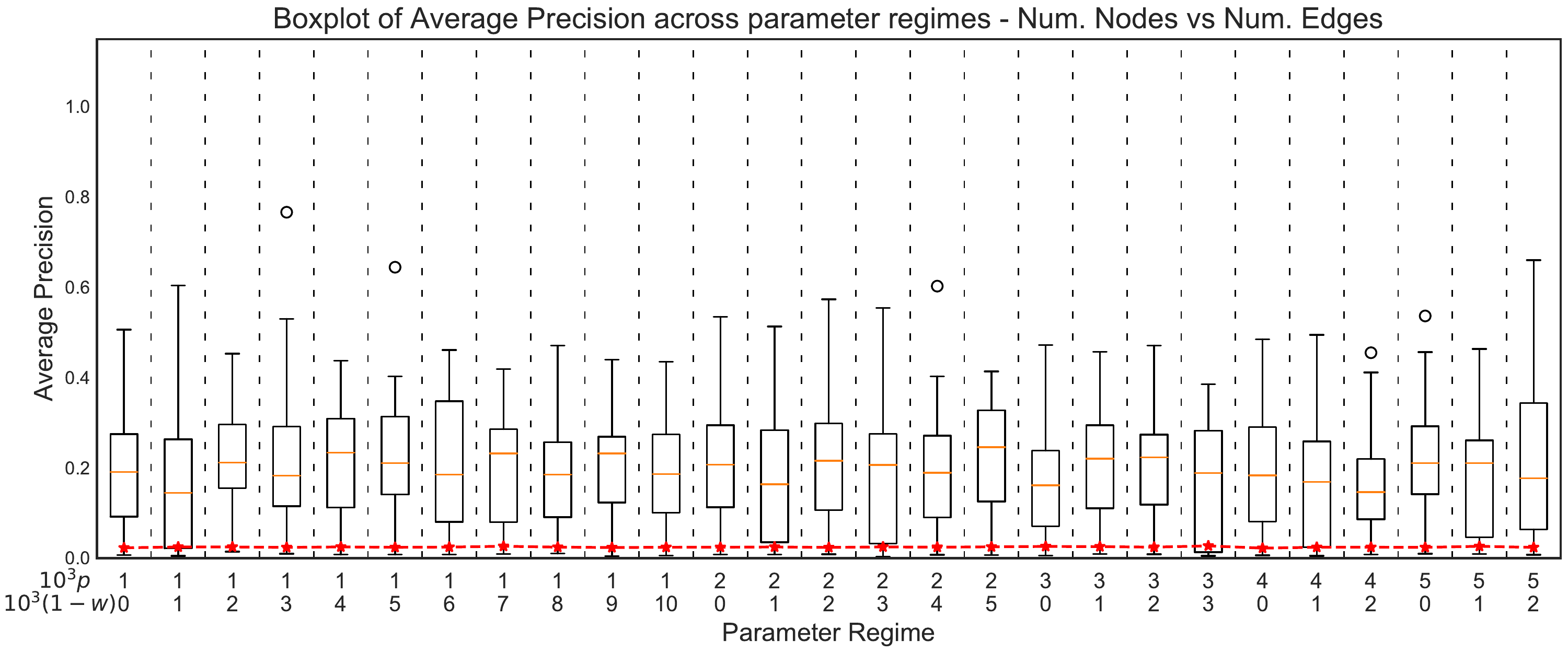}
\caption{\label{fig:oddball1_1_2}
Performance measurements for each of the parameter regimes using the 
Oddball Number of Nodes vs Number of Edges feature. 
The first pair of rows shows the mean precision and mean recall respectively with
error bars that are one 
one standard error over the relevant networks.
The final panel shows a boxplot of
the the average precision over the test networks.  The red starred line
indicates the largest percentage of anomalies expected at random using this
measure~\cite{sklearn}. 
    }
\end{figure*}

\begin{figure*}[h!]
\centering
\includegraphics[width=\textwidth]{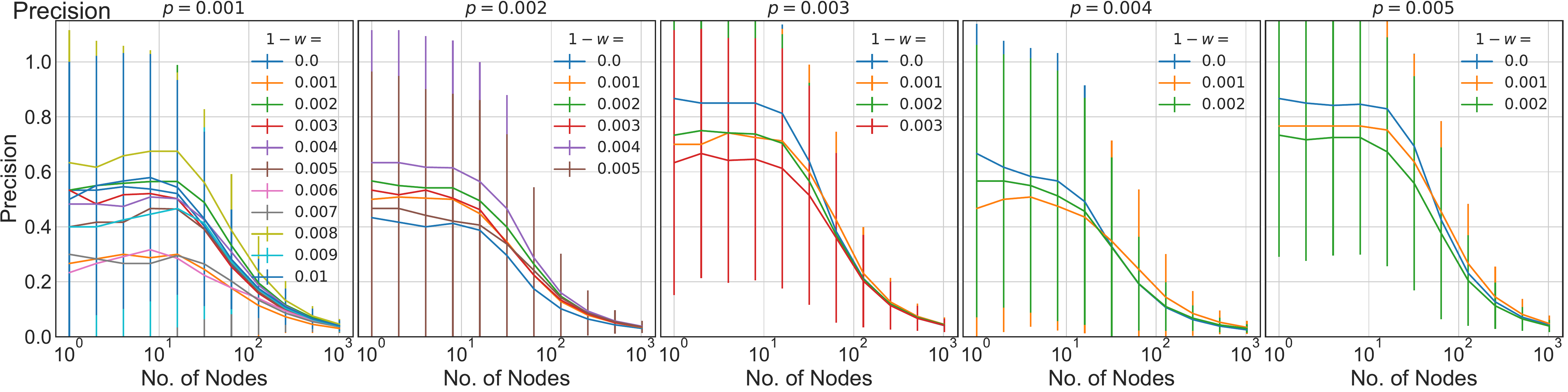}
\includegraphics[width=\textwidth]{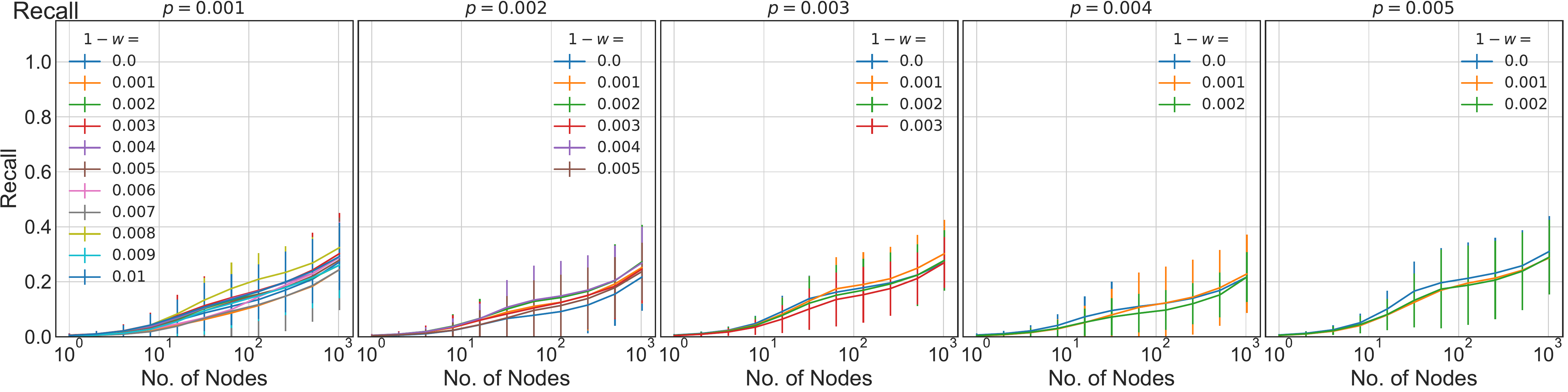}
\includegraphics[width=\textwidth]{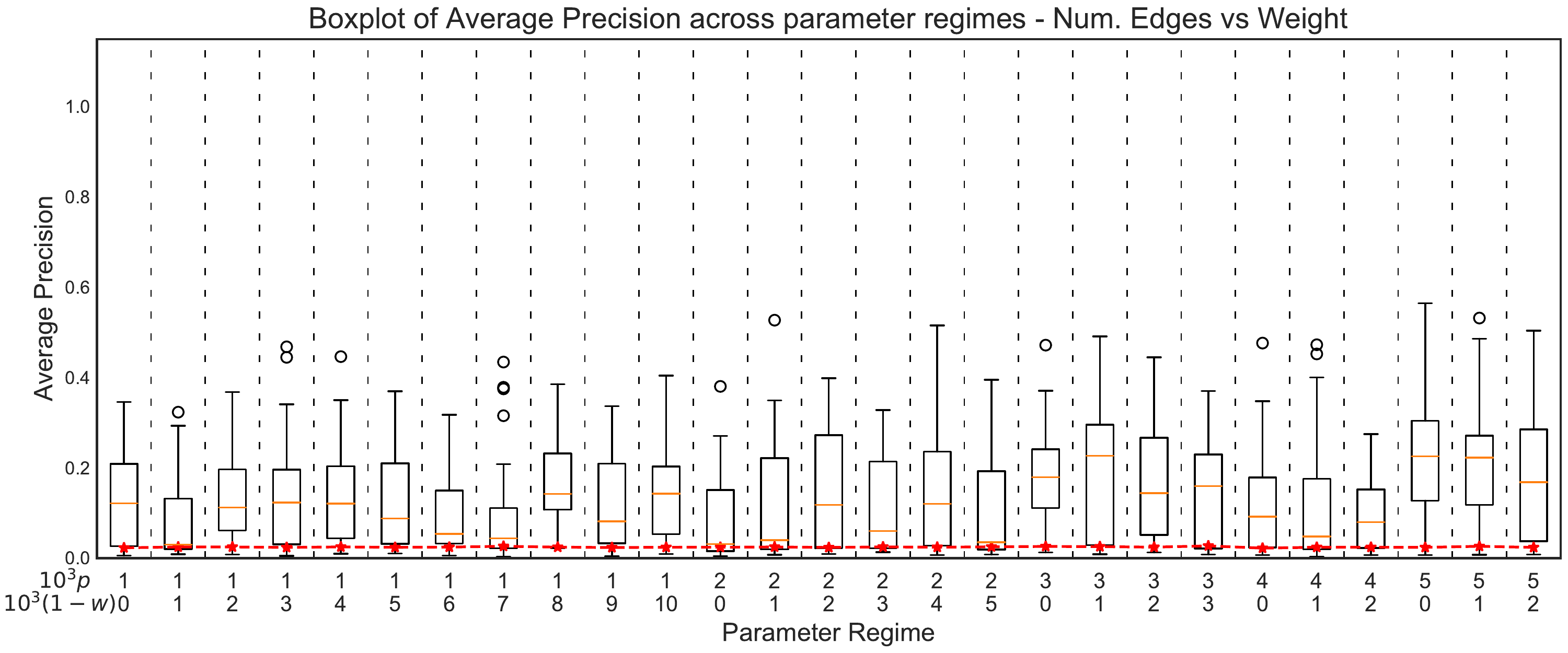}
\caption{\label{fig:oddball1_2_3} 
Performance measurements for each of the parameter regimes using the 
Oddball Number of  Edges vs Weight feature. 
The first pair of rows shows the mean precision and mean recall respectively with
error bars that are one 
one standard error over the relevant networks.
The final panel shows a boxplot of
the the average precision over the test networks.  The red starred line
indicates the largest percentage of anomalies expected at random using this
measure~\cite{sklearn}. 
    }
\end{figure*}

\begin{figure*}[h!]
\centering
\includegraphics[width=\textwidth]{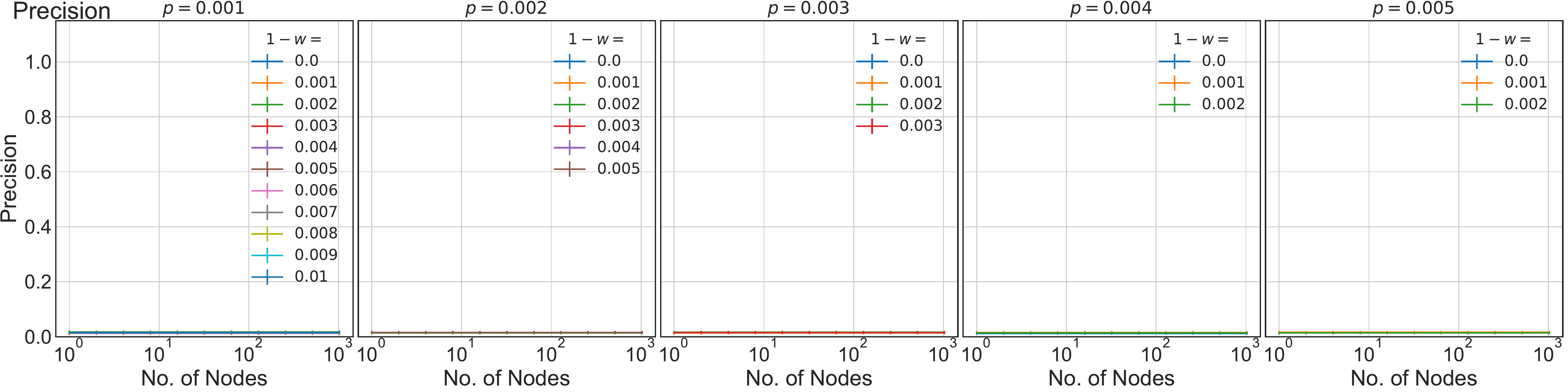}
\includegraphics[width=\textwidth]{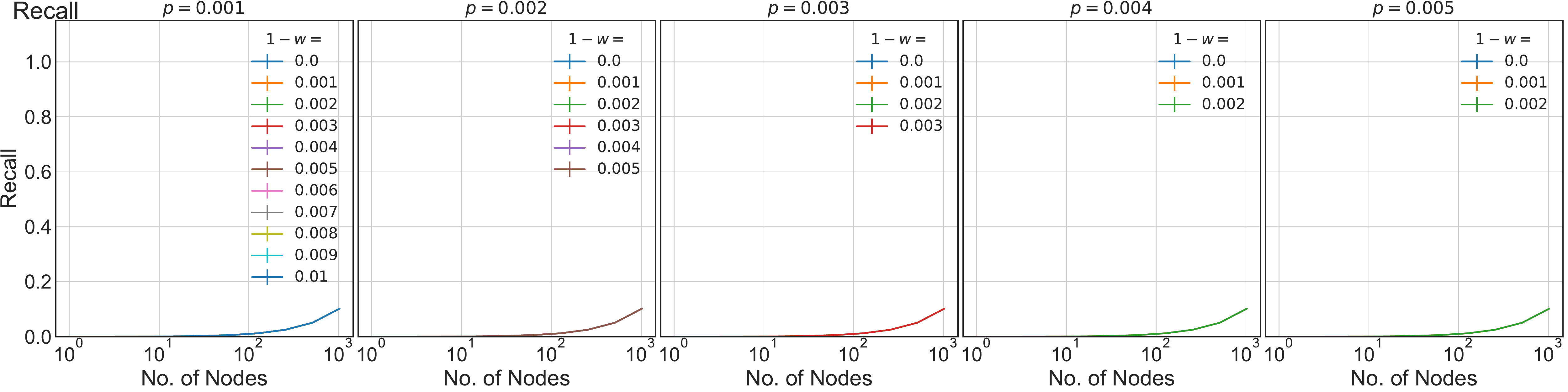}
\includegraphics[width=\textwidth]{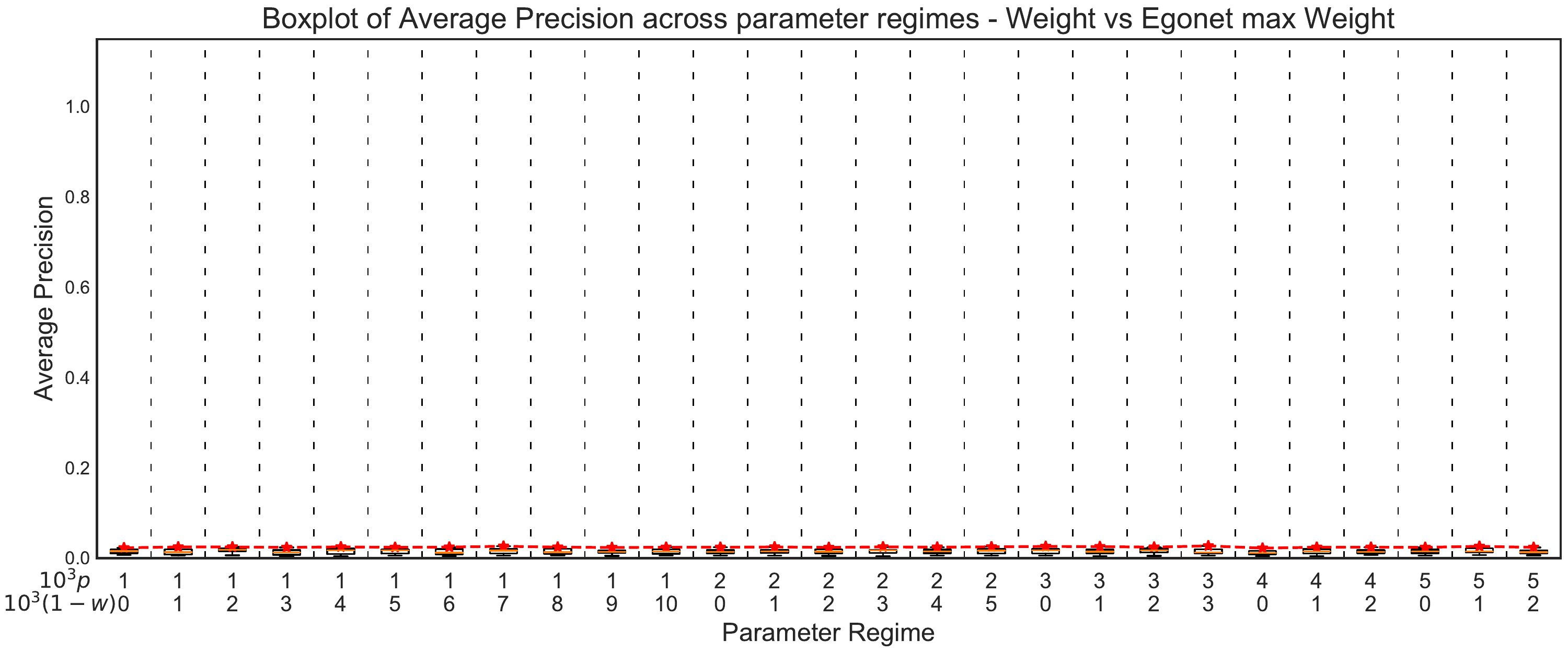}
\caption{\label{fig:oddball1_3_12} 
Performance measurements for each of the parameter regimes using the 
Oddball Weight vs Egonet Maximum Weight feature.
The first pair of rows shows the mean precision and mean recall respectively with
error bars that are one 
one standard error over the relevant networks.
The final panel shows a boxplot of
the the average precision over the test networks.  The red starred line
indicates the largest percentage of anomalies expected at random using this
measure~\cite{sklearn}. 
    }
\end{figure*}

\begin{figure*}[h!]
\centering
\includegraphics[width=\textwidth]{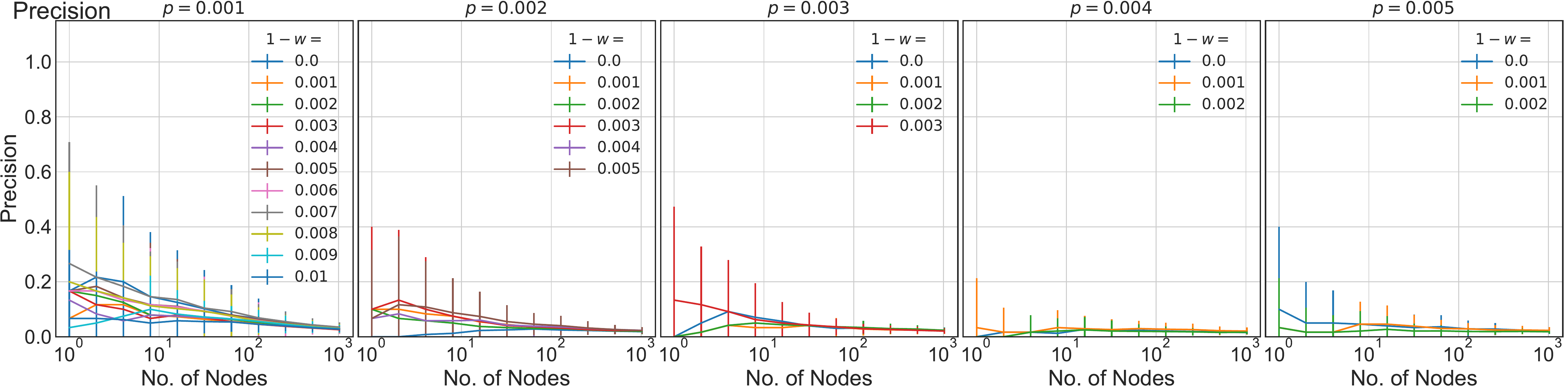}
\includegraphics[width=\textwidth]{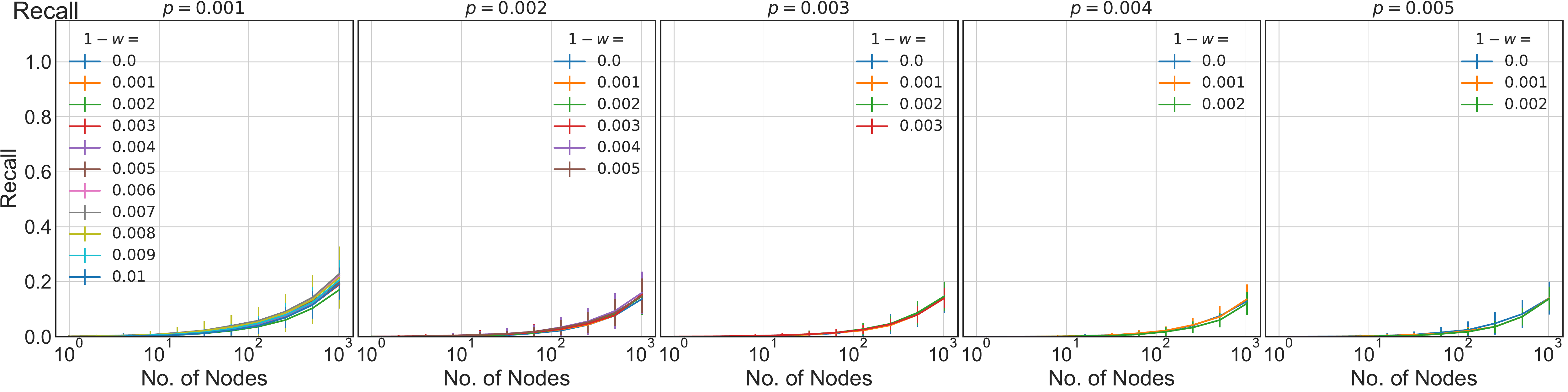}
\includegraphics[width=\textwidth]{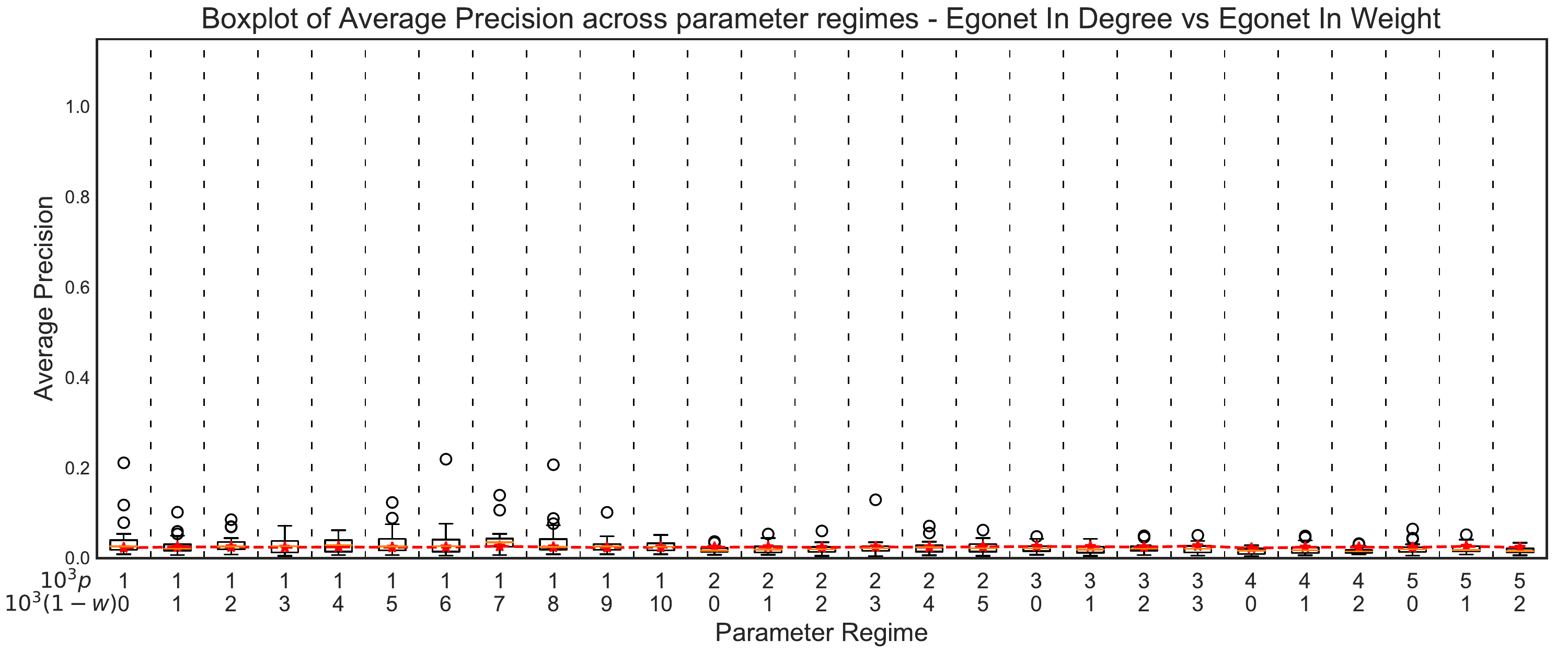}
\caption{\label{fig:oddball1_4_6} 
Performance measurements for each of the parameter regimes using the 
Oddball Egonet In-Degree vs Egonet In-Weight feature.
The first pair of rows shows the mean precision and mean recall respectively with
error bars that are one 
one standard error over the relevant networks.
The final panel shows a boxplot of
the the average precision over the test networks.  The red starred line
indicates the largest percentage of anomalies expected at random using this
measure~\cite{sklearn}. 
    }
\end{figure*}

\begin{figure*}[h!]
\centering
\includegraphics[width=\textwidth]{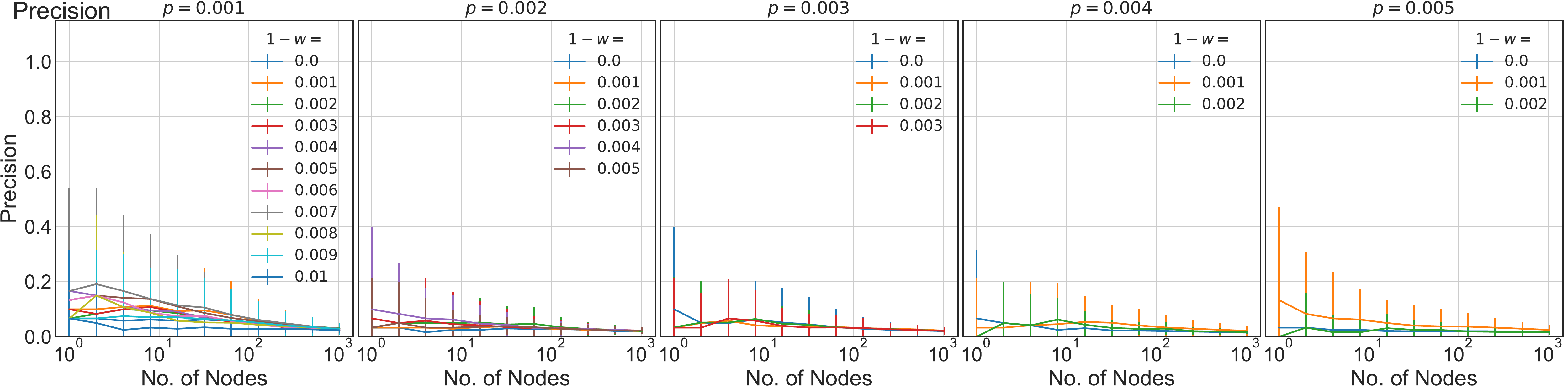}
\includegraphics[width=\textwidth]{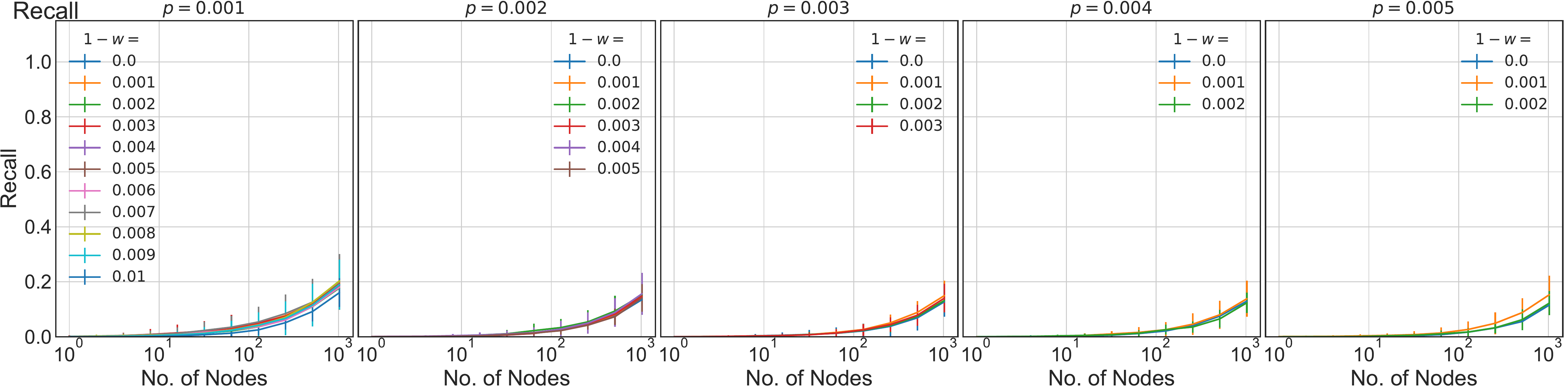}
\includegraphics[width=\textwidth]{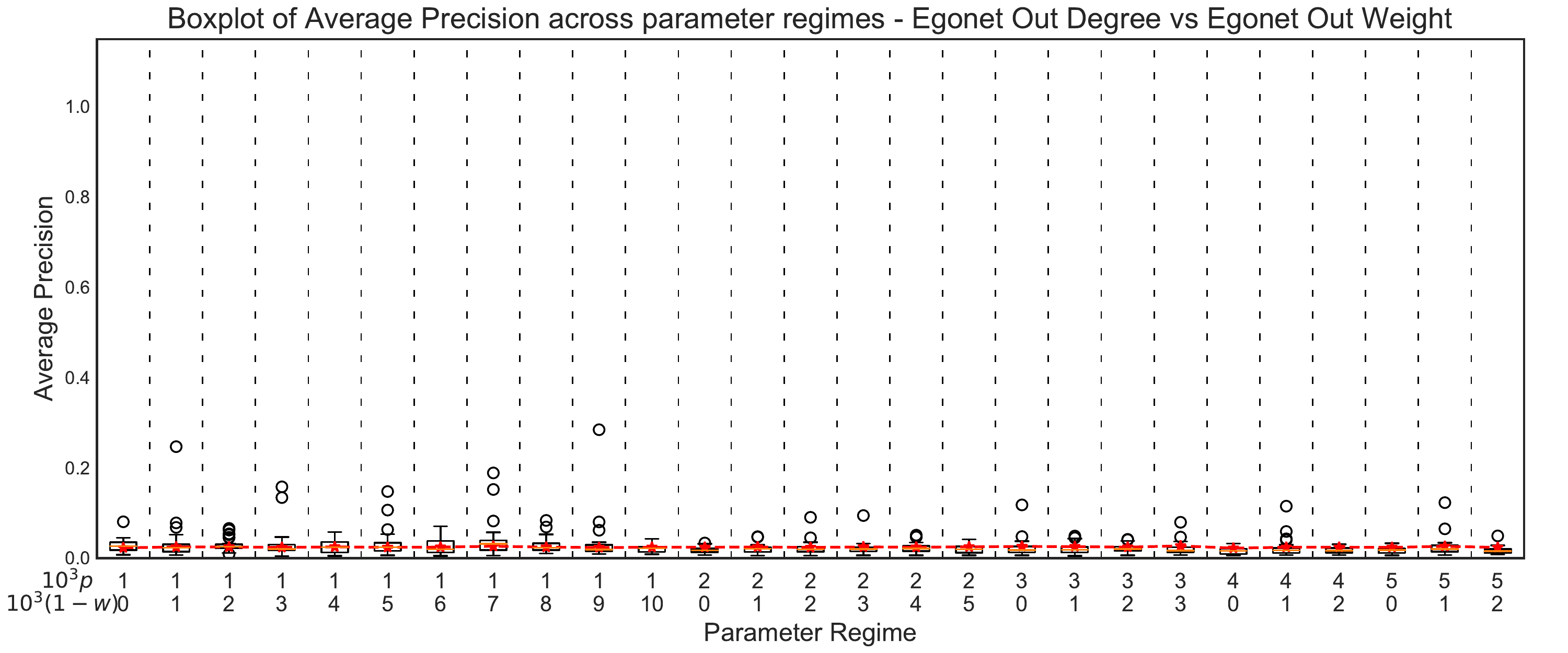}
\caption{\label{fig:oddball1_5_7} 
Performance measurements for each of the parameter regimes using the 
Oddball Egonet Out-Degree vs Egonet out-weight feature. 
The first pair of rows shows the mean precision and mean recall respectively with
error bars that are one 
one standard error over the relevant networks.
The final panel shows a boxplot of
the the average precision over the test networks.  The red starred line
indicates the largest percentage of anomalies expected at random using this
measure~\cite{sklearn}. 
    }
\end{figure*}

\begin{figure*}[h!]
\centering
\includegraphics[width=\textwidth]{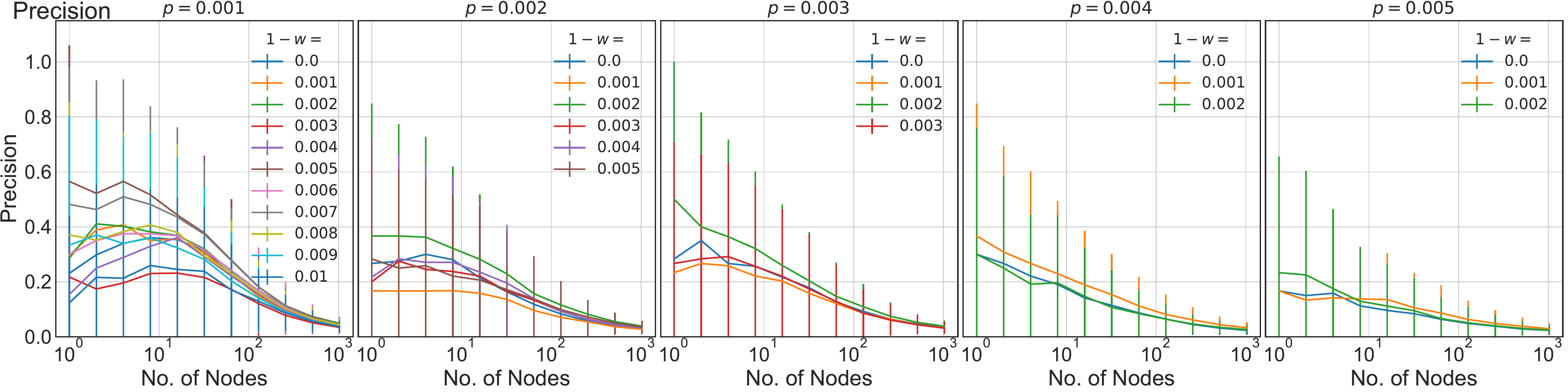}
\includegraphics[width=\textwidth]{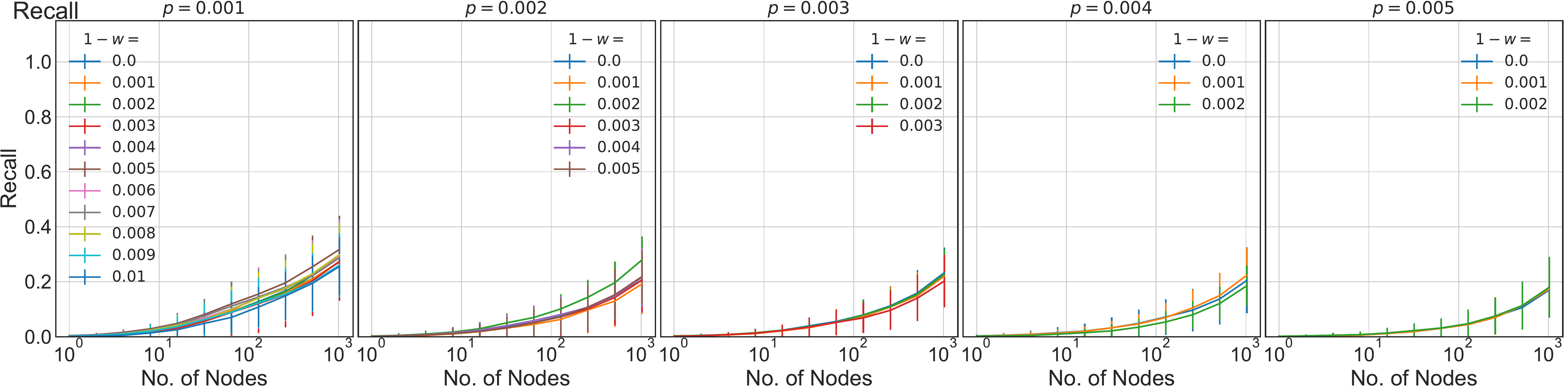}
\includegraphics[width=\textwidth]{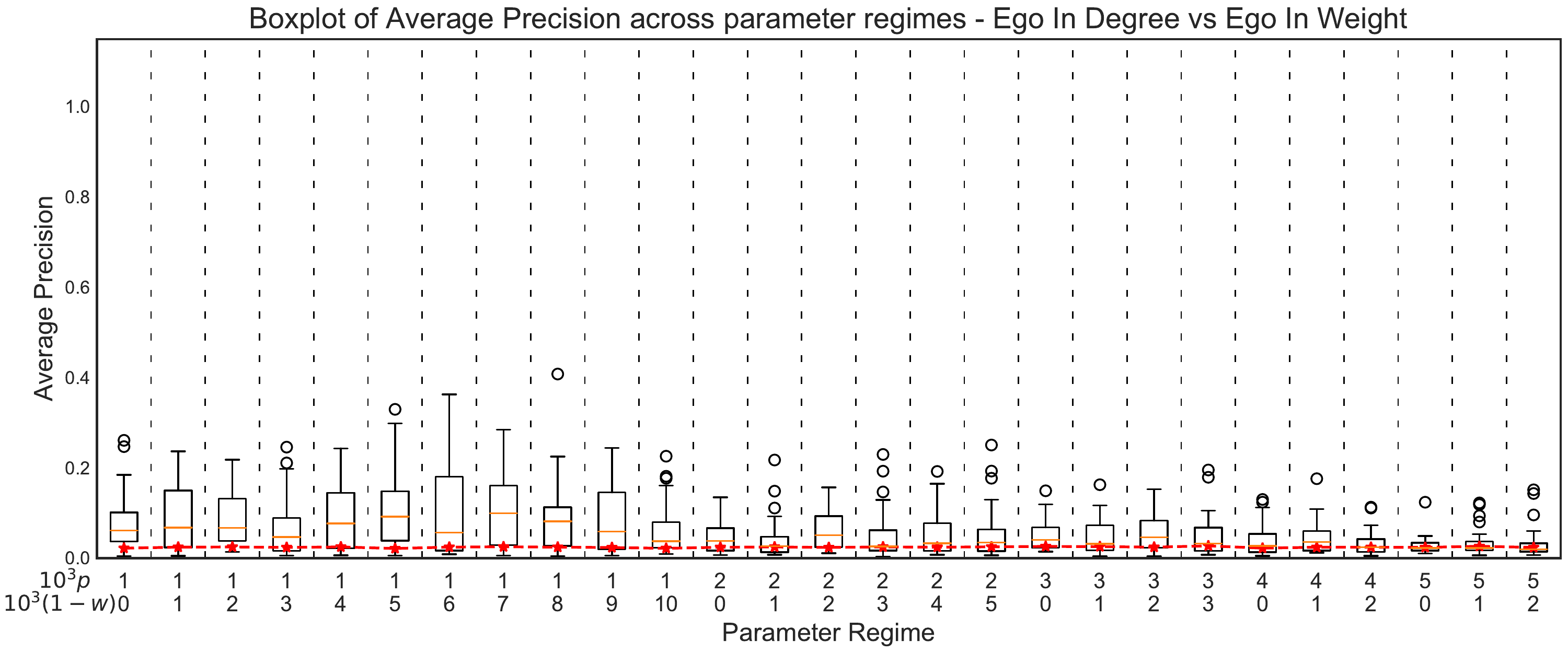}
\caption{\label{fig:oddball1_8_10} 
Performance measurements for each of the parameter regimes using the 
Oddball Ego In-Degree vs Ego In-Weight feature.
The first pair of rows shows the mean precision and mean recall respectively with
error bars that are one 
one standard error over the relevant networks.
The final panel shows a boxplot of
the the average precision over the test networks.  The red starred line
indicates the largest percentage of anomalies expected at random using this
measure~\cite{sklearn}. 
    }
\end{figure*}

\begin{figure*}[h!]
\centering
\includegraphics[width=\textwidth]{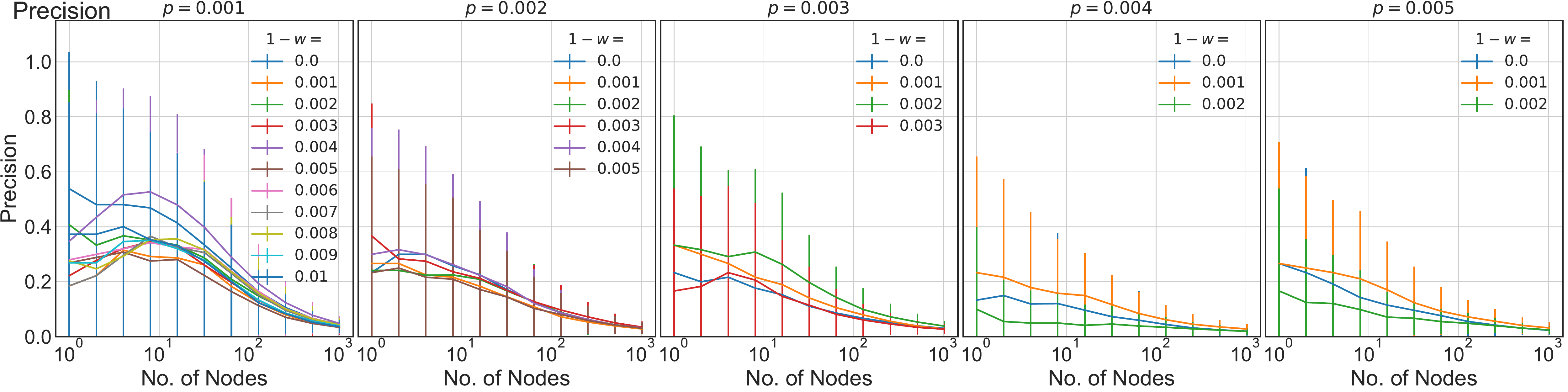}
\includegraphics[width=\textwidth]{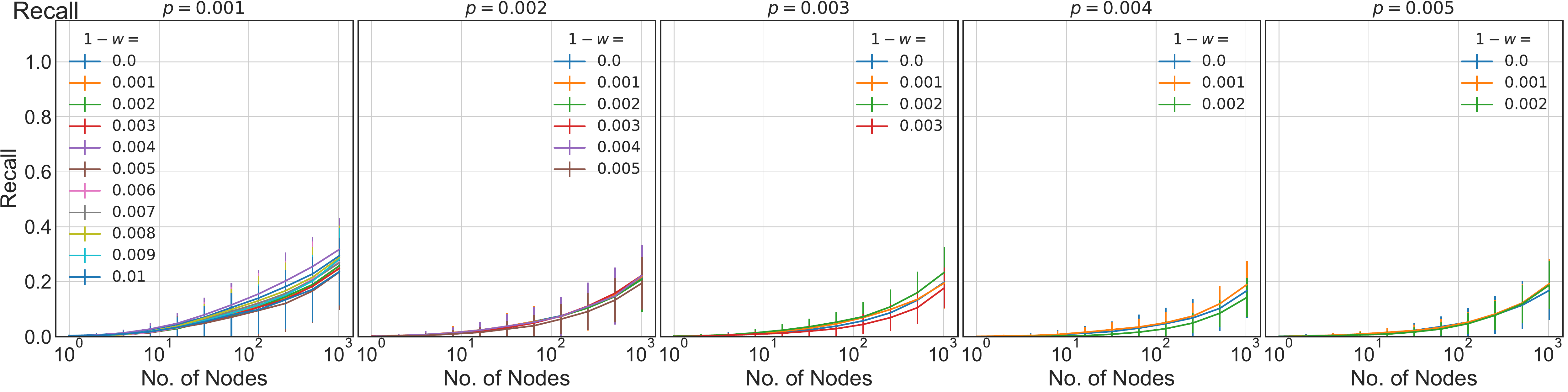}
\includegraphics[width=\textwidth]{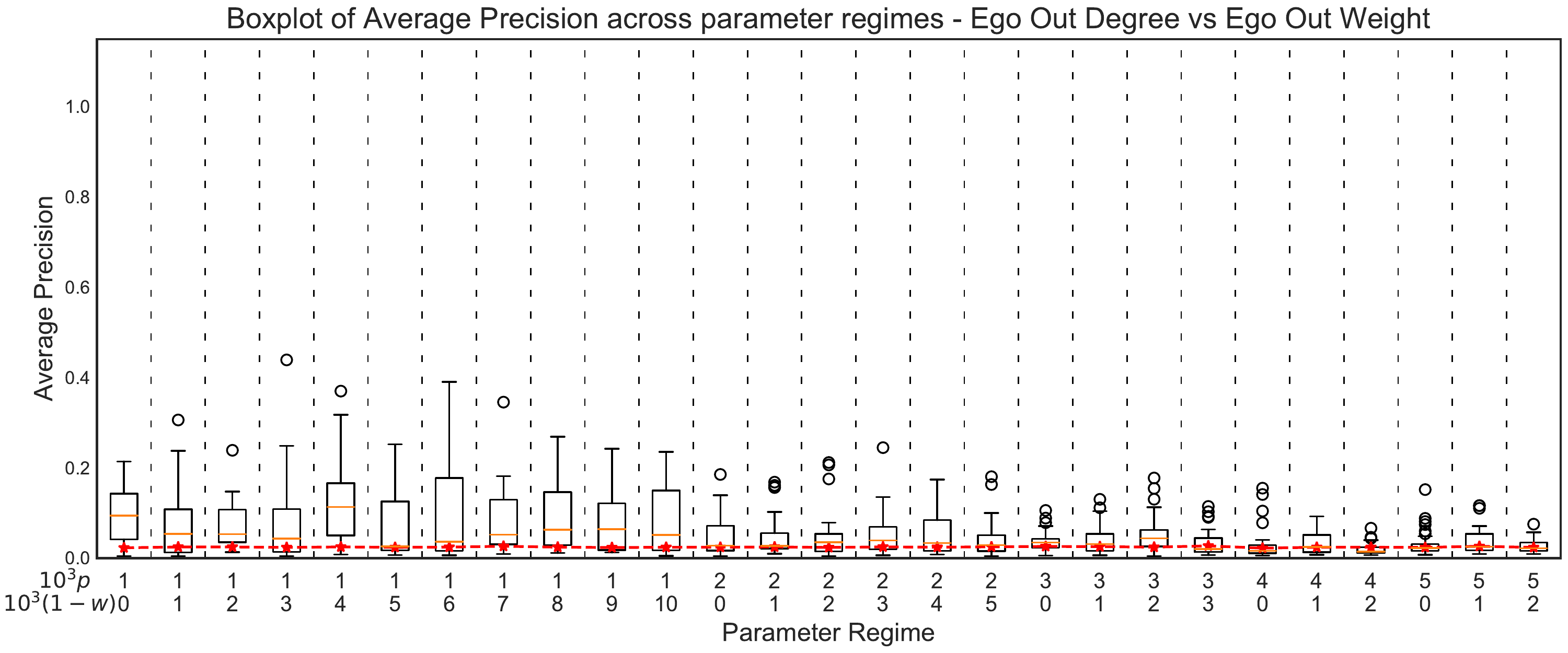}
\caption{\label{fig:oddball1_9_11}
Performance measurements for each of the parameter regimes using the 
Oddball Ego Out-Degree vs Ego Out-Weight feature. 
The first pair of rows shows the mean precision and mean recall respectively with
error bars that are one 
one standard error over the relevant networks.
The final panel shows a boxplot of
the the average precision over the test networks.  The red starred line
indicates the largest percentage of anomalies expected at random using this
measure~\cite{sklearn}. 
    }
\end{figure*}

\begin{figure*}[h!]
\centering
\includegraphics[width=\textwidth]{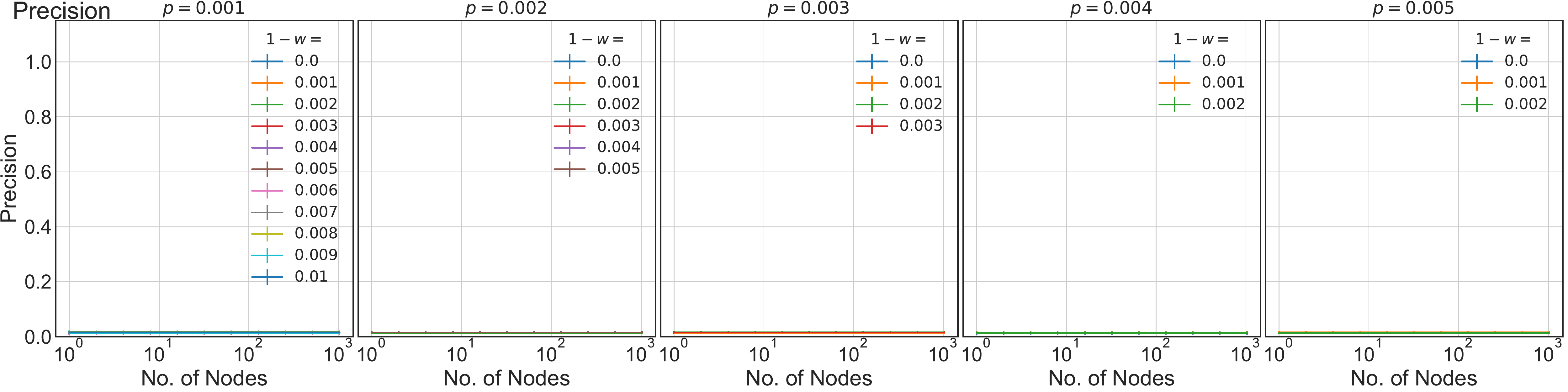}
\includegraphics[width=\textwidth]{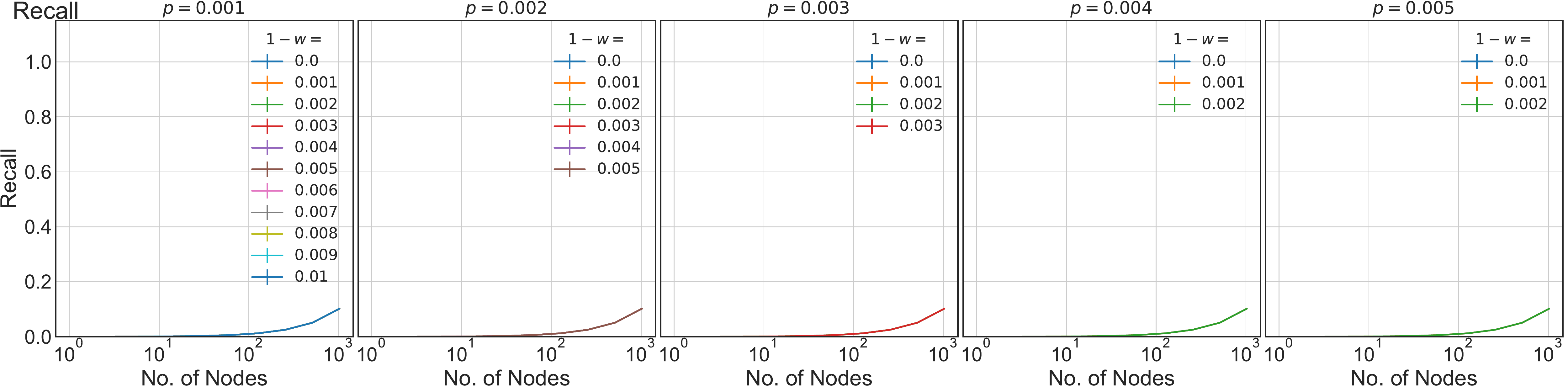}
\includegraphics[width=\textwidth]{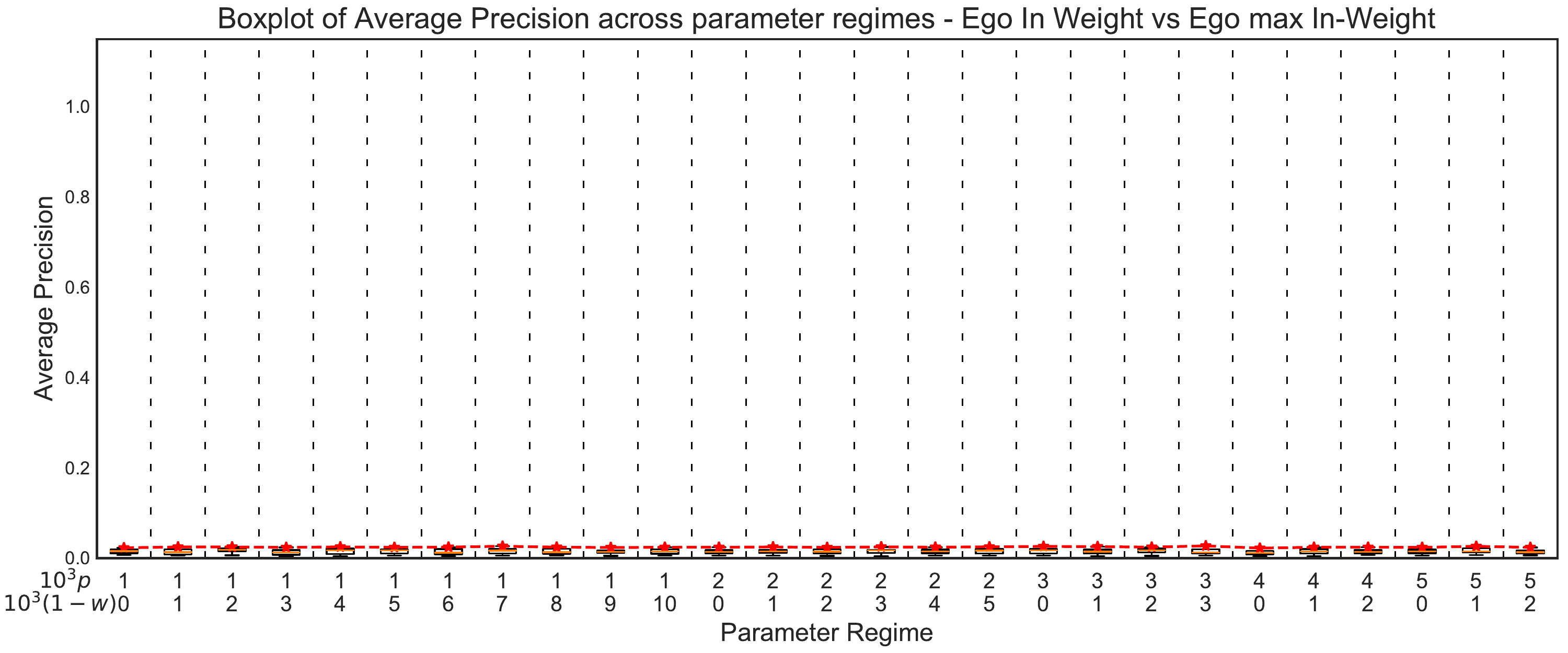}
\caption{\label{fig:oddball1_10_13} 
Performance measurements for each of the parameter regimes using the 
Oddball Ego Out-Weight vs Ego Maximal In-Weight feature.
The first pair of rows shows the mean precision and mean recall respectively with
error bars that are one 
one standard error over the relevant networks.
The final panel shows a boxplot of
the the average precision over the test networks.  The red starred line
indicates the largest percentage of anomalies expected at random using this
measure~\cite{sklearn}. 
    }
\end{figure*}

\begin{figure*}[h!]
\centering
\includegraphics[width=\textwidth]{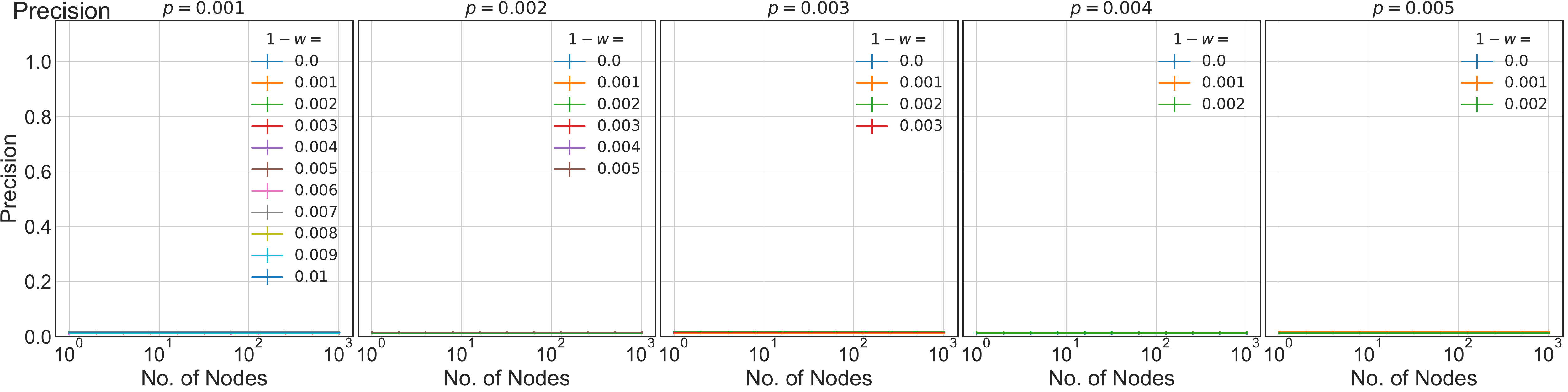}
\includegraphics[width=\textwidth]{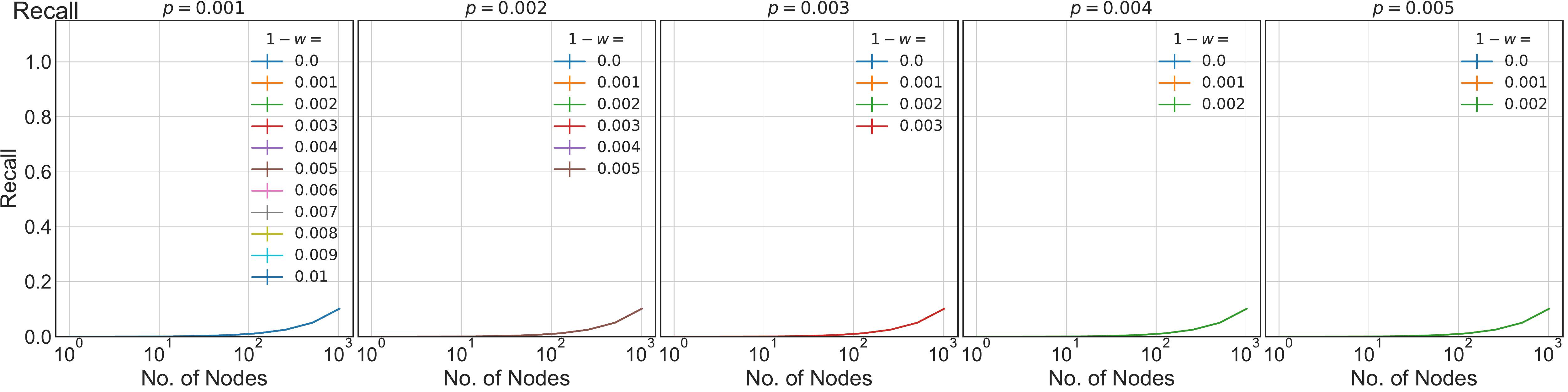}
\includegraphics[width=\textwidth]{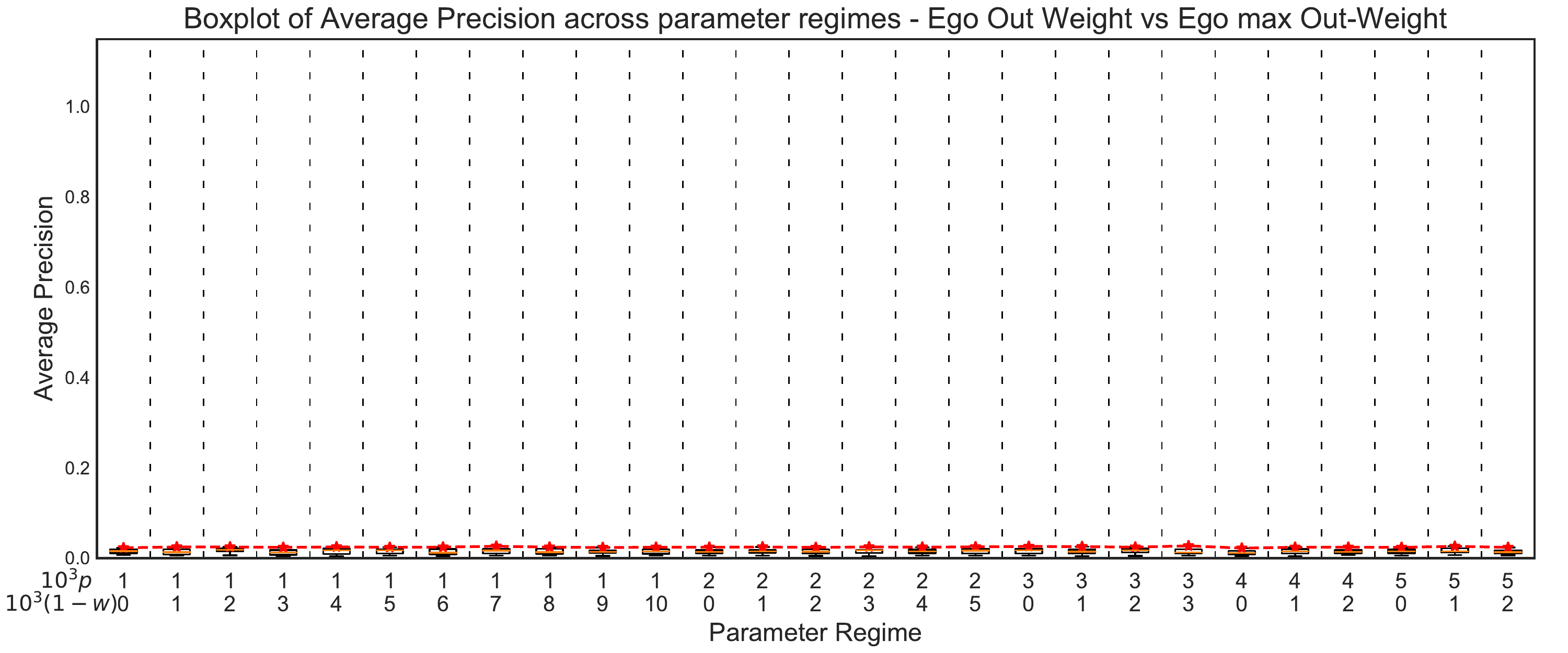}
\caption{\label{fig:oddball1_11_14} 
Performance measurements for each of the parameter regimes using the 
Oddball Ego out-Weight vs Ego maximal Out-Weight feature.
The first pair of rows shows the mean precision and mean recall respectively with
error bars that are one 
one standard error over the relevant networks.
The final panel shows a boxplot of
the the average precision over the test networks.  The red starred line
indicates the largest percentage of anomalies expected at random using this
measure~\cite{sklearn}. 
    }
\end{figure*}

For the Weighted ER network we only reported the performance of a statistic based
on the summation of each of the underlying Oddball statistics. In this section, 
we present the results for each of the statistics individually, in Figs.~\ref{fig:oddball1_1_2}, \ref{fig:oddball1_2_3},
\ref{fig:oddball1_3_12}, \ref{fig:oddball1_4_6}, \ref{fig:oddball1_5_7},
\ref{fig:oddball1_8_10}, \ref{fig:oddball1_9_11}, \ref{fig:oddball1_10_13}, and
\ref{fig:oddball1_11_14}.

None of the statistics outperform either the feature summation or
the random forest approach. A visual inspection indicates that the top four best-performing Oddball statistics are ego out-degree vs ego out-weight, ego in-degree vs ego in-weight, number of nodes vs number of edges, and number of edges vs weight.


\subsection{Oddball Individual Statistics - Accenture Model}
\label{app:oddballsplitAccenture}
\begin{figure*}[h!]
\centering
    \includegraphics[width=\textwidth]{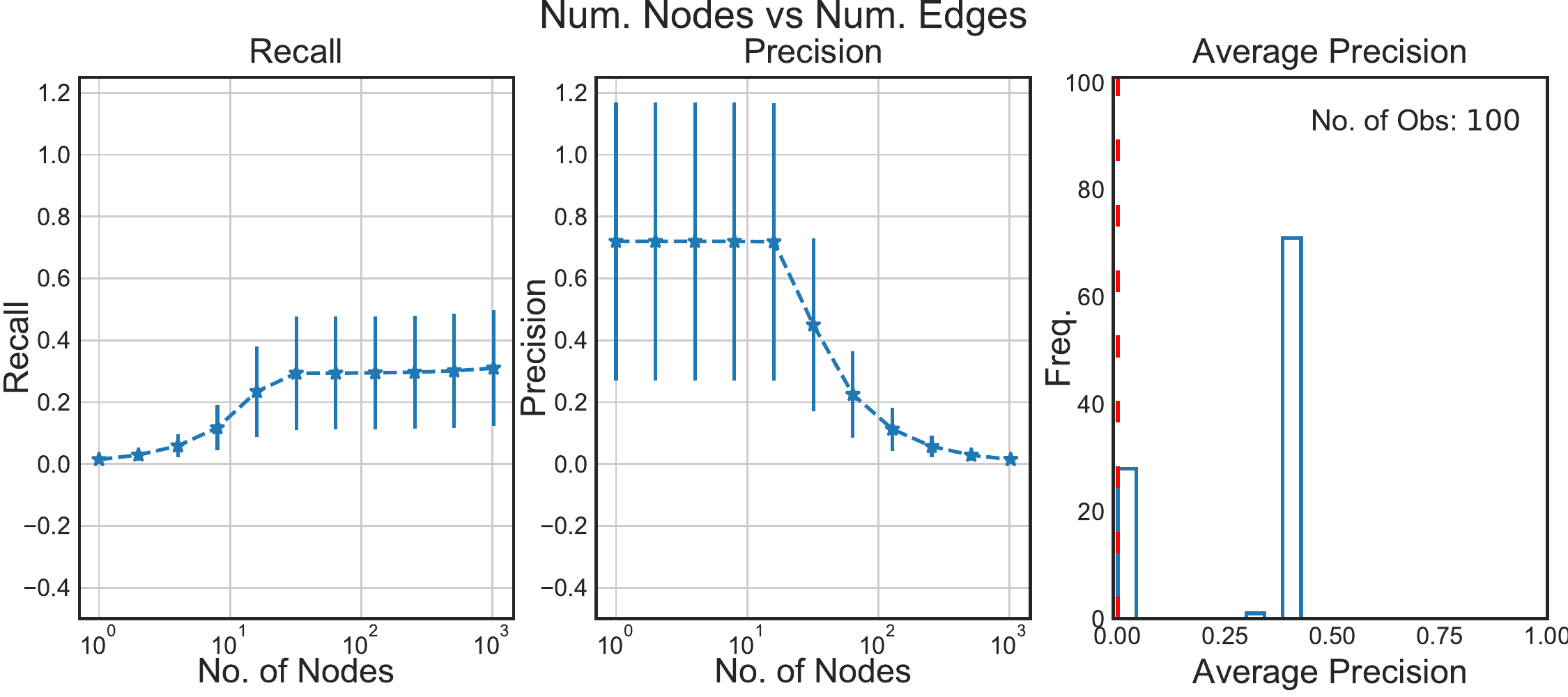}
\caption{\label{fig:AccentOddball1_2} 
    The performance, for each of the
    parameter regimes using the Number of Nodes vs Number of  Edges feature. 
    Performance is measured in three different ways.  First (far left):  
    the average 
    performance (error bars denoting 1 standard error) with respect to the recall
    when we consider the top nodes in our
    ranking. Second (middle): the average performance 
    (error bars denoting 1 standard error) 
    with respect to
    precision. Finally (far right):  the distribution of the average
    precision over our $100$ test networks. When the performance on a measure 
    resulted in Infs, then we removed the network from all of the comparisons.}
\end{figure*}

\begin{figure*}[h!]
\centering
    \includegraphics[width=\textwidth]{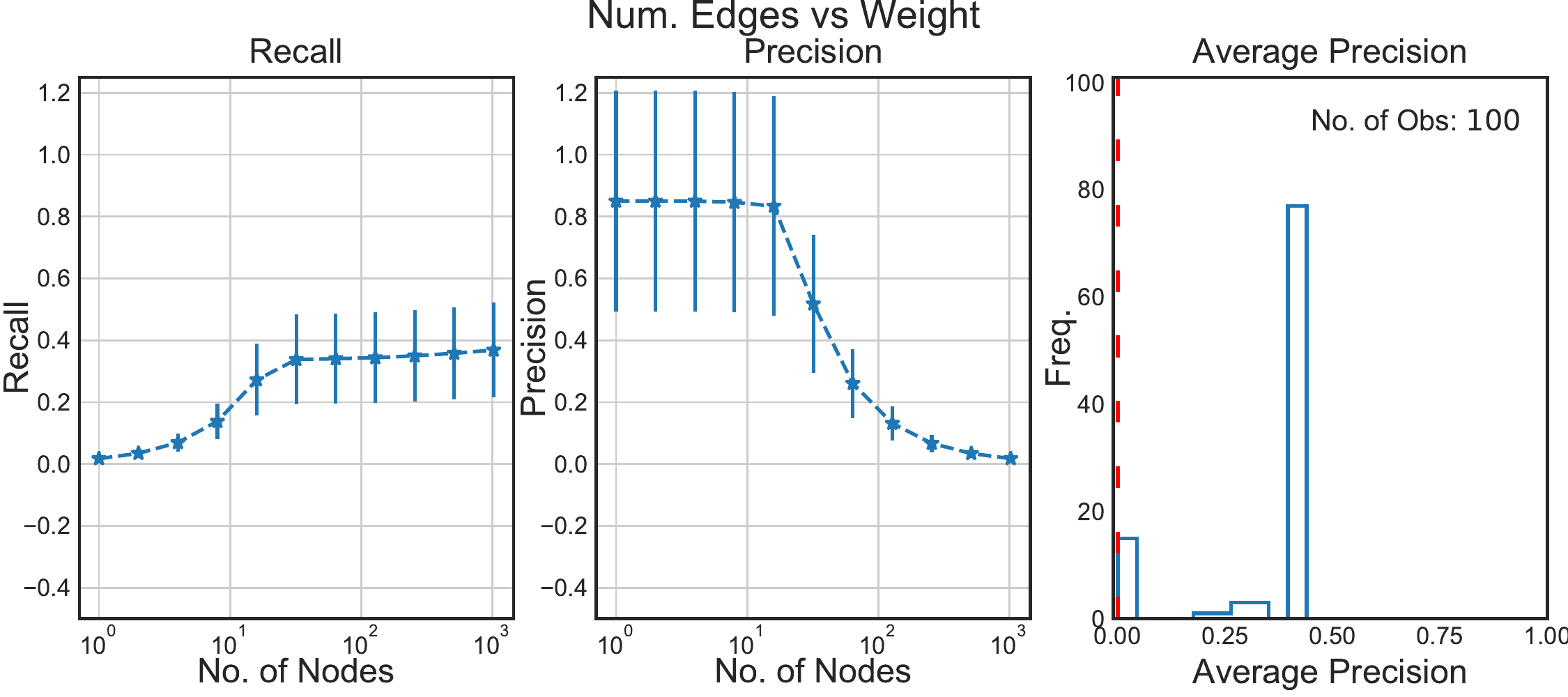}
\caption{\label{fig:AccentOddball2_3} 
The performance from scikit-learn~\cite{sklearn}, for each of the parameter regimes using the Number Edges and Weight feature. First (left): the average 
performance (error bars denoting 1 standard error) with respect to the recall when we consider the top nodes in our ranking. Second (middle):  the average performance (error bars denoting 1 standard error) with respect to precision. Finally (right): the distribution of the average precision over our $100$ test networks. When the performance on a measure resulted in Infs, then we removed the network from all of the comparisons. For ease we thus display the exact number of observations in the top right corner.}
\end{figure*}

\begin{figure*}[h!]
\centering
    \includegraphics[width=\textwidth]{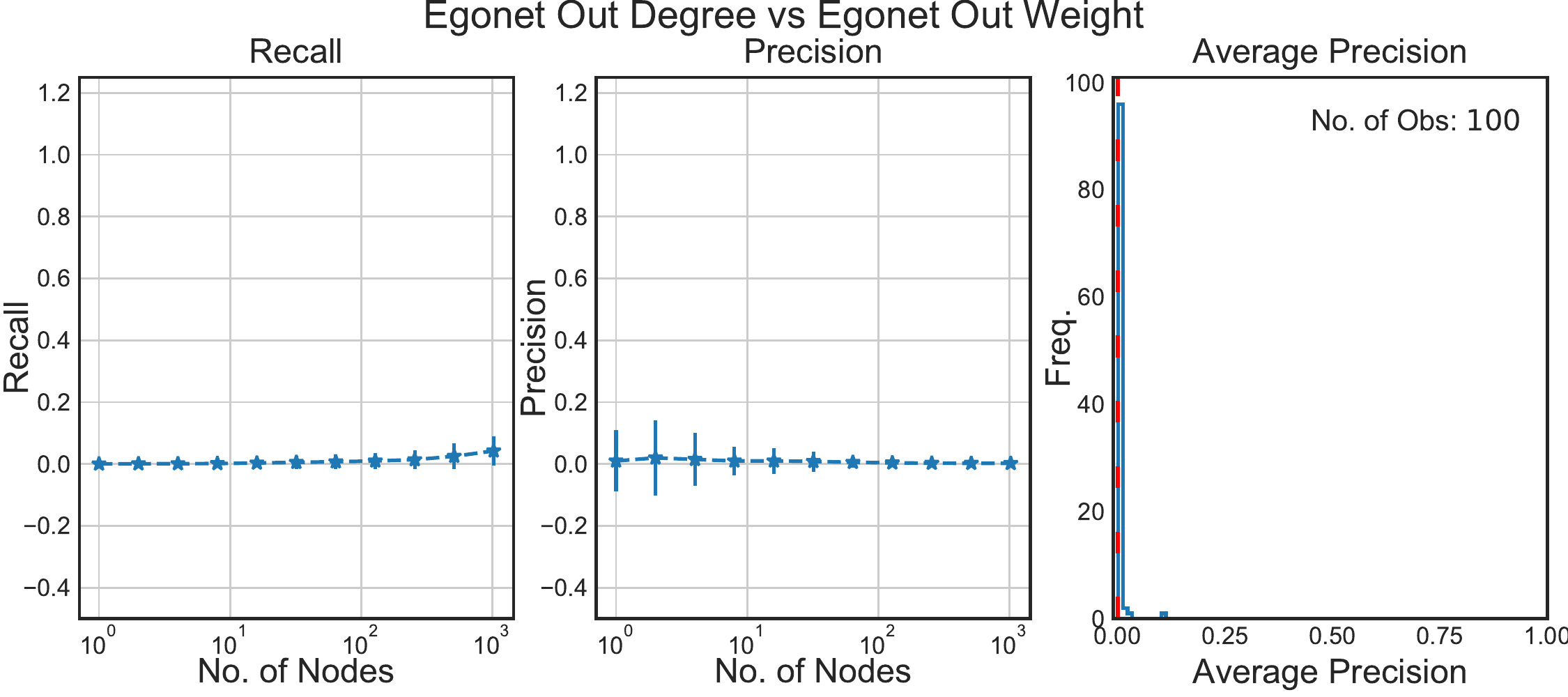}
\caption{\label{fig:AccentOddball5_7} 
    The performance from scikit-learn~\cite{sklearn}, for each of the
    parameter regimes using the Egonet Out-Degree and Egonet Out-Weight feature. 
   First (far left): 
    the average 
    performance (error bars denoting 1 standard error) with respect to the recall
    when we consider the top nodes in our
    ranking. Second (middle):  the average performance 
    (error bars denoting 1 standard error) 
    with respect to
    precision. Finally (far right): the distribution of the average
    precision over our $100$ test networks.
    When the performance on a measure 
    resulted in Infs, then we removed the network from all of the comparisons. For ease we thus 
display the exact number of observations in the top right corner.}
\end{figure*}

\begin{figure*}[h!]
\centering
    \includegraphics[width=\textwidth]{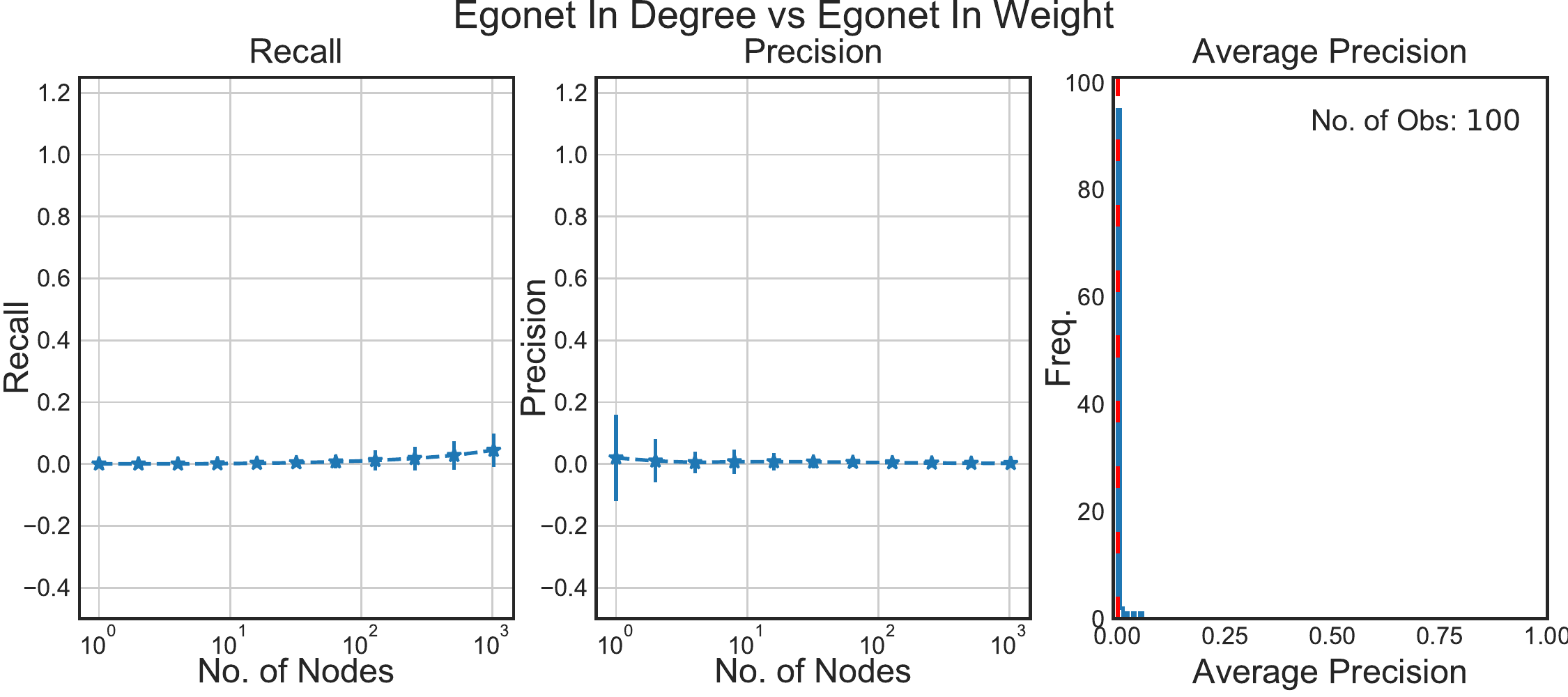}
\caption{\label{fig:AccentOddball4_6} 
    The performance from scikit-learn~\cite{sklearn}, for each of the
    parameter regimes using the Egonet In-Degree and Egonet In-Weight feature.
    First (far left): 
    the average 
    performance (error bars denoting 1 standard error) with respect to the recall
    when we consider the top nodes in our
    ranking. Second (middle):  the average performance 
    (error bars denoting 1 standard error) 
    with respect to
    precision. Finally (far right): the distribution of the average
    precision over our $100$ test networks.
    When the performance on a measure 
    resulted in Infs, then we removed the network from all of the comparisons. For ease we thus 
display the exact number of observations in the top right corner.}
\end{figure*}

\begin{figure*}[h!]
\centering
    \includegraphics[width=\textwidth]{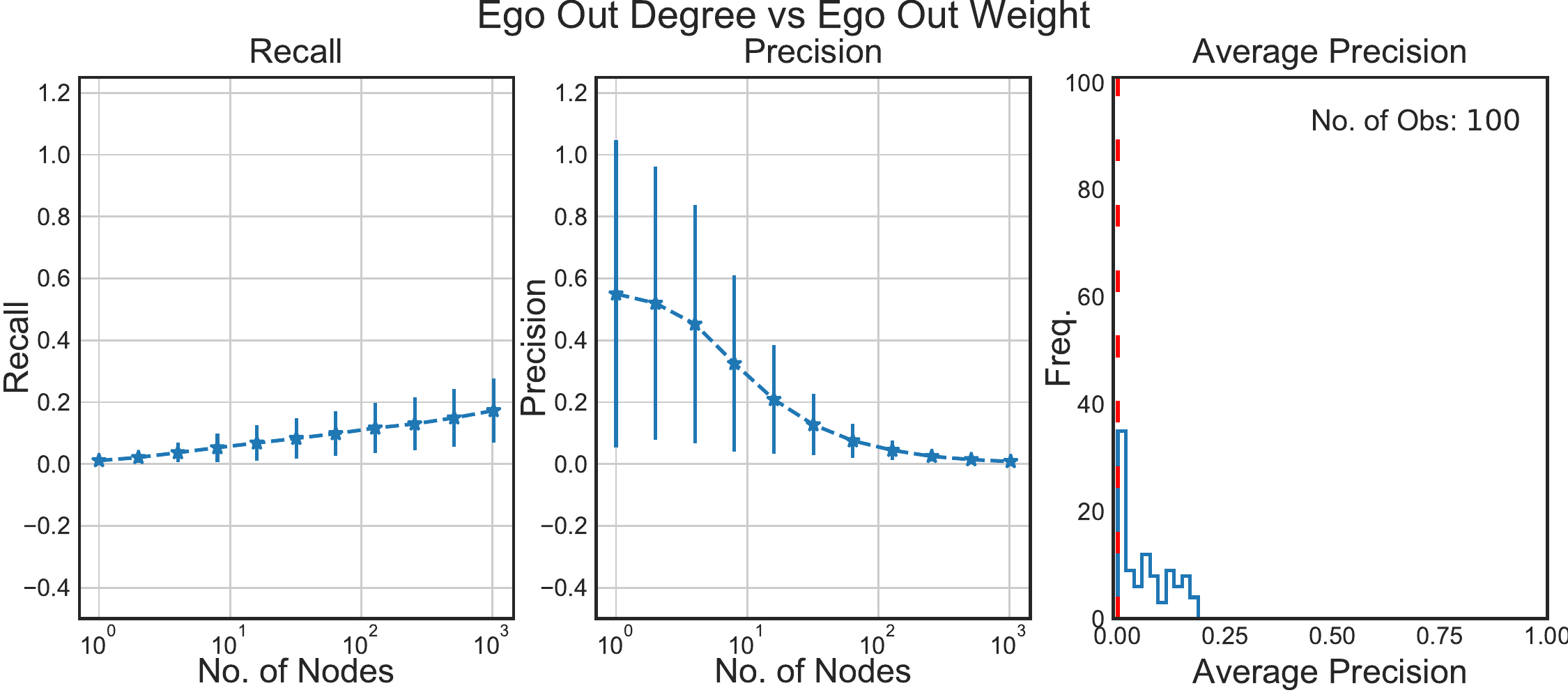}
\caption{\label{fig:AccentOddball9_11} 
    The performance from scikit-learn~\cite{sklearn}, for each of the
    parameter regimes using the Ego Out-Degree and Ego Out-Weight feature. First (far left): 
    the average 
    performance (error bars denoting 1 standard error) with respect to the recall
    when we consider the top nodes in our
    ranking. Second (middle):  the average performance 
    (error bars denoting 1 standard error) 
    with respect to
    precision. Finally (far right): the distribution of the average
    precision over our $100$ test networks.
    When the performance on a measure 
    resulted in Infs, then we removed the network from all of the comparisons.
    }
\end{figure*}

\begin{figure*}[h!]
\centering
    \includegraphics[width=\textwidth]{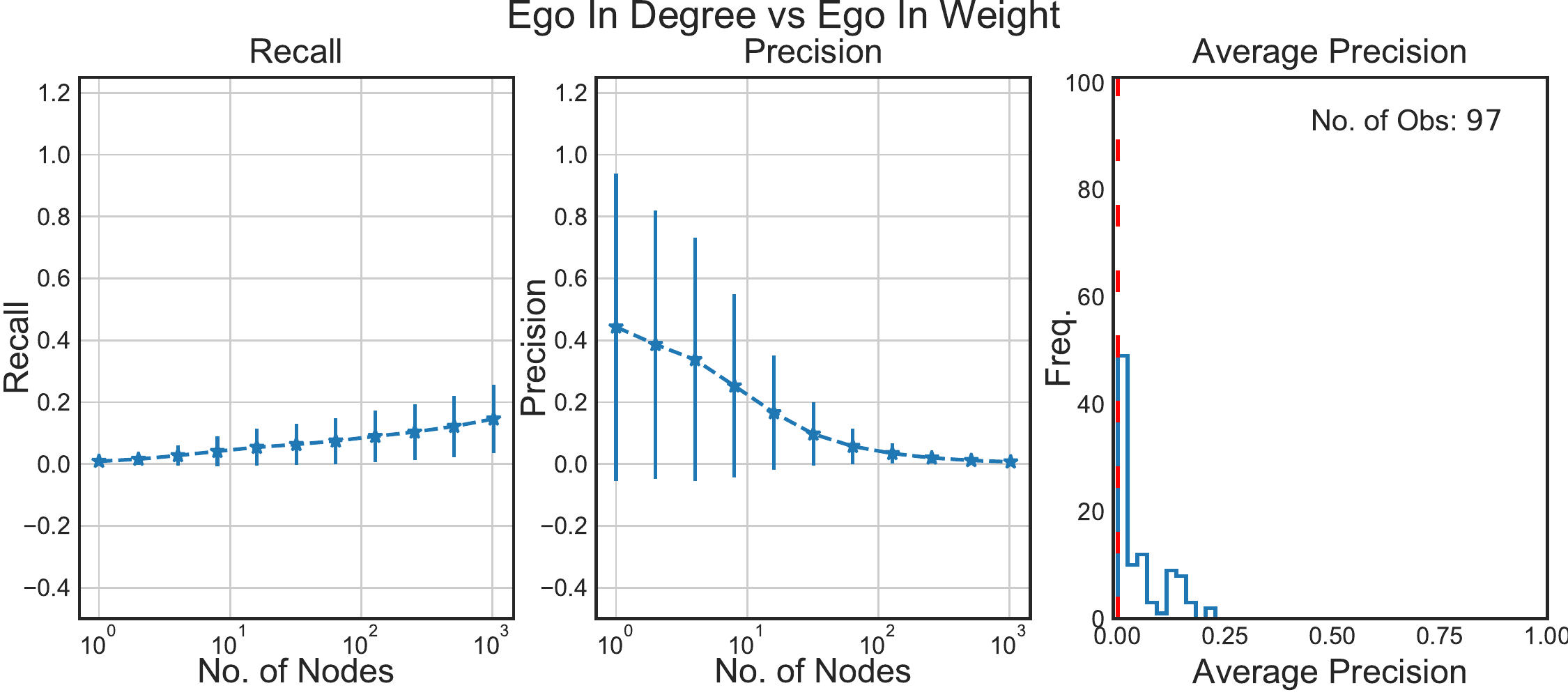}
\caption{\label{fig:AccentOddball8_10} 
    The performance from scikit-learn~\cite{sklearn}, for each of the
    parameter regimes using the Ego In-Degree and Ego In-Weight feature.
    First (far left): 
    the average 
    performance (error bars denoting 1 standard error) with respect to the recall
    when we consider the top nodes in our
    ranking. Second (middle):  the average performance 
    (error bars denoting 1 standard error) 
    with respect to
    precision. Finally (far right): the distribution of the average
    precision over our $100$ test networks.
    When the performance on a measure 
    resulted in Infs, then we removed the network from all of the comparisons. For ease we thus 
display the exact number of observations in the top right corner.}
\end{figure*}

\begin{figure*}[h!]
\centering
    \includegraphics[width=\textwidth]{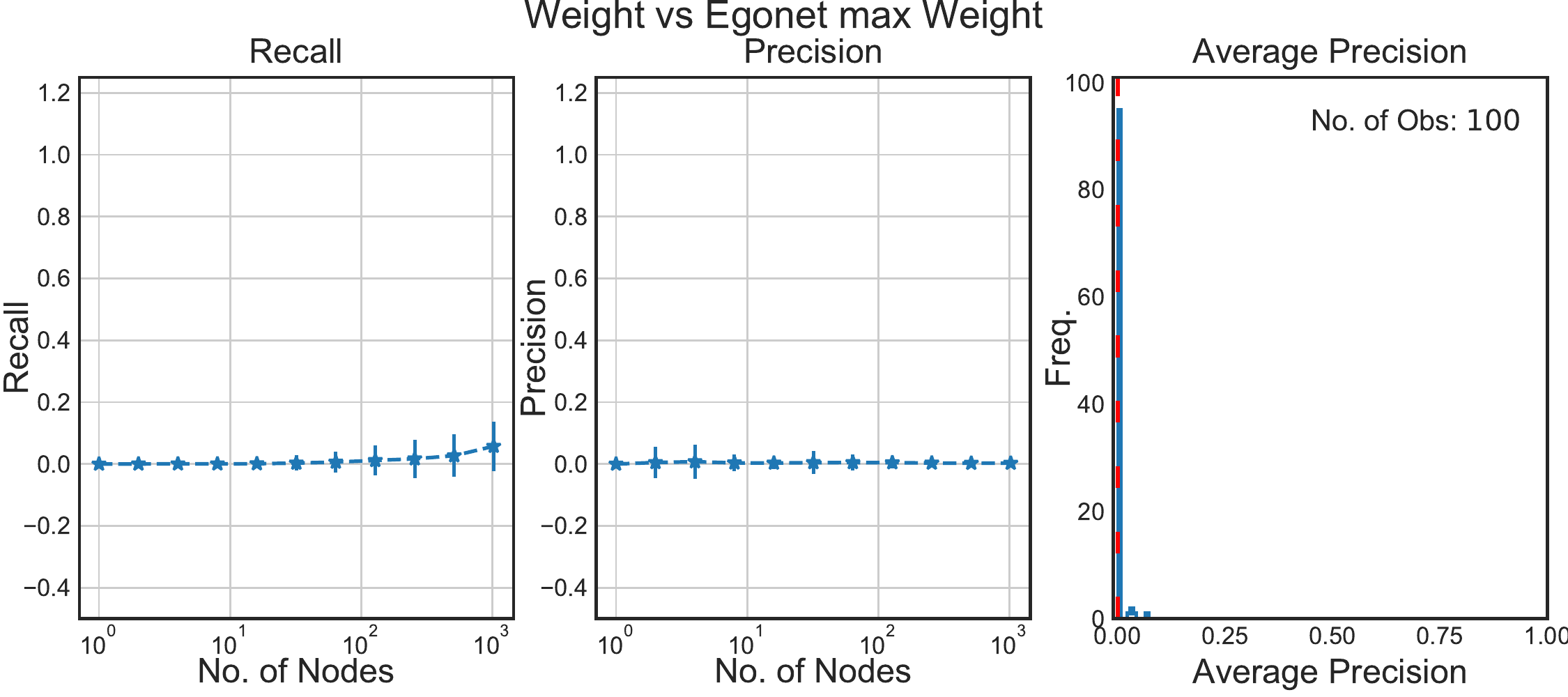}
\caption{\label{fig:AccentOddball3_12} 
   The performance from scikit-learn~\cite{sklearn}, for each of the
    parameter regimes using the Egonet weight and Egonet Maximum Weight. First (far left): 
    the average 
    performance (error bars denoting 1 standard error) with respect to the recall
    when we consider the top nodes in our
    ranking. Second (middle):  the average performance 
    (error bars denoting 1 standard error) 
    with respect to
    precision. Finally (far right): the distribution of the average
    precision over our $100$ test networks.
    When the performance on a measure 
    resulted in Infs, then we removed the network from all of the comparisons. For ease we thus 
display the exact number of observations in the top right corner.}
\end{figure*}

\begin{figure*}[h!]
\centering
    \includegraphics[width=\textwidth]{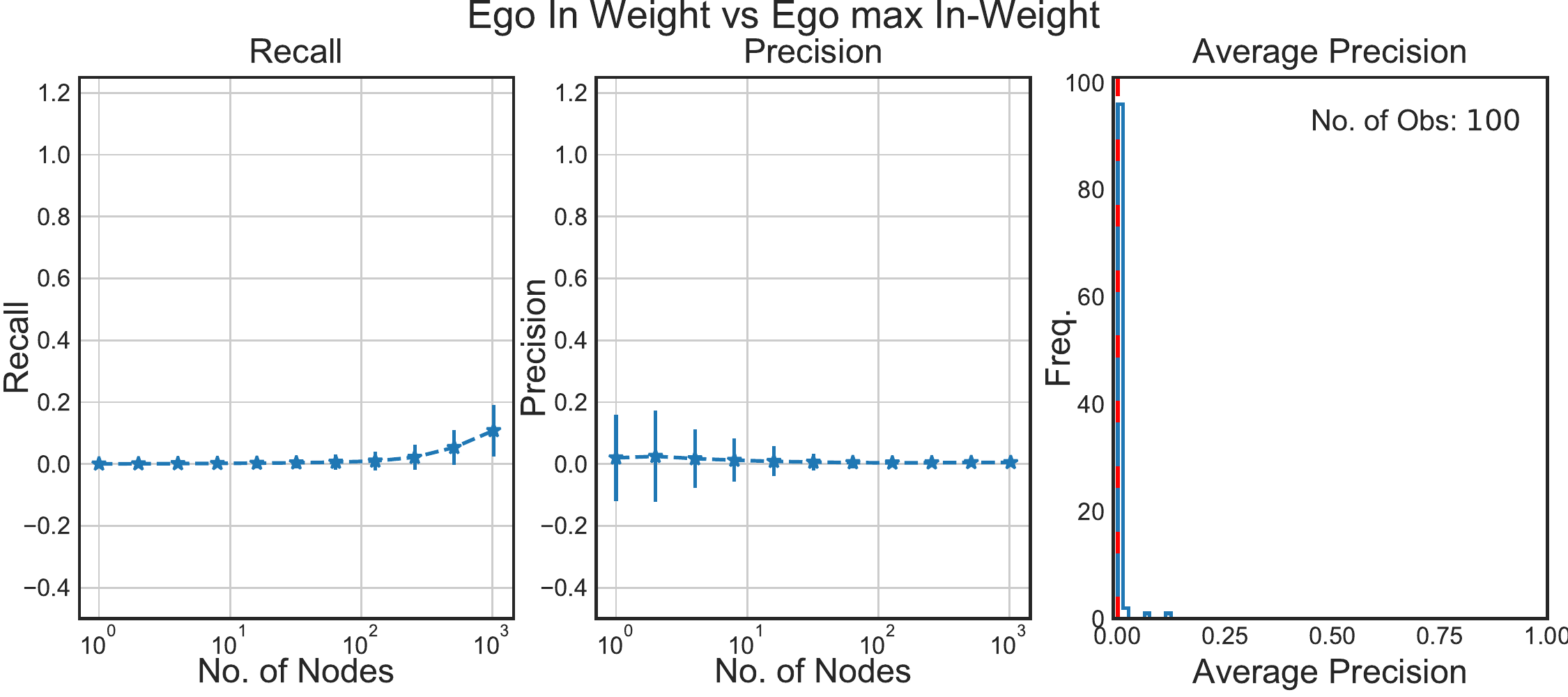}
\caption{\label{fig:AccentOddball10_13} 
    The performance from scikit-learn~\cite{sklearn}, for each of the
    parameter regimes using the Ego In-Weight and Ego Maximum In-Weight. First (far left): 
    the average 
    performance (error bars denoting 1 standard error) with respect to the recall
    when we consider the top nodes in our
    ranking. Second (middle):  the average performance 
    (error bars denoting 1 standard error) 
    with respect to
    precision. Finally (far right): the distribution of the average
    precision over our $100$ test networks.
    When the performance on a measure 
    resulted in Infs, then we removed the network from all of the comparisons. For ease we thus  display the exact number of observations in the top right corner.} 
\end{figure*}

\begin{figure*}[h!]
\centering 
    \includegraphics[width=\textwidth]{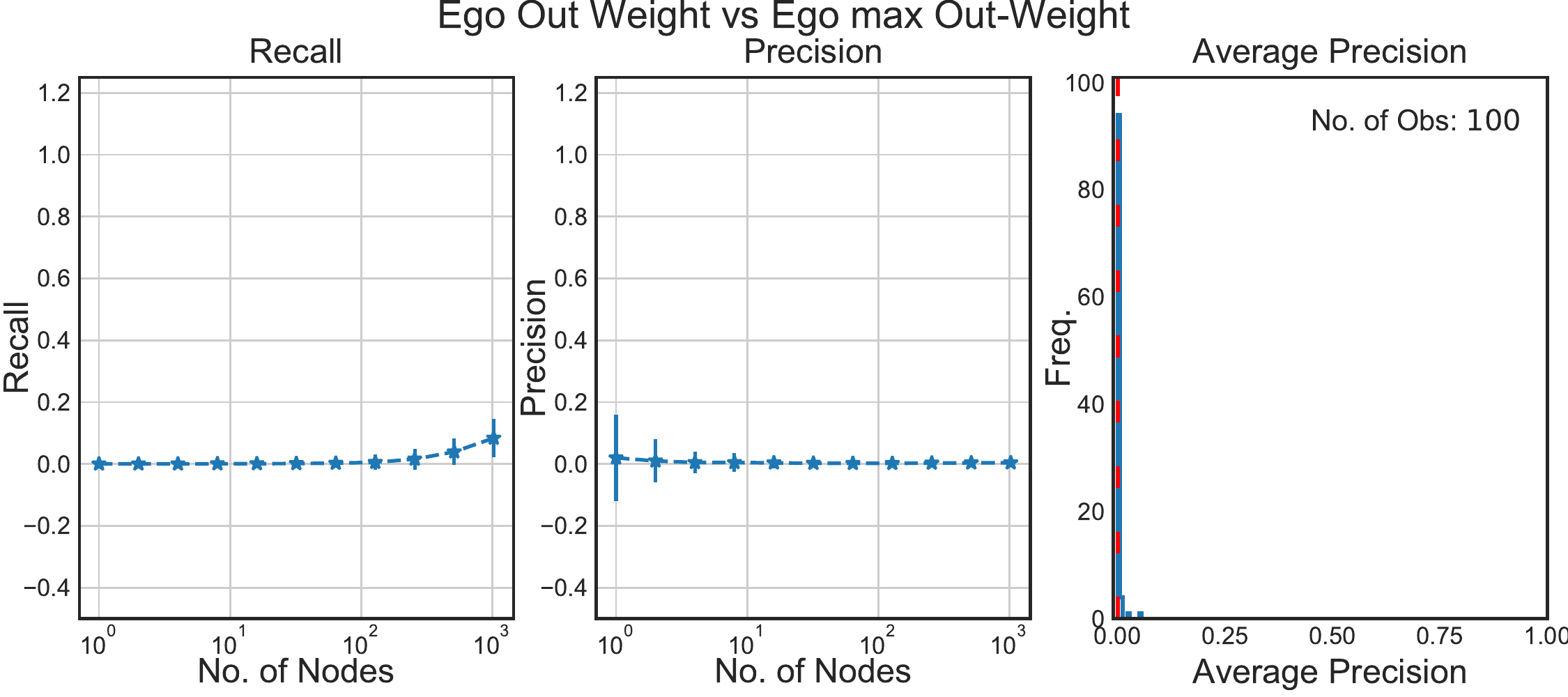}
\caption{\label{fig:AccentOddball11_14} 
    The performance from scikit-learn~\cite{sklearn}, for each of the
    parameter regimes using the Ego Out-Weight and Ego Maximal Out-Weight.
    First (far left): 
    the average 
    performance (error bars denoting 1 standard error) with respect to the recall
    when we consider the top nodes in our
    ranking. Second (middle):  the average performance 
    (error bars denoting 1 standard error) 
    with respect to
    precision. Finally (far right): the distribution of the average
    precision over our $100$ test networks.
    When the performance on a measure 
    resulted in Infs, then we removed the network from all of the comparisons. For ease we thus 
display the exact number of observations in the top right corner.}
\end{figure*}

For completeness, this section shows the recall, precision and average
precision plots for each of the underlying Oddball statistics over the
$100$ Accenture networks.  Note that, for some of the networks, some
of the underlying Oddball statistics are badly defined, in the sense that they contain Infs. The results can be found in 
Figs.~\ref{fig:AccentOddball1_2}, \ref{fig:AccentOddball2_3},
\ref{fig:AccentOddball5_7}, \ref{fig:AccentOddball4_6},
\ref{fig:AccentOddball9_11}, \ref{fig:AccentOddball8_10},
\ref{fig:AccentOddball3_12}, \ref{fig:AccentOddball10_13}, and 
\ref{fig:AccentOddball11_14}.

\clearpage

\section{Exploring Anomalous Non-Embedded Nodes}
\label{app:nonSelectedNodes}
\begin{figure}
\includegraphics[width=\linewidth]{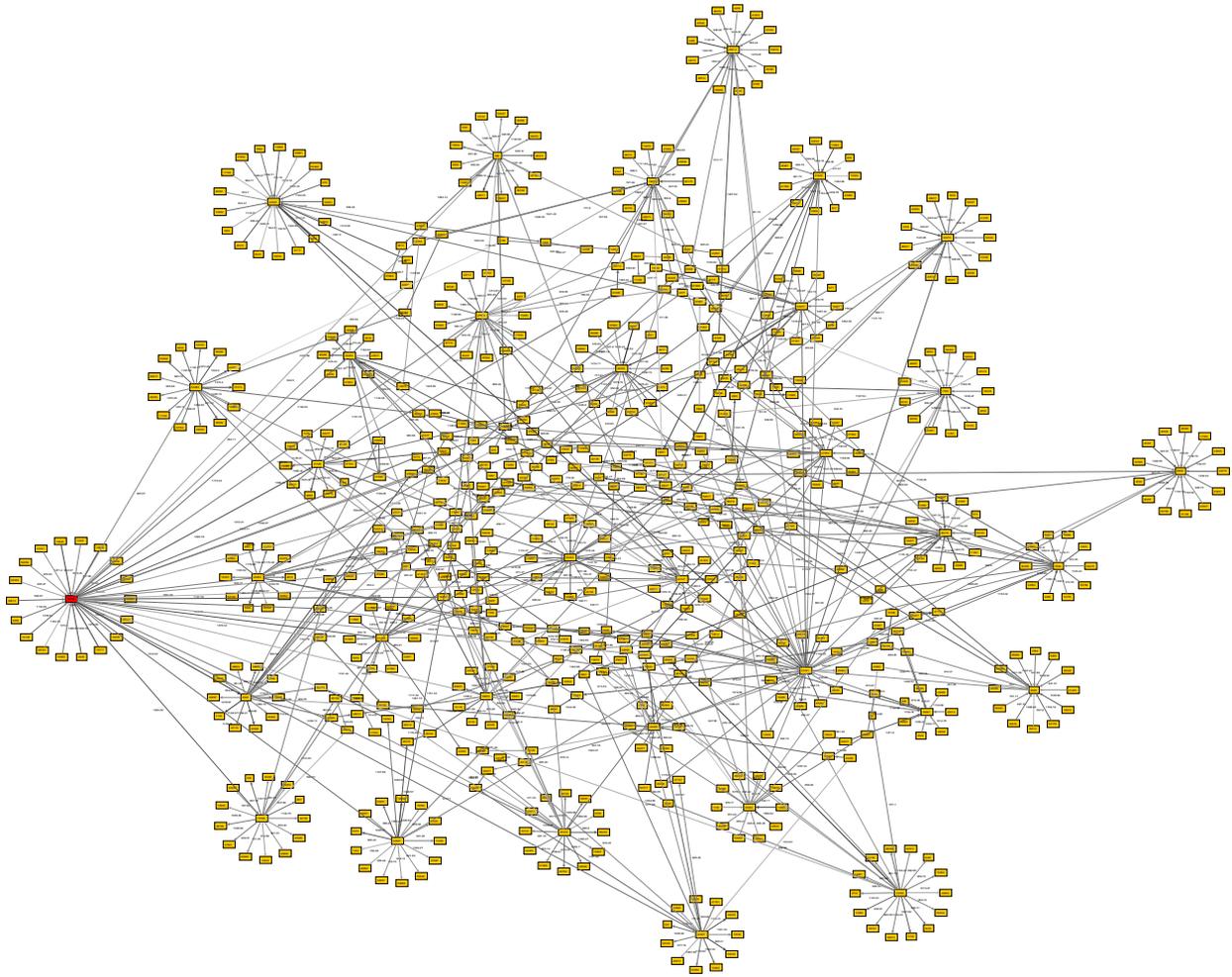}
\caption{Network plot of the 2-hop snowball sample (ignoring edge direction) of node 14,708 in a realisation of the Accenture synthetic model. This node was selected as it is identified as anomalous in the random forest but was not planted in the network.}
\label{fig:2hopNetwork14708}
\end{figure}

\begin{figure}
\includegraphics[width=\linewidth]{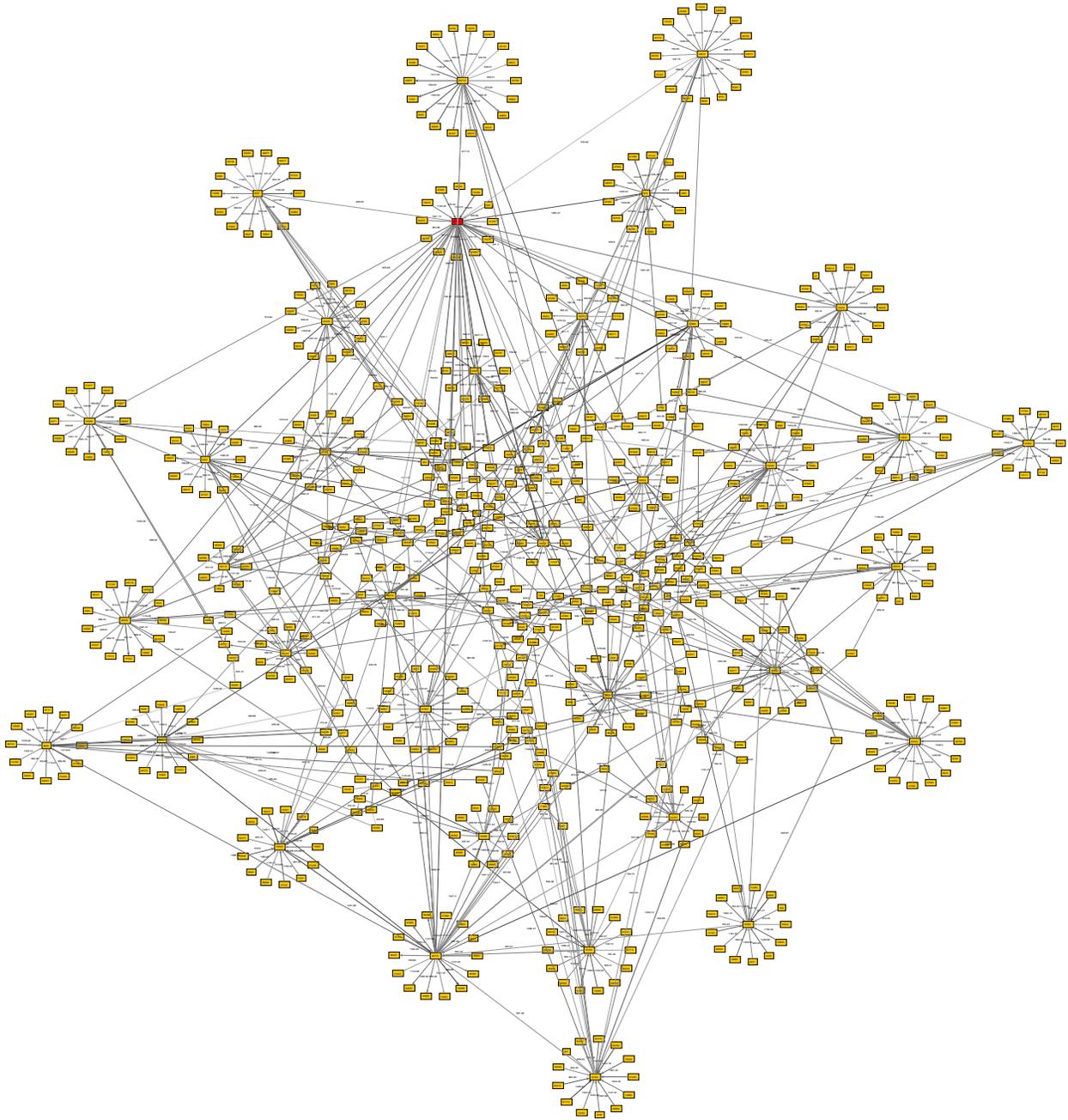}
\caption{Network plot of the 2 hop snowball sample (ignoring edge direction) of node 22,745 in a realisation of the Accenture synthetic model. This node was selected as it is identified as anomalous in the random forest but was not planted in the network.}
\label{fig:2hopNetwork22745}
\end{figure}

\begin{figure}
\includegraphics[width=\linewidth]{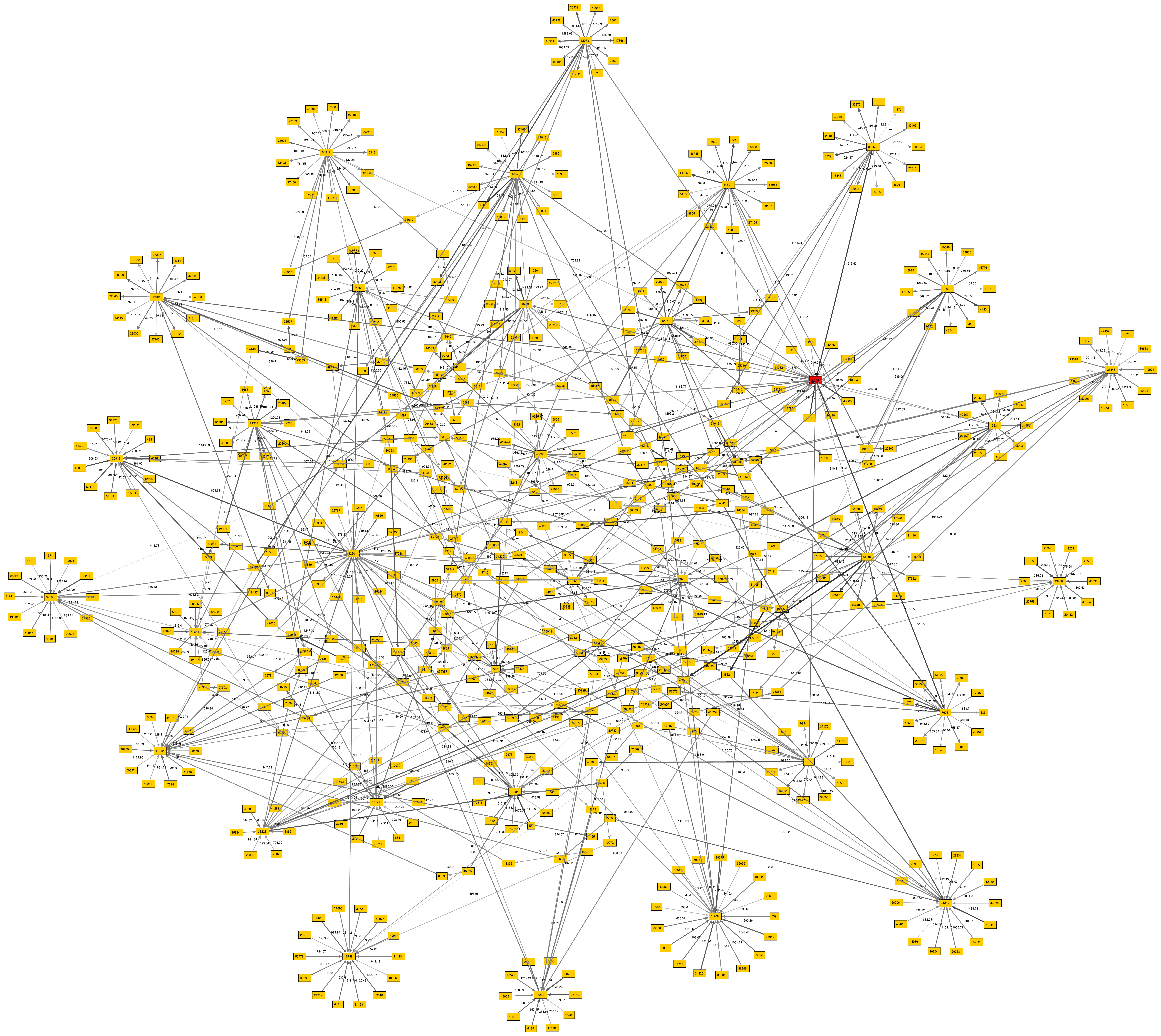}
\caption{Network plot of the 2 hop snowball sample (ignoring edge direction) of node 32,251 in a realisation of the Accenture synthetic model. This node was selected as it is identified as anomalous in the random forest but was not planted in the network.}
\label{fig:2hopNetwork32251}
\end{figure}

\begin{figure}
\includegraphics[width=\linewidth]{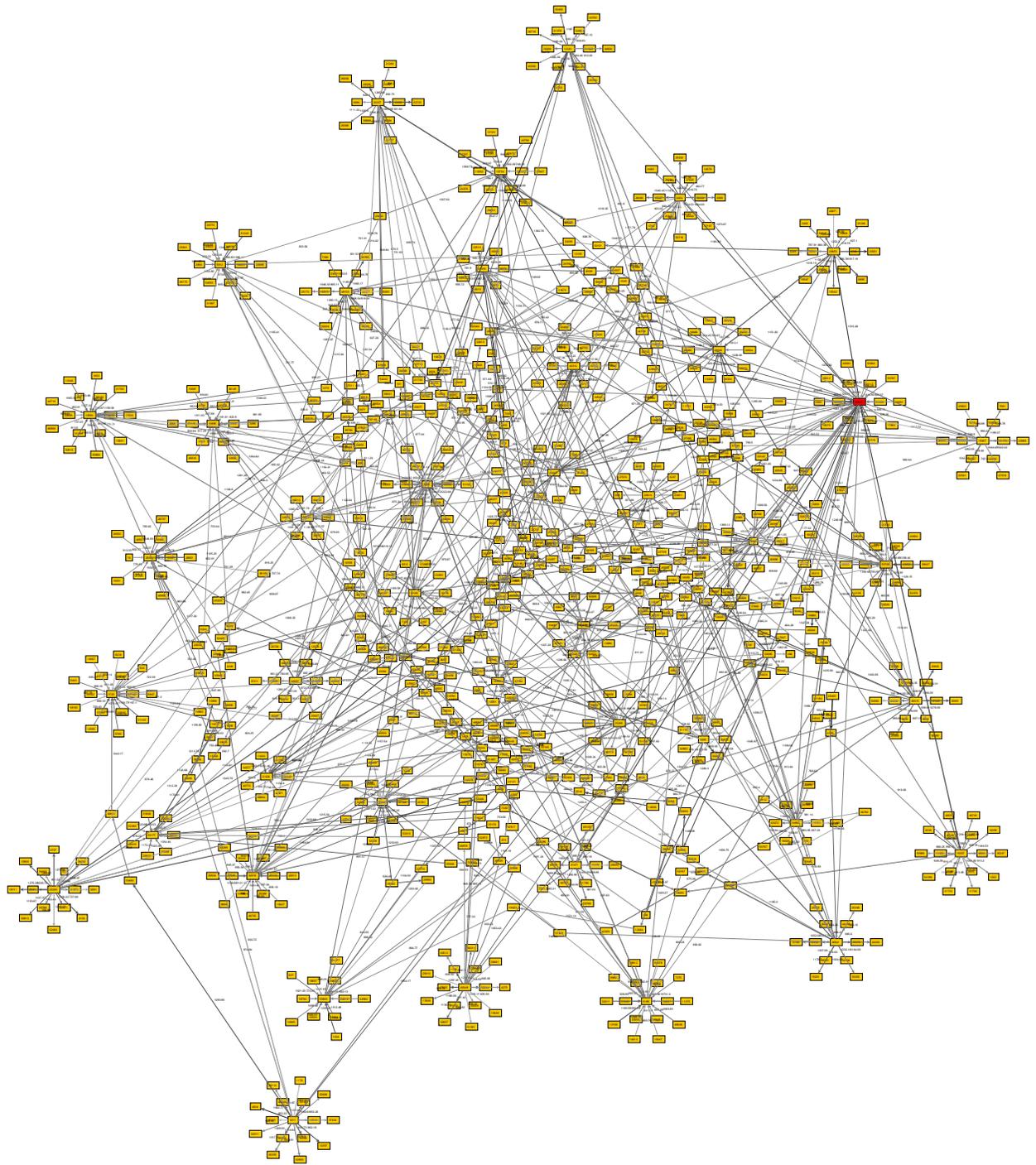}
\caption{Network plot of the 2 hop snowball sample (ignoring edge direction) of node 32,545 in a realisation of the Accenture synthetic model. This node was selected as it is identified as anomalous in the random forest but was not planted in the network.}
\label{fig:2hopNetwork32545}
\end{figure}

\begin{figure}
\includegraphics[width=\linewidth]{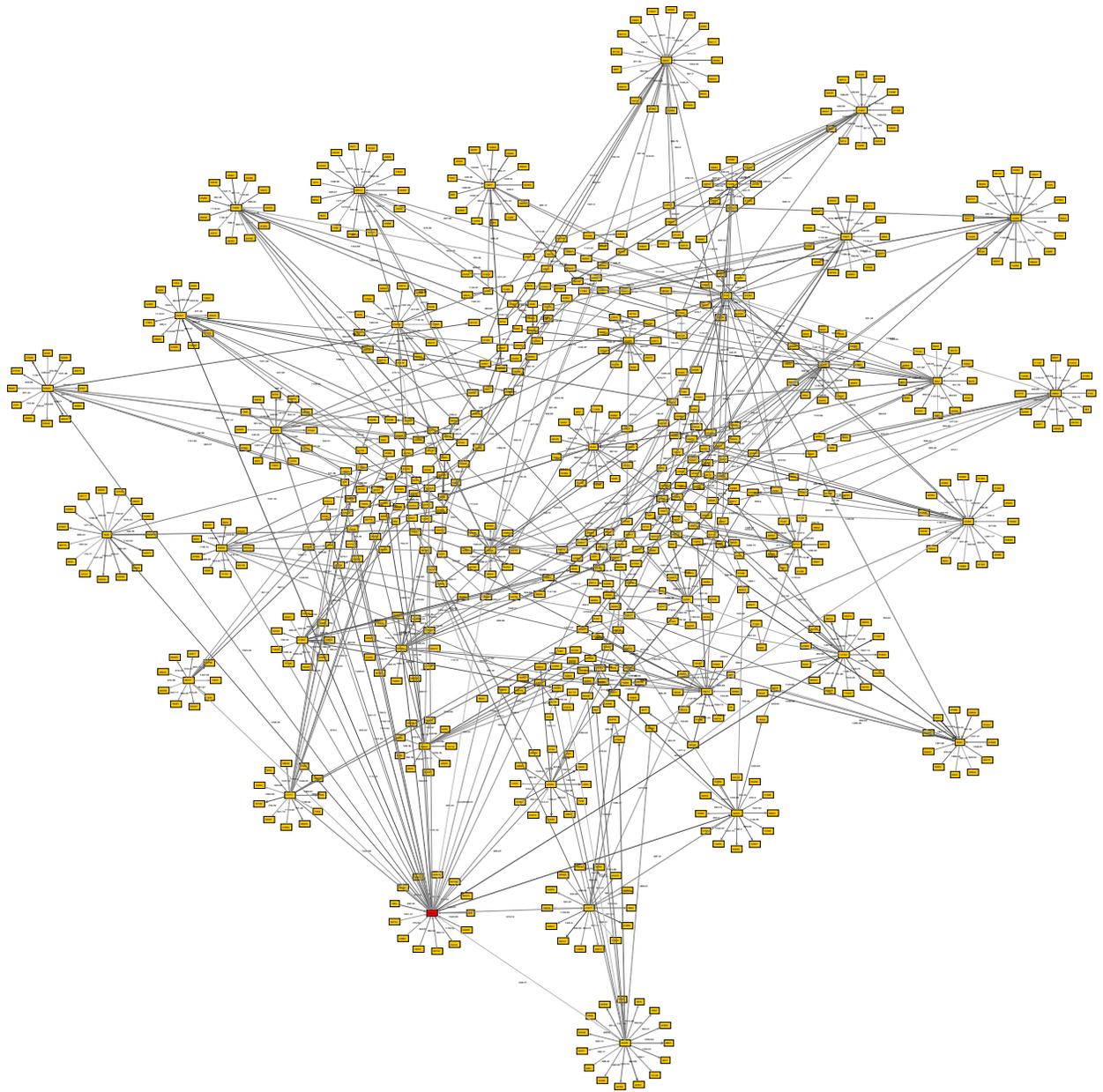}
\caption{Network plot of the 2 hop snowball sample (ignoring edge direction) of node 34,446 in a realisation of the Accenture synthetic model. This node was selected as it is identified as anomalous in the random forest but was not planted in the network.}
\label{fig:2hopNetwork34446}
\end{figure}

\begin{figure}
\includegraphics[width=\linewidth]{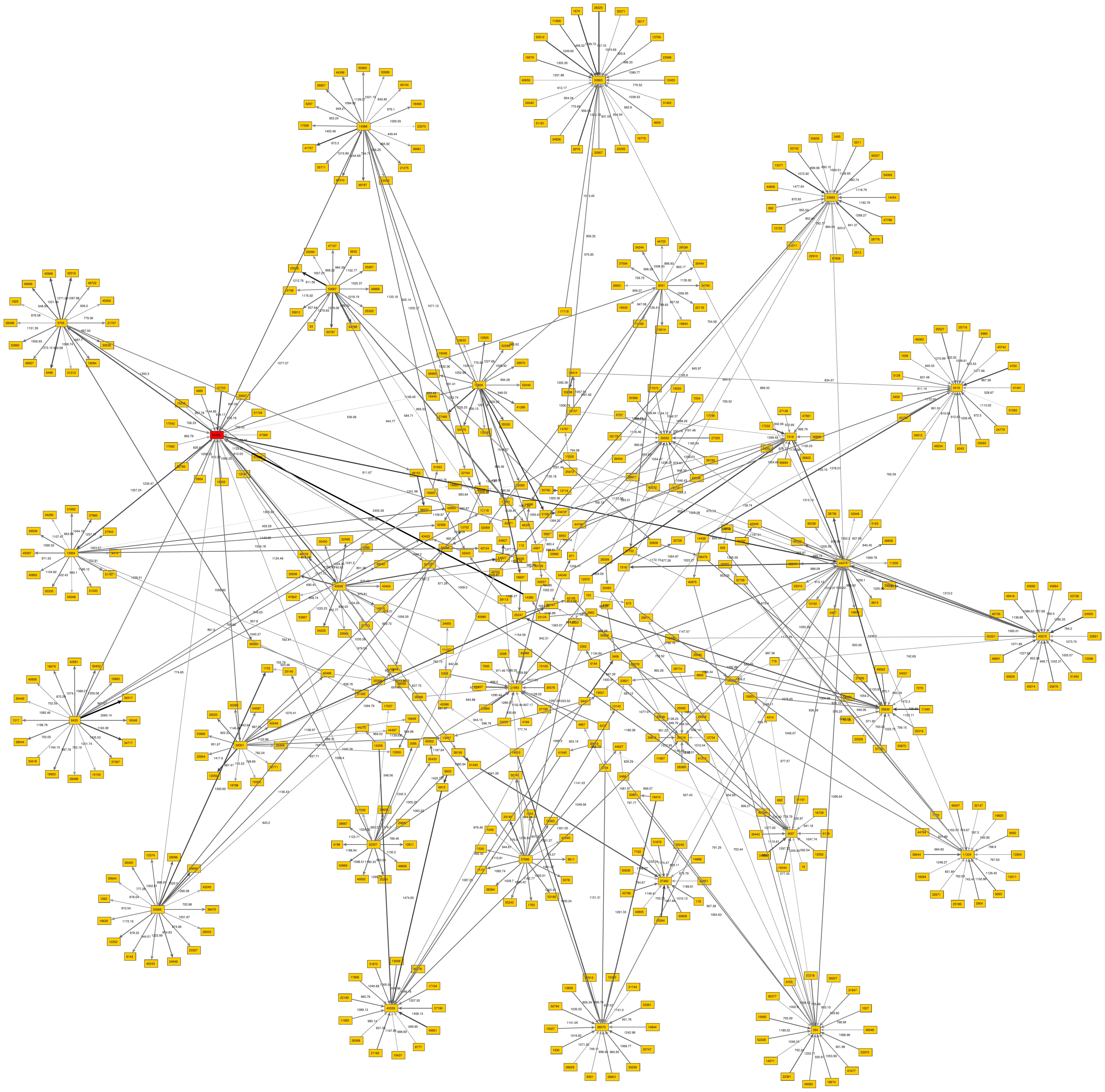}
\caption{Network plot of the 2 hop snowball sample (ignoring edge direction) of the node 35,493 in a realisation of the Accenture synthetic model. This node was selected as it is identified as anomalous in the random forest but was not planted in the network.}
\label{fig:2hopNetwork35493}
\end{figure}

\begin{figure}
\includegraphics[width=\linewidth]{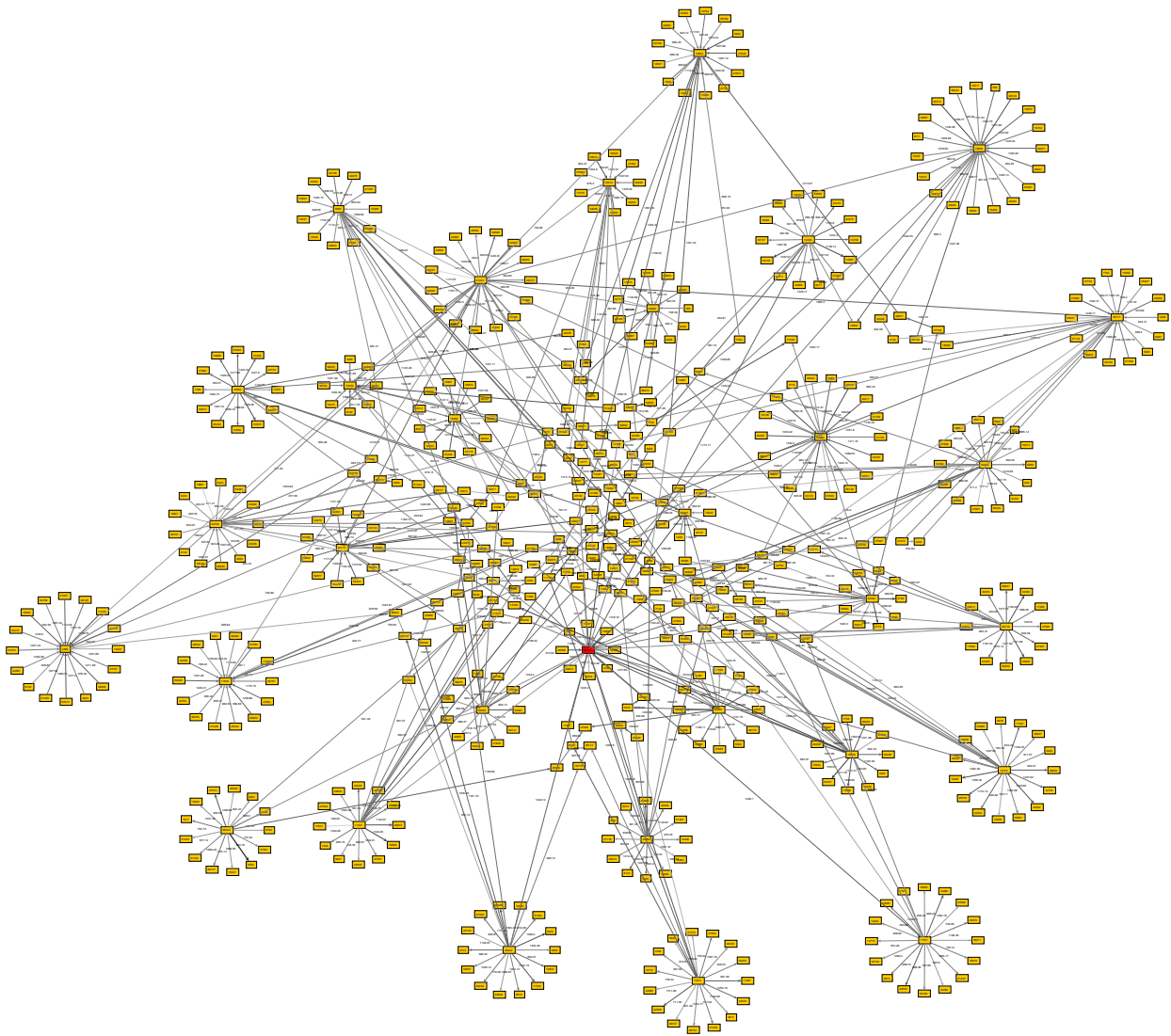}
\caption{Network plot of the 2 hop snowball sample (ignoring edge direction) of node 36,434 in a realisation of the Accenture synthetic model. This node was selected as it is identified as anomalous in the random forest but was not planted in the network.}
\label{fig:2hopNetwork36434}
\end{figure}

\setcounter{equation}{0}
\renewcommand\theequation{H.\arabic{equation}}
In the main body, we explored the ability of our method to detect anomalies that we did not embed. To this end, we visually inspected the nodes which are declared an anomaly by {\sc Random Forest} but were not part of a planted anomaly, in an example network generated by the synthetic network procedure constructed by Accenture. In this section, we present additional plots and statistics from this analysis. 
To that end, in Figures  
\ref{fig:2hopNetwork14708},
\ref{fig:2hopNetwork22745},
\ref{fig:2hopNetwork32251},
\ref{fig:2hopNetwork32545},
\ref{fig:2hopNetwork34446},
\ref{fig:2hopNetwork35493}
and
\ref{fig:2hopNetwork36434} 
we present the complete 2-hop snowball sampled plot for each network, rather than a restricted version presented in the main body. Finally, in Table~\ref{tab:nonSelectedStats} we present the statistics for the graphs.  

\ifnum\value{arXivVersion}>0 {
\begin{table}
\begin{center}
\begin{tabular}{|c|c|c | c| c| c|}
\hline
    \begin{tabular}{@{}c@{}}
    Node
    \end{tabular}
& 
    \begin{tabular}{@{}c@{}}
Total \\ Degree 
    \end{tabular}
& 
    \begin{tabular}{@{}c@{}}
Num. of \\ Heavy Edges 
    \end{tabular}
    &
    \begin{tabular}{@{}c@{}}
    Num. of \\ 
    Anom. nodes \\ 
    in 1 hop
    \end{tabular}
  & 
    \begin{tabular}{@{}c@{}}
    Num. of \\ Anom. nodes \\ in 2 hop
    \end{tabular}
    & 
    \begin{tabular}{@{}c@{}}
    Figure \\ name 
    \end{tabular}
\\
\hline
14708 &  43 & 4 & 0 & 0 & Fig.~\ref{fig:2hopNetwork14708} 
\\
\hline
22745 &  43 & 2 & 0 & 0  & Fig.~\ref{fig:2hopNetwork22745} 
\\
\hline
32251 &  31 & 4 & 0 & 0 & Fig.~\ref{fig:2hopNetwork32251} 
\\
\hline
32545 &  50 & 6 & 0 & 1 & Fig.~\ref{fig:2hopNetwork32545} 
\\
\hline
34446 &  42 & 3 & 0 & 1 & Fig.~\ref{fig:2hopNetwork34446} 
\\
\hline
35493 &  31 & 3 & 0 & 0 & Fig.~\ref{fig:2hopNetwork35493}  
\\
\hline
36434 &  33 & 2 & 0 & 0 & Fig.~\ref{fig:2hopNetwork36434} 
\\
\hline
\end{tabular}
\end{center}
\caption{Basic statistics for a subset of nodes that are identified as anomalous, but were not embedded in the Accenture network. Note that, in this context, heavy means a weight greater than $1400$ which is 2 standard deviations above the mean.} 
\label{tab:nonSelectedStats}
\end{table}
} \else {
\begin{table}
\begin{tabular}{|c|c|c | c| c| c|}
\hline
    \begin{tabular}{@{\hspace{-0.25cm}}c@{\hspace{-0.25cm}}}
    Node
    \end{tabular}
& 
    \begin{tabular}{@{\hspace{-0.25cm}}c@{\hspace{-0.25cm}}}
Total \\ Degree 
    \end{tabular}
& 
    \begin{tabular}{@{\hspace{-0.25cm}}c@{\hspace{-0.25cm}}}
Num. of \\ Heavy Edges 
    \end{tabular}
    &
    \begin{tabular}{@{\hspace{-0.25cm}}c@{\hspace{-0.25cm}}}
    Num. of \\ 
    Anom. nodes \\ 
    in 1 hop
    \end{tabular}
  & 
    \begin{tabular}{@{\hspace{-0.25cm}}c@{\hspace{-0.25cm}}}
    Num. of \\ Anom. nodes \\ in 2 hop
    \end{tabular}
    & 
    \begin{tabular}{@{\hspace{-0.25cm}}c@{\hspace{-0.25cm}}}
    Figure \\ name 
    \end{tabular}
\\
\hline
14708 &  43 & 4 & 0 & 0 & Fig.~\ref{fig:2hopNetwork14708} 
\\
\hline
22745 &  43 & 2 & 0 & 0  & Fig.~\ref{fig:2hopNetwork22745} 
\\
\hline
32251 &  31 & 4 & 0 & 0 & Fig.~\ref{fig:2hopNetwork32251} 
\\
\hline
32545 &  50 & 6 & 0 & 1 & Fig.~\ref{fig:2hopNetwork32545} 
\\
\hline
34446 &  42 & 3 & 0 & 1 & Fig.~\ref{fig:2hopNetwork34446} 
\\
\hline
35493 &  31 & 3 & 0 & 0 & Fig.~\ref{fig:2hopNetwork35493}  
\\
\hline
36434 &  33 & 2 & 0 & 0 & Fig.~\ref{fig:2hopNetwork36434} 
\\
\hline
\end{tabular}
\caption{Basic statistics for a subset of nodes that are identified as anomalous, but were not embedded in the Accenture network. Note that, in this context, heavy means a weight greater than $1400$ which is 2 standard deviations above the mean.} 
\label{tab:nonSelectedStats}
\end{table}
}\fi


\end{document}